\documentclass[preprint,aps,12pt,nofootinbib,tightenlines,amsmath,amssymb]{revtex4}
\usepackage{amsmath}
\usepackage{graphicx}
\usepackage{amssymb}
\usepackage{color}

\newcommand{\be}{\begin{eqnarray}}
\newcommand{\ee}{\end{eqnarray}}

\newcommand{\p}{\partial}

\newcommand{\dbar}{\lower0.1ex\hbox{$\mathchar'26$}\mkern-12mu d}
 \newcommand{\kbar}{\lower0.1ex\hbox{$\mathchar'26$}\mkern-10mu k}

\newcommand{\nn}{\nonumber}

\newcommand{\Tr}{\mathop{\rm Tr}\nolimits}
\newcommand{\diag}{\mathop{\rm diag}}
\newcommand{\cL}{{\mathcal L}}

\newcommand{\cO}{{\mathcal O}}
\newcommand{\cA}{{\mathcal A}}
\newcommand{\hcA}{ \hat{{\mathcal A}}}
\newcommand{\hcB}{ \hat{{\mathcal B}}}

\newcommand{\hcF}{ \hat{{\mathcal F}}}

\newcommand{\bsF}{\boldsymbol{F} }

\newcommand{\bs}{\boldsymbol}

\newcommand{\bscF}{ \boldsymbol{{\mathcal F}}}
\newcommand{\bscA}{\boldsymbol{{\mathcal A}}}
\newcommand{\bscD}{ \boldsymbol{{\mathcal D}}}

\newcommand{\bsD}{ \boldsymbol{D}}
\newcommand{\bshD}{ \boldsymbol{\hat{D}}}

\newcommand{\bsA}{ \boldsymbol{A}}
\newcommand{\bshA}{ \boldsymbol{\hat{A}}}

\newcommand{\bshF}{ \boldsymbol{\hat{F}}}

\newcommand{\hcW}{ \hat{{\mathcal W}}}
\newcommand{\hcV}{ \hat{{\mathcal V}}}

\newcommand{\hcS}{ \hat{{\mathcal S}}}

\newcommand{\bscM}{ \boldsymbol{{\mathcal M}}}

\newcommand{\hA}{\hat{A}}
\newcommand{\hphi}{\hat{\phi}}
\newcommand{\hB}{\hat{B}}

\newcommand{\bshcF}{ \boldsymbol{\hat{\mathcal F}}}
\newcommand{\bshcA}{\boldsymbol{\hat{\mathcal A}}}

\newcommand{\bshcW}{ \boldsymbol{\hat{\mathcal W}}}

\newcommand{\bshcG}{ \boldsymbol{\hat{\mathcal G}}}

\newcommand{\hF}{\hat{F}}

\newcommand{\cB}{\mathcal B }
\newcommand{\cF}{{\mathcal F}}

\newcommand{\hD}{\hat{D}}

\newcommand{\cM}{{\mathcal M}}

\newcommand{\sF}{{\mathsf F}}
\newcommand{\hsF}{{\hat{\sF}}}
\newcommand{\bshsF}{\boldsymbol{{\hat{\sF}}}}
\newcommand{\whsF}{{\widehat{\sF}}}

\newcommand{\cV}{{\mathcal V}}

\newcommand{\cD}{{\mathcal D}}

\newcommand{\hcD}{\hat{{\mathcal D}}}

\newcommand{\cG}{{\mathcal G}}
\newcommand{\fG}{{\mathbf G}}

\newcommand{\hcG}{\hat{{\mathcal G}}}

\newcommand{\whcG}{\widehat{{\mathcal G}}}

\newcommand{\mJ}{\mathrm J}
\newcommand{\fJ}{{\mathbf J}}

\newcommand{\hmJ}{\hat{\mJ}}

\newcommand{\whfJ}{\widehat{\fJ}}

\newcommand{\hT}{\hat{T}}

\newcommand{\cW}{{\mathcal W}}
\newcommand{\cS}{{\mathcal S}}

\newcommand{\bchi}{\bar{\chi} }

\newcommand{\wtjoin}{\,\widetilde{\Join}\, }

\newcommand{\1}{\mspace{1mu}}

\newcommand{\tGa}{\tilde{\Gamma }}
\newcommand{\hGa}{\hat{\Gamma }}

\newcommand{\hPhi}{\hat{\Phi}}

\newcommand{\mT}{\mathsf T}

\newcommand{\hR}{\hat{R}}

\newcommand{\adjoin}{\,\widetilde{\Join}\, }

\newcommand{\fM}{\mathbf M}

\newcommand{\fB}{\mathbf B}

\newcommand{\fE}{\mathbf E}

\newcommand{\mW}{\mathsf W}

\newcommand{\bM}{\bar{M}}

\newcommand{\mOm}{\mathsf \Omega }

\newcommand{\hmOm}{\hat{\mOm}}

\newcommand{\bshmOm}{\boldsymbol{\hmOm}}

\newcommand{\tla}{\tilde{\lambda}}
\newcommand{\hla}{\hat{\lambda}}

\newcommand{\chih}{\hat{\chi}}

\newcommand{\tchi}{\tilde{\chi}}

\newcommand{\hH}{\hat{H}}

\newcommand{\etat}{\tilde{\eta}}
\newcommand{\etab}{\bar{\eta}}

\newcommand{\etach}{\check{\eta}}

\newcommand{\ta}{\tilde{a}}
\newcommand{\tb}{\tilde{b}}
\newcommand{\tc}{\tilde{c}}
\newcommand{\td}{\tilde{d}}

\newcommand{\hW}{\hat{W}}

\newcommand{\hC}{\hat{C}}

\newcommand{\mG}{\mathsf{G}}

\newcommand{\cT}{\mathcal{T}}
\newcommand{\hcT}{\hat{\cT}}

\newcommand{\chib}{\bar{\chi}}

\begin{document}
\def\intdk{\int\frac{d^4k}{(2\pi)^4}}
\def\sla{\hspace{-0.22cm}\slash}
\hfill

%\begin{titlepage}

\title{Theoretical Foundations of the General Standard Model:  \\ A Unified Framework for Particle Physics and Cosmology}

\author{Yue-Liang Wu}\email{ylwu@itp.ac.cn, ylwu@ucas.ac.cn}
\affiliation{$^1$ Institute of Theoretical Physics, Chinese Academy of Sciences, Beijing 100190, China\\
$^2$ International Centre for Theoretical Physics Asia-Pacific (ICTP-AP), UCAS, Beijing 100190, China \\
$^3$ Taiji Laboratory for Gravitational Wave Universe (Beijing/Hangzhou), University of Chinese Academy of Sciences (UCAS), Beijing 100049, China \\
$^4$ School of Fundamental Physics and Mathematical Sciences, Hangzhou Institute for Advanced Study, UCAS, Hangzhou 310024, China }

%\date{\today}

\begin{abstract}
We present a comprehensive theoretical analysis of the General Standard Model (GSM), a recently proposed framework that unifies particle physics and cosmology within the Gravitational Quantum Field Theory (GQFT). Constructed from first principles based exclusively on the intrinsic properties of leptons and quarks, the GSM reveals an enlarged gauge symmetry structure, WS$_{c}$(1,3)$\times$GS(1)$\times$Z$_2$, which extends beyond the conventional U$_Y$(1)$\times$SU$_L$(2)$\times$SU$_C$(3) symmetry of the Standard Model. Here, WS$_{c}$(1,3) = SP(1,3)$\rtimes$W$^{1,3}\rtimes$SP$_c$(1,1) emerges as the conformal inhomogeneous spin gauge symmetry. Within GQFT, the GSM provides a consistent unification of the Standard Model of particle physics with cosmological models. It incorporates the four known fundamental interactions, electromagnetic, weak, strong, and gravitational, plus the Higgs scalar interaction, and also predicts novel interactions. These include spin gauge, chirality boost-spin gauge, chiral conformal-spin gauge, and scaling gauge forces, as well as additional scalar interactions. Furthermore, the GSM offers profound insights into the nature of gravity and spacetime and elucidates key mysteries of the dark side of the universe, such as the origins of dark matter, the dynamics of dark energy, and the physics of the early inflationary epoch. By establishing a new theoretical bridge between quantum field theory and general relativity, the GSM opens novel pathways for addressing long-standing challenges in fundamental physics. It provides a unified description of both fundamental interactions and cosmic evolution.
\end{abstract}

\maketitle

\begin{widetext}
\tableofcontents
\end{widetext}

\section{Introduction}

Quantum field theory (QFT)\cite{QFT1,QFT2a,QFT2b,QFT3a,QFT3b,QFT3c,QFT4a,QFT4b} has been successfully applied to describe the electromagnetic, weak, and strong interactions, governed by the gauge symmetries  U$_Y$(1) $\times$ SU$_L$(2) $\times$ SU$_C$(3) of known elementary particles (leptons and quarks), forming the Standard Model (SM) of particle physics\cite{SM1,SM2,SM3,SM4,SM5,SM6}. As a theoretical framework, QFT was developed based on relativistic quantum mechanics\cite{RQM} that unifies quantum mechanics\cite{QM1,QM2,QM3} and special relativity\cite{SR}, treating elementary particles as excitations of underlying quantum fields. This perspective supersedes Newton’s point-like particle hypothesis, which treated matter as dimensionless particles without internal structure. In QFT, elementary particles emerge as quanta of fundamental fields, each possessing intrinsic properties characterized by quantum numbers (e.g., electric charge, isospin, color-spin charge) that dictate their interactions under gauge symmetries\cite{GS1,GS2}, giving rise to the electroweak and strong forces in the SM.

By unifying space and time into a four-dimensional spacetime, Einstein's theories of special and general relativity \cite{SR, GR, FGR} provided a revolutionary framework for physics. General relativity(GR), in particular, replaced Newton's postulate of instantaneous action-at-a-distance with a description of gravity as the curvature of spacetime induced by mass and energy. Although this successfully geometrized the gravitational interaction, it did not address the constitution of matter, as it retained Newton's paradigm of point-like particles. Thus, the nature of matter itself lies beyond the explanatory reach of GR.

QFT provides an exceptionally successful framework for describing microscopic phenomena, while GR offers a powerful description of the macroscopic world, governing the dynamics of massive objects through the curvature of spacetime. A fundamental incompatibility arises, however, when attempting to merge these theories, due to their divergent foundations: QFT is built upon quantum fields with intrinsic structure, whereas GR relies on the concept of dimensionless point particles within a classical curved geometry.

GQFT\cite{GQFT1,GQFT2,GQFT3} resolves this tension by extending the gauge principle to gravity through the spin symmetry group SP(1,3) of elementary particles, treating gravitational interactions on equal footing with electroweak and strong forces. In GQFT, the reconciliation of GR and QFT hinges on a key principle: the laws of nature are governed by the intrinsic properties of matter’s fundamental constituents. This necessitates distinguishing between intrinsic symmetries, determined by quantum numbers of elementary particles as quantum fields, and external symmetries, describing particle motion in flat Minkowski spacetime. To reconcile these, GQFT introduces a bi-frame spacetime with a fiber bundle structure. In this framework, the base spacetime is a globally flat Minkowski spacetime, describing the free motion of basic fields, while the fiber is a local orthogonal gravigauge spacetime, governing their dynamic interactions.

This framework highlights the separation between external and intrinsic symmetries. The global Lorentz symmetry SO(1,3) and the intrinsic spin symmetry SP(1,3) unify as joint symmetries SO(1,3)$\Join$SP(1,3), with SP(1,3) localized as a gauge symmetry. To preserve the joint symmetries, a gravigauge field $\chi_{\mu}^{\; \, a}(x)$, as an invertible spin-associated bi-covariant field, is introduced to replace GR’s metric field. 

GQFT exhibits several key features that distinguish it from GR. The gravigauge field $\chi_{\mu}^{\; a}(x)$ behaves as a Goldstone-type boson, identified as a massless graviton. A spin-gauge invariant action involving this field and the spin gauge field $\cA_{\mu}^{ab}$ introduces non-geometric interactions that go beyond GR's purely geometric description, thereby invalidating the equivalence principle proposed by Einstein in GR.

The dynamics governed by $\chi_{\mu}^{\; a}(x)$ lead to a theory of gravidynamics that predicts five transverse polarizations for gravitational waves\cite{GQFT4}, in contrast to the two predicted by GR. These additional spin-0 transverse scalar and spin-1 transverse vector modes violate GR’s strong equivalence principle, providing a concrete, testable distinction between the two theories\cite{GQFT5}. The physical origin of these polarizations, linked to the spin properties of leptons and quarks, has been analyzed in detail, including their potential generation and detection\cite{GQFT6}. 

Furthermore, the general linear group symmetry GL(1,3,R), which underlies GR, emerges as a hidden symmetry within GQFT. This reveals that GQFT is a background-independent theory. A significant conceptual advancement is its consistent allowance for a flat Minkowski spacetime as a base manifold while preserving the full Poincar\'e symmetry PO(1,3)$\equiv$ P$^{1,3}\ltimes$SP(1,3). Within this framework, the translation subgroup symmetry P$^{1,3}$ gives rise to a zero energy-momentum tensor theorem via a cancellation law \cite{GQFT3}, constituting a formal extension of the conventional conservation law in QFT. Recently, the dynamics of a massive Dirac field outside a Schwarzschild black hole were revisited\cite{GQFT7} within the GQFT framework. This approach enabled an analytical solution to the coupled first-order Dirac equation and yielded fermionic quasibound states around Schwarzschild black holes with an improved analytic spectrum.

In our recent work \cite{GSM}, we developed a General Theory of the Standard Model (GSM), which offers a unified description of all fundamental forces within the framework of GQFT. The GSM successfully integrates the two established standard models in particle physics and cosmology, providing comprehensive insights into both the nature of gravity and the mysteries of the universe’s dark sector, including dark matter, dark energy, and the inflationary early universe.

Notably, these theoretical developments have implications for interpreting experimental data. 
The search for a gravitational-wave background in the NANOGrav 15-year dataset\cite{ST1,ST2} revealed strong evidence for correlations but did not decisively distinguish between quadrupolar (Hellings-Downs) and scalar-transverse signatures. The data offer a comparable fit to both tensor and scalar-transverse correlation models. This result suggests that scalar-transverse mode provides a competitive explanation for the observed signal, a scenario naturally predicted by GQFT, which incorporates precisely such an additional scalar-transverse polarization.

GQFT provides a unified framework for describing gravitational, electroweak, and strong interactions. This has motivated the development of the hyperunified field theory (HUFT) (see Refs. \cite{HUFT1,HUFT2,FHUFT1,FHUFT2,FHUFT} and references therein), which incorporates all fundamental interactions and elementary particles. HUFT requires a nineteen-dimensional hyper-spacetime, whose dimensionality is determined by the quantum numbers associated with the independent degrees of freedom of the fundamental building blocks.

This paper is going to present a detailed investigation of how gravitational and cosmological phenomena naturally emerge within the SM of particle physics through the maximal internal gauge symmetries of leptons and quarks. Our systematic analysis proceeds as follows: after a brief overview of the theoretical framework and key objectives in sect.1, we reformulate, in sect.2, the SM in a left-right symmetric chiral representation, revealing a conformal inhomogeneous spin symmetry, WS$_c$(1,3) = SP(1,3)$\rtimes$W$^{1,3}$$\rtimes$SP$_c$(1,1), and a chiral duality symmetry $Z_2$. Here, WS$_c$(1,3) represents a semi-direct product group combining the spin group SP(1,3) and translation-like chirality boost-spin group W$^{1,3}$ as well as chiral conformal-spin symmetry SP$_c$(1,1). In this representation, neutrinos become massive. In sect. 3, guided by the GQFT principle that nature's laws drive from the intrinsic properties of matter's fundamental constituents, we localize the conformal inhomogeneous spin symmetry WS$_c$(1,3) and scaling symmetry SG(1) as gauge symmetries. Alongside the gauge invariance principle, we introduce gauge fields with associated gravigauge field, which characterize novel gauge interactions along with the gravitational interaction. A local orthogonal gravigauge spacetime is constructed with dual gravigauge bases, yielding a spin-associated intrinsic non-commutative spacetime. In sect. 4, we explicitly construct a gauge-invariant and chiral-duality-invariant action of GSM in gravigauge spacetime. A constraint equation for gravigauge field strength is obtained as a gravitization equation in GSM, which demonstrates how the gravigauge field strength emerges from collective gauge field dynamics. We then derive, in sect. 5, the general gauge-type gravitational equation for the gravigauge field to describe the gravidynamics of GSM within the GQFT framework. By applying for the translational invariance in Minkowski spacetime, we establish, in sect. 6, cancellation laws for the energy-momentum tensor in GQFT instead of the conservation low in QFT, and prove the zero energy-momentum tensor theorem in GSM. In sect. 7, we present detailed gauge fixing conditions for WS$_c$(1,3)$\times$SG(1) gauge symmetries, and make a comprehensive analysis of mass generation mechanisms via spontaneous symmetry breaking. In sect. 8, we identify the chirality boost-spin gauge field as a stable massive dark graviton, serving as a dark matter candidate. Its interactions with all leptons and quarks are mediated via heavy spin gauge field. The dark gravitons get effective self-interactions through spin gauge field and chiral conformal-spin gauge field as well as inflaton scalar field. In sect.9, we explore cosmological implications.  An inflaton scalar field whose potential satisfies slow-roll conditions provides a primordial potential energy for early universe inflation. A cosmological constant is generated by a second scalar field referred to as dark cosmino. Such a dark cosmino yields a tiny cosmic mass, serving as canonical dark energy. Our conclusions and discussion are presented in the last section.

\section{Chiral Duality and Conformal Inhomogeneous Spin Symmetry in the Standard Model}

In the SM built within the framework of QFT, the electroweak and strong interactions among leptons and quarks are characterized by the gauge symmetries,
\be
\fG_{SM} = U_Y(1)\times SU_L(2)\times SU_C(3) ,
\ee
and the four-dimensional spacetime and four-component spinor field describing the motion and intrinsic spin of leptons and quarks possess the associated symmetry,
\be
G_{SM} = SO(1,3) \adjoin SP(1,3) ,
\ee 
where it requires that the global transformations under the group symmetry SO(1,3) for coordinate system and the group symmetry SP(1,3) for spinor field must be made coincidentally to preserve the so-called Lorentz symmetry in special relativity. This lays the foundation of relativistic quantum mechanics\cite{RQM}, which successfully unites special relativity with quantum mechanics via the so-called Lorentz symmetry, without explicitly distinguishing between these two group structures. Although SO(1,3) and SP(1,3) are isomorphic groups (with the latter being the double cover of the former), it is important to denote them separately using distinct notations. This distinction emphasizes that they act on different spaces: one acts on coordinate spacetime, and the other acts on the Hilbert space of spinor fields, which has been shown to be essential to establish the GQFT\cite{GQFT1,GQFT2,GQFT3}.

The leptons and quarks in SM are all Weyl fermions due to maximal parity violation\cite{PV1,PV2,VA1,VA2} in the weak interaction governed by the gauge symmetry SU$_L$(2). On the other hand, the charged leptons and quarks behave as Dirac fermions\cite{RQM} in the electromagnetic interactions, where the left-handed and right-handed components appear to be symmetric and their superposition forms a Dirac spinor. Therefore, it is natural to deal with the left-handed and right-handed leptons and quarks on the same footing. For that, it is useful to  introduce the following conventional chiral spinor representations for leptons and quarks:
\be \label{CR1}
\Psi_{-}^{i} & \equiv & \binom{\Psi_{L}^{i}}{\Psi_{R}^{i}} \quad \mbox{or} \quad  \Psi_{+}^{i}\equiv \binom{\Psi_{R}^{i}}{\Psi_{L}^{i}}  , 
\ee
where $\Psi_{-}^i$ and $\Psi_{+}^i$ denote two equivalent chirality representations of leptons and quarks, $\Psi_{\mp}^i \equiv l_{\mp}^i$, $q_{\mp}^i$, and $\Psi_{L,R}^{i}$ represent the doublets of left-handed and right-handed leptons and quarks, $\Psi_{L,R}^{i}=  l_{L,R}^{i}, q_{L,R}^{i }$. Explicitly, we have, 
\be \label{CR2}
\Psi_{-}^{i} & : & l_{-}^i \equiv \binom{l_{L}^{i}}{l_{R}^{i}}, \;\; q_{-}^i \equiv \binom{q_{L}^{i}}{q_{R}^{i}},  \nn \\
\Psi_{+}^{i} & : &  l_{+}^i \equiv \binom{l_{R}^{i}}{l_{L}^{i}}, \;\; q_{+}^i \equiv \binom{q_{R}^{i}}{q_{L}^{i}},  , \nn \\
\Psi_{L,R}^{i} & : &  l_{L,R}^{i} \equiv \binom{\nu_{L,R}^{i}}{e_{L,R}^{i }} ,  \;\;  q_{L,R}^{i }\equiv \binom{u_{L,R}^{i}}{d_{L,R}^{i}}  , 
\ee
where $\psi_{L,R}^{i} \equiv \nu_{L,R}^{i}, e_{L,R}^{i}, u_{L,R}^{i}, d_{L,R}^{i}$ with $i=1,2,3$ denoting three families of left-handed and right-handed neutrinos, charged leptons, up-type quarks, and down-type quarks:
\be
\psi_{L,R}^{i} \equiv \gamma_{\mp} \psi^{i} \equiv \frac{1}{2}(1 \mp \gamma_5) \psi^{i} ,
\ee
where $\gamma_5$ matrix is defined in four dimensional spinor representations with $\gamma_5^2 =1$. 

The definition of two 16-dimensional chirality representations assigned to leptons and quarks in Eqs. (\ref{CR1}) and (\ref{CR2}) is merely a matter of convention, they are distinguished just from their opposite chirality:
\be
\gamma_9 \Psi_{\pm}^{i} = \pm \Psi_{\pm}^{i}  ,\;\; \gamma^9 =\sigma_3\otimes \sigma_0 \otimes\gamma_5,
\ee
with chirality $\gamma$-matrix $\gamma^9$ defined in sixteen dimensional spinor representation. The negative and positive chirality representations are interconnected through a chiral duality operation, denoted as ${\cal C}_{d}$, i.e.:
\be
& & \Psi_{\mp}^{c_d} \equiv {\cal C}_{d} \Psi_{\mp} {\cal C}_{d}^{-1} = C_d\Psi_{\mp} = \Psi_{\pm}  , \nn \\
& & C_d = \sigma_1\otimes \sigma_0 \otimes \sigma_0 \otimes \sigma_0 .
\ee
Where $\sigma_i$($i=1,2,3$) are Pauli matrices and $\sigma_0$ denotes the $2\times 2$ unit matrix. 

The fundamental laws of nature ought to be unaffected by the specific selection of the two chirality representations, $\Psi_{-}^i$ and $\Psi_{+}^i$, which brings on the principle of chirality independence. Namely, the two chirality representations should be equivalent. This chirality independence principle demands the invariance of theory under the chiral duality operation ${\cal C}_{d}$. Namely, the theory should be invariant under the exchange between two chirality representations,
\be
Z_2 : \quad \Psi_{-}^i \leftrightarrow \Psi_{+}^i, \;\;  \mbox{i.e.} \;\; {\cal C}_{d}  \left(\Psi_{\mp}^{c_d}\right) {\cal C}_{d}^{-1} = \Psi_{\mp} , 
\ee
which exhibits a $Z_2$ discrete symmetry due to $C_d^2 = 1$. 

For our current purpose, let us first reformulate the SM in light of the left-right symmetric chiral spinor representations of leptons and quarks. Based on the  gauge symmetries U$_Y$(1) $\times$ SU$_L$(2) $\times$ SU$_C$(3), and in terms of the equivalent chirality representations of leptons and quarks, we are able to express the SM action into the following chiral duality invariant formalism:
\be  \label{SMaction}
\cS_{SM} & = &\int [dx]  \frac{1}{2} \sum_{s=\mp}  \{  \sum_{\Psi=l,q}  [ \frac{1}{2} \bar{\Psi}_{s}^{i} \Sigma_{s}^{a}   \delta_{a}^{\; \mu} i D_{\mu}^{(\Psi_{s})}\, \Psi_{s}^{i}  + H.c. 
 \nn \\
& + & \bar{\Psi}_{s}^{i} \Phi_{s} ( \tilde{\Gamma}_{s} \tilde{\lambda}^{\Psi}_{ij}  +  \hGa_{s}  \hat{\lambda}^{\Psi}_{ij} ) \, \Psi_{s}^{j} ] + \frac{1}{8} \eta^{\mu\nu} \Tr (\cD_{\mu}\Phi_{s} )^{\dagger} \cD_{\nu}\Phi_{s} \nn \\
& - & \frac{1}{4} \eta^{\mu\mu'}\eta^{\nu\nu'} ( \frac{1}{2g_1^2} \Tr F_{\mu\nu}^{(l_{s})} F_{\mu'\nu'}^{(l_{s})}  + \frac{1}{2g_2^2} \Tr F_{\mu\nu}^{(L_{s})} F_{\mu'\nu'}^{(L_{s})} )  \} \nn \\
& - & \eta^{\mu\mu'}\eta^{\nu\nu'}  \frac{1}{g_3^2} \Tr F_{\mu\nu}^{(C)} F_{\mu'\nu'}^{(C)}  -  \cV_{h}(\Phi_{-}, \Phi_{+}) ,
\ee
where $\delta_{a}^{\;\;\mu} $ is the Kronecker symbol with Greek alphabet ($\mu,\nu = 0,1,2,3$) and Latin alphabet ($a,b,=0,1,2,3$) introduced to distinguish four-vector indices in coordinate spacetime and spin-associated intrinsic spacetime, respectively. Both the Greek and Latin indices are raised and lowered as well as contracted by the constant metric matrices $\eta^{\mu\nu} $ and $\eta^{ab}$ with $\eta^{\mu\nu}$($\eta_{\mu\nu}$)=$\diag.$(1,-1,-1,-1) and $\eta^{ab}$($\eta_{ab}$)=$\diag.$(1,-1,-1,-1). The summations over $s=\mp$ and $\Psi=l,q$ are in respective to the negative and positive chiral duality representations and the spinor fields of leptons and quarks. 

The alternative formalism for the SM presented above can be reduced to its standard form through straightforward algebraic operations. This requires the explicit definitions of the covariant derivatives, gauge fields and their corresponding group generators, the 16-dimensional gamma matrices obeying the Clifford algebra, and the Yukawa coupling matrices, which are provided below. Relevant algebraic techniques are elaborated upon in the supplementary materials.

In the above chiral duality invariant formulation of SM action, we have introduced the following definitions for $\Sigma$-matrices $\Sigma_{\mp}^{A}\equiv(\Sigma_{\mp}^a , \Sigma_{\mp}^{I})$ ($a=0,1,2,3, I=5,6,7,8$) with projection matrices $\Gamma_{\mp}$, $\tilde{\Gamma}_{\mp}$ and $\hat{\Gamma}_{\mp}$,
\be 
& & \Sigma_{\mp}^A =  \Gamma^{A} \Gamma_{\mp}, \quad \Gamma_{\mp} = \frac{1}{2} ( 1 \mp\gamma_9) , \nn \\
& & \tilde{\Gamma}_{\mp} = \frac{1}{2} ( 1 \mp \tilde{\gamma}_9) , \quad \hat{\Gamma}_{\mp} =  \hat{\gamma}_9 \tilde{\Gamma}_{\mp} , 
\ee
where the eight $\Gamma$-matrices $\Gamma^{A}=(\Gamma^a , \Gamma^{I})$ ($a=0,1,2,3, I=5,6,7,8$) in the 16-dimensional spinor representation are explicitly given by, 
\be
& & \Gamma^{a} = \, \sigma_0 \otimes \sigma_0 \otimes \gamma^{a} ,  \; \;  \Gamma^5 =  i\sigma_2\otimes \sigma_0\otimes \gamma_5 , \nn\\
& &  \Gamma^6 = i\sigma_1\otimes \sigma_1 \otimes\gamma_5, \;\;  \gamma^9 = \sigma_3\otimes \sigma_0 \otimes\gamma_5, \nn \\
& & \Gamma^7 = i\sigma_1\otimes \sigma_2 \otimes\gamma_5,\;\; \tilde{\gamma}_9 =  \sigma_3\otimes \sigma_0 \otimes I_4, \nn \\
& & \Gamma^8 = i\sigma_1\otimes \sigma_3 \otimes\gamma_5, \;\;  \hat{\gamma}_9 = \sigma_0\otimes \sigma_3 \otimes I_4, 
\ee
with $\gamma^{a} $ and $\gamma_5$ being Dirac-type $\gamma$-matrices. The eight $\Gamma$-matrices $\Gamma^{A}$ form a Clifford algebra and satisfy the anti-commutation relations:
\be
\{ \Gamma^{A}, \Gamma^{B} \} = 2 \eta^{AB}, \;\; \{ \Gamma^{A}, \gamma^9 \} = 0,\;\; 
\{ \Gamma^{I}, \tilde{\gamma}^9 \} = 0.
\ee

The covariant derivatives $D_{\mu}^{(\Psi_{\mp})}$ ($\Psi=l, q$) for leptons and quarks are defined as follows:
\be \label{GFSM}
& & i D_{\mu}^{(\Psi_{\mp})} \equiv i\p_{\mu} + A_{\mu} ^{(\Psi_{\mp})} , \nn \\
& &  A_{\mu} ^{(l_{\mp})}\equiv B_{\mu} ^{(l_{\mp})} + W_{\mu}^{(L_{\mp})} \equiv  B_{\mu} \Sigma_{Y \mp}^{(l)} + W_{\mu}^{i} \Sigma_{L \mp}^i , \nn \\
& & A_{\mu} ^{(q_{\mp})}\equiv B_{\mu} ^{(q_{\mp})} + W_{\mu}^{(L_{\mp})} + A_{\mu}^{(C)} \equiv B_{\mu} \Sigma_{Y \mp}^{(q)} + W_{\mu}^{i} \Sigma_{L\mp}^i + A_{\mu}^{\alpha} T^{\alpha}  .
\ee
The gauge fields $B_{\mu}$, $W_{\mu}^{i}$ ($i=1,2,3$) and $A_{\mu}^{\alpha}$ ($\alpha=1,\cdots,8$) are assigned to the electroweak and strong gauge symmetries U$_Y$(1) $\times$ SU$_L$(2) $\times$ SU$_C$(3). Their field strengths are defined as follows:
\be \label{SMGFS}
& & F_{\mu\nu}^{(\Psi_{\mp})} \equiv F_{\mu\nu}\Sigma_{Y\mp}^{(\Psi)}, \quad F_{\mu\nu} \equiv \p_{\mu}B_{\nu} -  \p_{\nu}B_{\mu},   \nn \\
& & F_{\mu\nu}^{(L_{\mp})} \equiv F_{\mu\nu}^{i}\Sigma_{L\mp}^i, \quad F_{\mu\nu}^{i} \equiv \p_{\mu}W_{\nu}^i -  \p_{\nu}W_{\mu}^i + \epsilon^{ijk}W_{\mu}^jW_{\nu}^k  ,\nn \\
& & F_{\mu\nu}^{(C)} \equiv F_{\mu\nu}^{\alpha}T^{\alpha}, \quad F_{\mu\nu}^{\alpha} \equiv \p_{\mu}A_{\nu}^{\alpha} -  \p_{\nu}A_{\mu}^{\alpha} + C^{\alpha\beta\gamma}A_{\mu}^{\beta}A_{\nu}^{\gamma}, 
\ee
where $F_{\mu\nu}$, $F_{\mu\nu}^{i}$ and $F_{\mu\nu}^{\alpha}$ correspond to the field strengths of electromagnetic gauge ($B_{\mu}$), weak gauge ($W_{\mu}^i$) and strong gauge ($A_{\mu}^{\alpha}$) interactions.

The associated group generators of U$_Y$(1)$\times$SU$_L$(2)$\times$SU$_C$(3) are given by:
\be
& & \Sigma_{Y \mp}^{(\Psi)} \equiv \frac{1}{2}N^{(\Psi)} +  \frac{1}{2}\hat{\Gamma}_{\mp} , \quad N^{(l)}= -1,\;  N^{(q)}=\frac{1}{3}, \nn \\
& & \Sigma_{L\mp}^i \equiv \Sigma^i  \tilde{\Gamma}_{\pm} , \;\; \Sigma^i = \epsilon^{iJK}\Sigma^{JK},  \; \Sigma^{JK} = \frac{i}{4}[ \Gamma^J, \Gamma^K ] , \nn \\
& & T^{\alpha}= \lambda^{\alpha}/2, \quad [\lambda^{\alpha}, \lambda^{\beta}] = 2i C^{\alpha\beta \gamma} \lambda^{\gamma}, \;\; \Tr  \lambda^{\alpha} \lambda^{\beta} = \delta^{\alpha\beta}, 
\ee
with $i\equiv I-5$ and $I, J,K = 6,7,8$. 

The chiral project operators $\Gamma_{\mp}$, $\tilde{\Gamma}_{\mp}$ and $\hat{\Gamma}_{\mp}$ defined in sixteen-dimensional spinor representation are used to produce gauge symmetries and Yukawa couplings in the SM. This is because we are not going to extend the SM to be left-right symmetric gauge model, which is beyond the scope of our present consideration. Namely, the electroweak gauge interactions remain the same as in SM governed by the gauge symmetries SU$_L$(2)$\times$U$_Y$(1), although the right-handed leptons and quarks are formally expressed as doublets in the chiral spinor representations. Actually, the U$_Y$(1) symmetry in SM is a subgroup of SU$_R$(2)$\times$U$_{B-L}$(1).  

The covariant derivative of scalar field $\Phi_{\mp}$ is given by 
\be
& & \cD_{\mu}\Phi_{\mp} \equiv (\p_{\mu} - i B_{\mu} \tGa_{\pm}/2  - i W_{\mu}^{i}\Sigma_{L \mp}^{i} ) \Phi_{\mp} , 
\nn \\
& &  \frac{1}{8} \eta^{\mu\nu} \Tr (\cD_{\mu}\Phi_{\mp} )^{\dagger} \cD_{\nu}\Phi_{\mp}  = \eta^{\mu\nu} (\cD_{\mu}H)^{\dagger} \cD_{\nu}H ,
\ee
with $\Phi_{\mp}$ defined as follows:
\be
\Phi_{\mp} \equiv \phi_I \Sigma_{h \mp}^{I} , \quad \Sigma_{h -}^{I}\equiv \Gamma^{I}\Gamma_{-}, \; \Sigma_{h +}^{I}\equiv -(\Gamma^{I})^{\ast}\Gamma_{+},
\ee
where the four scalar fields $\phi_{I}$ ($I=5,6,7,8$) constitute Higgs doublet in SM via the following relations:
\be \label{HDSM}
H = \binom{H^+}{H^0} , \quad H^{+} = \phi_7 + i \phi_6  , \quad H^{0} = \phi_5 - i \phi_8 . 
\ee 
The effective Higgs potential of SM is presented as follows:
\be \label{HPSM}
\cV_{h}(\Phi_{-}, \Phi_{+}) \equiv \frac{1}{4} \lambda_h^2 [ \frac{1}{16} \Tr (\Phi_{-}^{\dagger}\Phi_{-} + \Phi_{+}^{\dagger}\Phi_{+}) - v_{h}^2 ]^2 
= \frac{1}{4} \lambda_h^2 ( H^{\dagger}H - v_{h}^2 )^2 ,
\ee
with $v_{h}$ the vacuum expectation value of neutral Higgs boson $H^0$ and $\lambda_h$ the coupling constant. 

The chiral duality invariance is maintained via the following relations:
\be
D_{\mu}^{(\Psi_{\mp})c_d}\equiv C_{d}D_{\mu}^{(\Psi_{\mp})} C_{d}^{-1}  \equiv D_{\mu}^{(\Psi_{\pm})}, \quad \Phi_{\mp}^{c_d}\equiv C_{d}\Phi_{\mp} C_{d}^{-1}  \equiv \Phi_{\pm} .
\ee

It is noticed that the right-handed neutrinos are introduced and treated equally to the left-handed neutrinos as the neutrinos are considered to be massive due to the observations of neutrino oscillations, which is already beyond the SM. Just in SM, only three families of chirality leptons and quarks are introduced, which is indicated from the present experiments around TeV energy scales.

The Yukawa couplings of scalars concern four Yukawa coupling matrices $\hla^{l}_{ij}$, $\tla^{l}_{ij}$, and $\hla^{q}_{ij}$, $\tla^{q}_{ij}$ in association to the leptons and quarks. Their combinations give rise to four mass matrices in correspondence to the neutrinos and charged leptons as well as up-type and down-type quarks,
\be
& & \lambda^{\nu} = \tla^{l} + \hla^{l}, \quad \lambda^e = \tla^{l} - \hla^{l} , \nn \\
& &  \lambda^{u} = \tla^{q} + \hla^{q}, \quad \lambda^d = \tla^{q} - \hla^{q}  .
\ee  

Based on the above reformulated chiral duality invariant SM action, we arrive at the following observations: 

(i) The neutrinos become massive as the Yukawa coupling matrix of neutrinos is in general no longer zero;

(ii) All four Yukawa coupling matrices are hermitian matrices,
\be
& & \tla^{\Psi} = (\tla^{\Psi})^{\dagger}, \quad  \hla^{\Psi} = (\hla^{\Psi})^{\dagger},  \quad \Psi= (l, q), \nn \\
& & \lambda^{l} = (\lambda^{l})^{\dagger}, \quad \lambda^{q} = (\lambda^{q})^{\dagger}, \quad l=(\nu, e), \, q=(u, d) .
\ee

These properties differ from those in the SM, where neutrinos are massless and the Yukawa coupling matrices are generally arbitrary complex matrices rather than Hermitian. Specifically, each complex Yukawa coupling matrix in the SM contains 18 parameters, twice as many as the Hermitian matrices discussed here.

It is well known that a complex Yukawa coupling matrix is diagonalized via bi-unitary transformations (using separate left-handed and right-handed unitary matrices), whereas a Hermitian Yukawa coupling matrix has only 9 parameters and is diagonalized by a single unitary matrix.

Notably, the Hermitian nature of the Yukawa coupling matrices implies that the strong CP-violating phase can safely be set to a small value, as it no longer receives corrections from weak CP-violating phases. This is because Hermitian Yukawa matrices are diagonalized by the same unitary transformation for both left-handed and right-handed quarks, unlike the arbitrary complex Yukawa matrices in the conventional SM framework.

(iii) By normalizing the kinetic terms of the electromagnetic and weak gauge fields using the gauge coupling constants 
$g'$ and $g$ as defined in the SM, 
\be
& & B_{\mu} \to g' B_{\mu}, \quad W_{\mu}^{i} \to g W_{\mu}^{i} , 
\ee 
we relate them to the gauge couplings $g_1$ and $g_2$ via the following relations:
\be
 g^{' 2} = g_1^2/3,\quad g^2 = g_2^2 .
 \ee
 
It should be noted that these relations are not an assumption. The couplings $g'$ and $g$ are the standard SM gauge couplings associated with the hypercharge gauge symmetry U$_Y$(1) and the left-handed weak isospin symmetry SU$_L$(2), respectively. In contrast, $g_1$ and $g_2$ are the gauge couplings for the same symmetries, but with the group generators expressed in terms of the 16-dimensional gamma matrix algebra, following a unified definition. The relations above thus reflect the unified description of the electroweak theory. This supports the postulate that at some high energy scale, the hypercharge U$_Y$(1) and weak isospin SU$_L$(2) symmetries unify, i.e. $g_1 = g_2$. Thus we are led to the following prediction for the weak-mixing angle $\theta_w$ by taking $g_1 = g_2$ at a unified energy scale:
\be
\sin^2\theta_w = \frac{g^{' 2}}{g^2 + g^{' 2}} =  \frac{g_1^{2}/3}{g_2^2 + g_1^{2}/3} = \frac{1}{4} = 0.25 ,
\ee
which is comparable to the effective weak-mixing angle $\sin^2\theta = 0.23153$\cite{PDG} at low energies. This implies that the chiral duality invariant formulation of SM may be regarded as a framework of unified electroweak interaction for leptons.

(iv) The chiral duality invariant formulation of SM allows to make a chiral scaling transformation for the chiral representations of leptons and quarks, i.e.:
\be
\Psi_{\mp}(x) \to \Psi_{\mp}'(x) = e^{\pm i\varpi\Gamma_9/2} \Psi_{\mp}(x) , 
\ee 
which is viewed as a transformation of chiral conformal-spin group symmetry SP$_{c}$(1,1) with group generator $\Gamma_9$. 

(v) It is interesting to demonstrate that in addition to the usual gauge symmetries U$_Y$(1)$\times$ SU$_L$(2)$\times$SU$_C$(3) in SM, the chiral duality invariant formulation of SM in Eq.(\ref{SMaction}) possesses the following associated global symmetry:
\be \label{AS}
\mG_{GSM} & = &  \mbox{SC}(1)\times \mbox{PO}(1,3) \adjoin \mbox{WS}_c(1,3) \times \mbox{SG}(1) \times Z_2, \nn \\
& \equiv & \mbox{SC}(1)\times \mbox{P}^{1,3}\ltimes \mbox{SO}(1,3) \adjoin \mbox{SP}(1,3)\rtimes \mbox{W}^{1,3} \rtimes \mbox{SP}_c(1,1) \times \mbox{SG}(1) \times Z_2, 
\ee
with the notations,
\be
& & \mbox{PO}(1,3) \equiv  \mbox{P}^{1,3}\ltimes \mbox{SO}(1,3)  , \nn \\
& &  \mbox{WS}_c(1,3) \equiv \mbox{SP}(1,3)\rtimes \mbox{W}^{1,3}\rtimes \mbox{SP}_c(1,1) .
\ee
Here, PO(1,3) represents Poincar\'e group (inhomogeneous Lorentz group) symmetry in Minkowski spacetime with rotational Lorentz symmetry SO(1,3), and translational symmetry P$^{1,3}$. WS$_c$(1,3) stands for the inhomogeneous spin group symmetry in Hilbert space of leptons and quarks with rotation-like spin symmetry SP(1,3), and translation-like chirality boost-spin symmetry W$^{1,3}$ as well as scaling-like chiral conformal-spin symmetry SP$_c$(1,1). Except for the Higgs potential, SC(1) and SG(1) stand for the associated homogeneous scaling symmetries operating on the coordinates and basic fields, respectively. $Z_2$ denotes the discrete symmetry linked to the chiral duality invariance.

The symbol ``$\adjoin$" is used to designate the associated symmetry in which the transformations of group symmetries SP(1,3)$\times$SG(1) in Hilbert space should be coincidental to those of group symmetries SO(1,3)$\times$SC(1) in Minkowski spacetime. Though these two groups are isomorphic and form an associated symmetry, it is crucial to distinguish them conceptually. This is because one symmetry acts on an external coordinate spacetime and another operates on a spin-associated intrinsic spacetime.  

Let us explicitly write down the group generators of conformal inhomogeneous spin symmetry WS$_c$(1,3) as follows: 
\be \label{GG1}
\Sigma^{a b} = \frac{i}{4}  [\Gamma^{a}, \Gamma^{b} ] ,  \quad \Sigma_{\mp}^{a}=\Gamma^{a} \Gamma_{\mp} , \quad \Sigma_{\mp} = \pm \frac{1}{2}\Gamma_9 \equiv \pm \frac{1}{2}i\gamma_9,
\ee
where either $\Sigma_{-}^{\ta\tb} \equiv (\Sigma^{a b}, \Sigma_{-}^{a}, \Sigma_{-})$ or $\Sigma_{+}^{\ta\tb} \equiv (\Sigma^{a b}, \Sigma_{+}^{a}, \Sigma_{+})$ provides the group generators of WS$_c$(1,3) in correspondence to group symmetries SP(1,3), W$^{1,3}$ and SP$_c$(1,1). The chiral duality symmetry implies that $\Sigma_{-}^{\ta\tb}$ and $\Sigma_{+}^{\ta\tb}$ are chiral dual group generators, they are equivalent and related via the chiral duality operation, 
\be
\Sigma_{+}^{\ta\tb}  \equiv (\Sigma^{a b}, \Sigma_{+}^{a}, \Sigma_{+})  = C_d\Sigma_{-}^{\ta\tb}C_d^{-1} \equiv C_d(\Sigma^{a b}, \Sigma_{-}^{a}, \Sigma_{-}) C_d^{-1} .
\ee
It can be checked that the above generators satisfy the following group algebra:
\be
& & [\Sigma^{ab}, \Sigma^{cd}] =  i (\Sigma^{ad}\eta^{bc} -\Sigma^{bd}  \eta^{ac} - \Sigma^{ac} \eta^{bd} + \Sigma^{bc} \eta^{ad}) , \nn \\
& &  [\Sigma^{ab}, \Sigma_{\mp}^{c}] = i( \Sigma_{\mp}^{a}\eta^{bc} -\Sigma_{\mp}^{b}  \eta^{ac}  ) , \quad [\Sigma_{\mp}, \Sigma_{\mp}^{a}] = i \Sigma_{\mp}^{a} , \nn \\
& &  [\Sigma^{ab}, \Sigma_{\mp} ]  = 0,  \quad [\Sigma_{\mp}, \Sigma_{\mp} ]=0, \quad [\Sigma_{\mp}^{a}, \Sigma_{\mp}^{b}] = 0 .
\ee

The associated symmetry SC(1)$\ltimes$SO(1,3)$\adjoin$SP(1,3)$\rtimes$SG(1) is well known in the context of Lorentz symmetry and scaling symmetry for massless spinor fields. Specifically, this symmetry imposes the following transformation relations:
\be
& & x^{\mu} \to x^{' \mu} = L^{\mu}_{\; \nu} x^{\nu}, \quad  \Psi_{\mp}(x) \to  \Psi'_{\mp}(x') = e^{i \omega_{ab} \Sigma^{ab}} \Psi_{\mp}(x), \nn \\
& &  e^{- i \omega_{ab} \Sigma^{ab}} \Gamma^{a}  e^{i \omega_{ab} \Sigma^{ab}} = \Lambda^{a}_{\; b} \Gamma^b, \quad   \Lambda^{a}_{\; b}  \equiv \eta^{a}_{\; \mu} L^{\mu}_{\; \nu} \eta^{\nu}_{\; b}  \in SO(1,3) , \nn \\
& & x^{\mu} \to x^{' \mu} = \lambda^{-1} x^{\nu}, \quad  \Psi_{\mp}(x) \to  \Psi'_{\mp}(x') = \lambda^{3/2}\Psi_{\mp}(x),
\ee
where $L^{\mu}_{\; \nu}$ and $\Lambda^{a}_{\; b}  \equiv \eta^{a}_{\; \mu} L^{\mu}_{\; \nu} \eta^{\nu}_{\; b}$ are treated as constant matrices, and $\lambda$ is a constant parameter.

From the global transformations of the translation symmetry P$^{1,3}$ for coordinates and the chirality boost-spin symmetry W$^{1,3}$ for spinor fields, 
\be
x^{\mu}\to x^{' \mu} = x^{\mu} + \omega^{\mu} , \quad \Psi_{\mp}(x) \to  \Psi'_{\mp}(x') = e^{i \varpi_{a} \Sigma^{a}_{\mp}} \Psi_{\mp}(x) , 
\ee
it is clear that these transformations do not necessarily form an associated symmetry.  Specifically, $\omega^{\mu}$ and $\varpi^{a}$ can be arbitrary independent constant vectors. This is due to the fact that both the derivative operator $\p_{\mu}$ and the displacement vector $dx^{\mu}$ are translational invariant in globally flat Minkowski spacetime of coordinates, while the Abelian-type group generator $\Sigma^{a}_{\mp}$ is nilpotent, satisfying $(\Sigma^{a}_{\mp})^2=0$.

In summary, we begin by recasting the SM into a left-right symmetric chiral duality formulation. This reveals a hidden conformal inhomogeneous spin symmetry WS$_c$(1,3) as an internal symmetry, a semi-direct product structure that unifies conventional spin, chirality boosts-spin, and chiral conformal-spin symmetries, along with chiral duality symmetry. A significant outcome of this formulation is that the Yukawa coupling matrices become Hermitian, and neutrinos naturally emerge as massive particles. Furthermore, it indicates a unified description of electroweak interactions for leptons.

The associated symmetry (Eq. \ref{AS}) for the chiral duality invariant formulation (Eq. \ref{SMaction}) and the reduction to the Standard Model are detailed in the Supplementary Materials. Specifically, Supplementary Material A discusses the enlarged symmetries, while Supplementary Material B covers the reduction to the standard SM Lagrangian formalism.

\section{WS$_c$(1,3)$\times$SG(1) Gauge Symmetries and the Spin-fiber Non-commutative Gravigauge Spacetime }

Let us now apply the gauge symmetry principle to the internal symmetries that act on leptons and quarks as the fundamental constituents of matter. This principle asserts that the laws of nature should be independent of the choice of local field configurations, implying that all internal symmetries must be localized as gauge symmetries. This reasoning motivates us to localize the internal conformal inhomogeneous spin symmetry WS$_c$(1,3) and the scaling symmetry SG(1) into gauge symmetries.

It is then a necessity to introduce conformal inhomogeneous spin gauge field $\bscA_{\mu}(x)$ and local scaling gauge field $\cW_{\mu}(x)$. Meanwhile, the covariant derivative $D_{\mu}^{(\Psi_{\mp})}$ ($\Psi = l, q$) operating on the chiral spinor representations of leptons and quarks in the SM should be extended to a general one $\bscD_{\mu}^{(\Psi_{\mp})}$ by including gauge fields $\bscA_{\mu}(x)$ and $\cW_{\mu}(x)$, i.e.:
\be \label{WSGF}
& & i\bscD_{\mu}^{(\Psi_{\mp})} \Psi_{\mp} = ( i D_{\mu}^{(\Psi_{\mp})} + \frac{3}{2} i\cW_{\mu} + \bscA_{\mu}^{(\mp)}  )\Psi_{\mp} \nn \\
%& & \qquad \qquad =  (i\p_{\mu} + A_{\mu}^{(\Psi_{\mp})} + \frac{3}{2} i\cW_{\mu} +  \bscA_{\mu}^{(\mp)}  )\Psi_{\mp}, \nn \\
& & \bscA_{\mu}^{(\mp)}  \equiv  \cA_{\mu} + \cW_{\mu}^{(\mp)}  + \cB_{\mu}^{(\mp)}, \nn \\
& & \cA_{\mu} \equiv  \cA_{\mu}^{ab} \frac{1}{2} \Sigma_{ab} ,\;  \cW_{\mu}^{(\mp)} \equiv \cW_{\mu}^{a}  \frac{1}{2}\Sigma_{a \mp } , \; \cB_{\mu}^{(\mp)} \equiv \cB_{\mu} \Sigma_{\mp} ,
\ee
where $\cA_{\mu}^{\; ab}(x)$ represents the spin gauge field, $\cW_{\mu}^{\; a}(x)$ and $\cB_{\mu}(x)$ are referred to as chirality boost-spin gauge field and chiral conformal-spin gauge field, respectively, and $\cW_{\mu}(x)$ is called as scaling gauge field (or Weyl gauge field).

In light of the general covariant derivative $\bscD_{\mu}$, we are able to define the field strength via the following commutation of covariant derivative operator: 
\be \label{GFS}
 i [\bscD_{\mu}^{(\Psi_{\mp})}, \bscD_{\nu}^{(\Psi_{\mp})} ] \equiv \cF_{\mu\nu}^{(\Psi_{\mp})} + \bscF_{\mu\nu}^{(\mp)} + i\frac{3}{2} \cW_{\mu\nu} , 
\ee
with the definitions for $\cF_{\mu\nu}^{(\Psi_{\mp})}$ ($\Psi=l,q$) and $\bscF_{\mu\nu}^{(\mp)}$ as follows:
\be \label{GSMGFS}
& & \cF_{\mu\nu}^{(l_{\mp})} \equiv F_{\mu\nu}^{(l_{\mp})}  + F_{\mu\nu}^{(L_{\mp})} , \nn \\
& &  \cF_{\mu\nu}^{(q_{\mp})} \equiv F_{\mu\nu}^{(q_{\mp})}  + F_{\mu\nu}^{(L_{\mp})} + F_{\mu\nu}^{(C)},  \nn \\
& & \bscF_{\mu\nu}^{(\mp)} \equiv \cF_{\mu\nu} + \cF_{\mu\nu}^{(W\mp)} + \cF_{\mu\nu}^{(\mp)} , 
\ee
where $F_{\mu\nu}^{(\Psi_{\mp})}$ ($\Psi=l,q$), $F_{\mu\nu}^{(L_{\mp})}$ and $F_{\mu\nu}^{(C)}$ correspond to the field strengths of electromagnetic gauge ($B_{\mu}$), weak gauge ($W_{\mu}^i$) and strong gauge ($A_{\mu}^{\alpha}$) interactions as shown in Eq.(\ref{SMGFS}). The gauge field strengths $\cF_{\mu\nu}$, $\cF_{\mu\nu}^{(W \mp)}$, $\cF_{\mu\nu}^{(\mp)}$ and $\cW_{\mu\nu}$ are explicitly defined as follows:  
\be \label{WSGFS}
\cF_{\mu\nu} & \equiv &  \cF_{\mu\nu}^{ab}\frac{1}{2}\Sigma_{ab} , \;  \; \cF_{\mu\nu}^{(W\mp)} \equiv \cF_{\mu\nu}^{a} \frac{1}{2}\Sigma_{a \mp}, \;\; \cF_{\mu\nu}^{(\mp)} \equiv \cF_{\mu\nu}\Sigma_{\mp}, \nn \\
 \cF_{\mu\nu}^{ab} & = & \p_{\mu} \cA_{\nu}^{ab} - \p_{\nu} \cA_{\mu}^{ab} + \cA_{\mu c}^{a} \cA_{\nu}^{cb} - \cA_{\nu c}^{a} \cA_{\mu}^{cb}, \nn \\
\cF_{\mu\nu}^{a} & = & (\p_{\mu} + \cB_{\mu})\cW_{\nu}^{\; a} - (\p_{\nu} + \cB_{\nu})\cW_{\mu}^{\; a} + \cA_{\mu c}^{a} \cW_{\nu}^{\; c} - \cA_{\nu c}^{a} \cW_{\mu}^{\; c} , \nn \\
\cF_{\mu\nu} & = & \p_{\mu} \cB_{\nu} - \p_{\nu}\cB_{\mu} ,\quad \cW_{\mu\nu} = \p_{\mu} \cW_{\nu} - \p_{\nu} \cW_{\mu} , 
 \ee 
where $\cF_{\mu\nu}^{ab}$, $\cF_{\mu\nu}^{a}$, $\cF_{\mu\nu}$ and $\cW_{\mu\nu}$ represent the field strengths of the spin gauge ($\cA_{\mu}^{ab}$), the chirality boost-spin gauge ($\cW_{\mu}^{\; a}$), the chiral conformal-spin gauge ($\cB_{\mu}$) and the scaling gauge ($\cW_{\mu}$) interactions.

The gauge transformations of the conformal inhomogeneous spin gauge field and the scaling gauge field can be shown to have the following properties:
\be
& & \Psi_{\mp}(x)\to \Psi'_{\mp}(x) = S(\varpi^{ab})  \Psi_{\mp}(x),  \nn \\
& & \cA_{\mu}(x) \to \cA'_{\mu}(x) =  S(\varpi^{ab}) \cA_{\mu}S^{-1}(\varpi^{ab}) + S(\varpi^{ab}) i\p_{\mu} S^{-1}(\varpi^{ab}) , \nn \\
& & \qquad  \quad \equiv [ \Lambda^{a}_{\; c}(x)  \cA_{\mu}^{cd} \Lambda_{d}^{\; b}(x) + 
  \frac{1}{2}(\Lambda^{a}_{\; c}(x)  \p_{\mu} \Lambda^{ c b}(x)- \Lambda^{b}_{\; c}(x)  \p_{\mu} \Lambda^{c a}(x) ) ]  \frac{1}{2}\Sigma_{ab} , \nn \\
& & \cW_{\mu}^{(\mp)}(x) \to \cW_{\mu}^{'(\mp)}(x) =  S(\varpi^{ab}) \cW_{\mu}^{(\mp)} S^{-1}(\varpi^{ab}) \equiv \Lambda^{a}_{\; c}(x) \cW_{\mu}^{\; c}  \frac{1}{2}\Sigma_{a \mp} , 
\ee
under the spin gauge transformation of SP(1,3), and 
\be
& & \Psi'_{\mp}(x) =S_{\mp}(\varpi^{a})  \Psi_{\mp}(x) = e^{i\varpi^{a}(x)\Sigma_{a \mp}}  \Psi_{\mp}(x), \nn \\
& & \cW_{\mu}^{(\mp)}(x) \to \cW_{\mu}^{'(\mp)}(x)  = (\cW_{\mu}^{\; a} + \cD_{\mu}\varpi^{a}(x) ) \frac{1}{2}\Sigma_{a \mp }, \nn \\
& & \cD_{\mu}\varpi^{a}(x) \equiv (\p_{\mu} +\cB_{\mu}) \varpi^{a}(x) + \cA_{\mu\, b}^{a}\varpi^{b}(x) , 
\ee
under the chirality boost-spin gauge transformation of W$^{1,3}$, and
\be
& & \Psi_{\mp}(x)\to \Psi'_{\mp}(x) = S_{\mp}(\varpi)  \Psi_{\mp}(x) \equiv  e^{i\varpi(x)\Sigma_{\mp}} \Psi_{\mp}(x) , \nn \\
& & \cB_{\mu}^{(\mp)}(x) \to  \cB_{\mu}^{'(\mp)}(x) =   (\cB_{\mu}(x) - \p_{\mu}\varpi(x) ) \Sigma_{\mp},  \nn \\
& & \cW_{\mu}^{(\mp)}(x) \to \cW_{\mu}^{'(\mp)}(x) =  S_{\mp}(\varpi) \cW_{\mu}^{(\mp)} S_{\mp}^{-1}(\varpi)  \equiv e^{-\varpi(x)} \cW_{\mu}^{\; a}  \frac{1}{2}\Sigma_{a \mp} , 
\ee
under the chiral conformal-spin gauge transformation of SP$_c$(1,1), as well as 
\be
& & \Psi_{\mp}(x)\to \Psi'_{\mp}(x) = S(\xi) \Psi_{\mp}(x) \equiv \xi^{3/2}(x) \Psi_{\mp}(x) , \nn \\
& & \cW_{\mu} \to  \cW'_{\mu}=  \cW_{\mu} - \p_{\mu}\ln \xi(x) , 
\ee
under the scaling gauge transformation of SG(1). The group elements of WS$_c$(1,3)$\times$ SG(1) are explicitly presented as follows:
\be
& & S(\varpi^{ab}) = e^{i\varpi^{ab}(x)\frac{1}{2}\Sigma_{ab} } \in \mbox{SP}(1,3) ,\nn \\
& & S_{\mp}(\varpi^{a}) = e^{i\varpi^{a}(x)\Sigma_{a \mp}} \in \mbox{W}^{1,3},\nn \\
& &  S_{\mp}(\varpi) = e^{i\varpi(x)\Sigma_{\mp}} \in \mbox{SP}_c(1,1), \nn \\
& & S(\xi) \in \mbox{SG}(1), \quad \Lambda^{a}_{\; b}(x)\in \mbox{SP}(1,3) \cong \mbox{SO}(1,3).
\ee

These gauge transformations distinguish the local internal gauge symmetries of fundamental fields in Hilbert space from the global external symmetries of coordinates in Minkowski spacetime. Specifically, the spin gauge symmetry SP(1,3) and Lorentz symmetry SO(1,3), as well as the scaling gauge symmetry SG(1) and global scaling symmetry SC(1), no longer appear as associated symmetries required in QFT. 

To preserve both symmetries of SP(1,3) and SO(1,3), it is necessary to generalize the SM. First, it becomes essential to introduce a bi-covariant vector field $\chih_{a}^{\; \mu}(x)$ to replace the Kronecker delta symbol $\delta_{a}^{\;\;\mu}$. Consequently, the coordinate derivative operator $\p_{\mu}$, which is multiplied by the Kronecker delta symbol $\delta_{a}^{\;\;\mu}$ in QFT, becomes associated with the bi-covariant vector field $\chih_{a}^{\; \mu}(x)$, 
\be
\delta_{a}^{\;\;\mu} \to \chih_{a}^{\; \mu}(x), \quad \delta_{a}^{\;\;\mu}  \p_{\mu}  \to \chih_{a}^{\; \mu}(x)\p_{\mu} .
\ee
 This leads to the introduction of the spin-associated intrinsic derivative operator $\eth_{a}$ and the displacement vector $\dbar\zeta^{a}$ through the bi-covariant vector field via the following relationships:
\be
 \eth_{a}  \equiv  \chih_{a}^{\; \mu}(x)\p_{\mu}, \quad \dbar\zeta^{a} \equiv  \chi_{\mu}^{\; a}(x) dx^{\mu} ,
\ee
with $\chih_{a}^{\; \mu}(x)$ and $\chi_{\mu}^{\; a}(x)$ satisfying the following dual conditions:
\be
\chi_{\mu}^{\; a}(x)  \eta_{a}^{\; b} \chih_{b}^{\; \nu}(x) = \eta_{\mu}^{\; \nu} , \quad 
\chih_{b}^{\; \nu}(x) \eta_{\nu}^{\; \mu} \chi_{\mu}^{\; a}(x)  = \eta_{b}^{\; a} ,
\ee
where $\chi_{\mu}^{\; a}(x)$ and $\chih_{a}^{\; \mu}(x)$ are regarded as dual bi-covariant vector fields.

It is evident that $\chih_{a}^{\; \mu}(x)$ must be an SP(1,3)-valued invertible vector field with a non-zero determination $\det \chih^{\; \mu}_{a}\neq 0$. Under the transformations of SP(1,3) and SO(1,3) group symmetries, both $\chih_{a}^{\; \mu}(x)$ and $\chi_{\mu}^{\; a}(x)$ exhibit the behavior of bi-covariant vector fields. Specifically, they transform as follows:
\be
& & \chi_{\mu}^{\; a}(x) \to \chi_{\mu}^{'\; a}(x) =  \Lambda^{a}_{\; b}(x)\chi_{\mu}^{\; b}(x), \nn \\
& &  \chih_{a}^{\; \mu}(x) \to \chih_{a}^{'\; \mu}(x) =  \Lambda_{a}^{\; b}(x) \chih_{b}^{\; \mu}(x), \nn \\
& & \Lambda^{a}_{\; b}(x) \in \mbox{SP}(1,3) \cong \mbox{SO}(1,3) ,
\ee
under the transformation of the spin gauge symmetry SP(1,3) in the spin-associated intrinsic spacetime, and
\be
& & \chi_{\mu}^{\; a}(x) \to \chi_{\mu}^{'\; a}(x') =  L_{\mu}^{\; \nu}\, \chi_{\nu}^{\; a}(x), \nn \\
& & \chih_{a}^{\; \mu}(x) \to \chih_{a}^{'\; \mu}(x') =  L^{\mu}_{\; \nu}\, \chih_{a}^{\; \nu}(x) ,\nn \\
& & x^{' \mu} = L^{\mu}_{\; \nu}\, x^{\nu} , \quad L^{\mu}_{\; \nu} \in \mbox{SO}(1,3), 
\ee
under the transformation of the global Lorentz symmetry SO(1,3) in the coordinate spacetime.

Furthermore, under the transformations of the scaling gauge symmetry SG(1) and the global scaling symmetry SC(1), $\chi_{\mu}^{\; a}(x)$ and $\chih_{a}^{\; \mu}(x)$ have the following transformation properties:
\be
& & \chi_{\mu}^{\; a}(x) \to \chi_{\mu}^{'\; a}(x) = \xi^{-1}(x) \, \chi_{\nu}^{\; a}(x), \;\; \chih_{a}^{\; \mu}(x) \to \chih_{a}^{'\; \mu}(x) =  \xi(x) \, \chih_{a}^{\; \nu}(x) , \nn \\
& & \chi_{\mu}^{\; a}(x) \to \chi_{\mu}^{'\; a}(x') = \lambda \, \chi_{\nu}^{\; a}(x), \;\; \chih_{a}^{\; \mu}(x) \to \chih_{a}^{'\; \mu}(x') =  \lambda^{-1} \, \chih_{a}^{\; \nu}(x) ,\nn \\
& &  x^{' \mu} = \lambda^{-1}\, x^{\mu} ,\quad \lambda \in \mbox{SC}(1), \;\; \xi(x) \in \mbox{SG}(1) ,
\ee
where $\xi(x)$ is an arbitrary local scaling parameter and $\lambda$ denotes an arbitrary constant. 

It can be demonstrated that the local displacement vector $\dbar \zeta^{a}$ and the derivative operator $\eth_{a}$ achieve global scaling invariance with a zero global scaling charge $C_s=0$, while they undergo local scaling gauge transformations characterized by a non-zero local scaling charge $\hat{C}_s = \mp 1$. Notably, the dual bi-covariant vector fields $\chi_{\mu}^{\; a}(x)$ and $\chih_{a}^{\; \mu}(x)$ are invariant under the transformations of the chirality boost-spin gauge symmetry group $W^{1.3}$ and the chiral conformal-spin gauge symmetry group SP$_c$(1,1). 

In coordinate spacetime, the derivative operator $\p_{\mu}$ and the displacement vector $dx^{\mu}$ constitute dual bases $\{\p_{\mu}\}\equiv \{\partial/\partial x^{\mu}\}$ and $\{dx^{\mu}\}$, which span the tangent spacetime $\mT_{4}$ and the cotangent spacetime $\mT^{*}_{4}$, respectively. They are commonly referred to as external spacetime of coordinates. Correspondingly, the spin-associated intrinsic derivative operator $\eth_a$ and displacement vector $\dbar \zeta^{a}$ define dual bases $\{\eth_a\}$ and $\{\dbar \zeta^{a}\}$, establishing a locally orthogonal intrinsic spacetime $\mG_{4}$ and its dual spacetime $\mG^{*}_{4}$. 

It can be verified that the basis $\{\eth_a \}$ is non-commutative and satisfies the following commutation relation:
\be \label{NCR}
& & [ \eth_c ,\; \eth_d] = \hsF_{cd}^a \eth_a ,
\ee
which demonstrates that the spin-associated intrinsic derivative operator $\eth_a$ constitutes a non-Abelian group algebra with local group structure factor $\hsF_{ab}^c$ defined as follows:
\be \label{LGSF}
& &  \hsF_{cd}^a \equiv (\eth_c \chih_{d}^{\; \nu} - \eth_d\chih_{c}^{\; \nu} ) \chi_{\nu}^{\; a} \equiv - \chih_{c}^{\; \mu} \chih_{d}^{\; \nu} \sF_{\mu\nu}^{a} , \nn \\
& &  \sF_{\mu\nu}^{a} \equiv \p_{\mu}\chi_{\nu}^{\; a}(x) - \p_{\nu}\chi_{\mu}^{\; a}(x) .
\ee 

It is straightforward to verify that when the spin-fiber gravigauge exterior derivative operator $\dbar \equiv \dbar \zeta^{c}\wedge \eth_c$ is applied to the gravigauge displacement vector $\dbar \zeta^{a}$, a non-vanishing tensor structure arises, determined by the local group structure factor, i.e.:
\be \label{EDO}
\dbar (\dbar \zeta^{a}) \equiv \dbar \zeta^{c} \wedge \eth_c (\dbar \zeta^{a}) = \frac{1}{2} 
\hsF_{cd}^a \; \dbar \zeta^{c} \wedge \dbar \zeta^{d} \equiv - \frac{1}{2} 
\sF_{\mu\nu}^a\, dx^{\mu}\wedge dx^{\nu} .
\ee

The local group structure factor reveals the non-commutative nature of the intrinsic derivative operator, thereby defining the geometric properties of spin-associated intrinsic spacetime. There exists a relationship between the group structure factor $\hsF_{cd}^a$ and the spin connection $\hmOm_{c}^{ab}$ of the intrinsic spacetime,
\be \label{GGFS1}
& & \hsF_{cd}^{a}  = \hmOm_{cd}^{a}  - \hmOm_{dc}^{a} \equiv \hmOm_{[cd]}^{a},\;\; \hmOm_{cd}^{a} = \hmOm_{c}^{ab} \eta_{bd}, \nn \\
& & \hmOm_{c}^{ab} = \etach_{ca'}^{[ab] c'd'} \hsF_{c'd'}^{a'} \equiv \etach_{ca'}^{[ab] c'd'} \hmOm_{[c'd']}^{a'} \equiv \chih_{c}^{\; \mu} \mOm_{\mu}^{ab} , 
\ee
with the constant tensor $\etach_{ca'}^{[ab] c'd'}$ defined by,
\be 
& & \etach_{ca'}^{[ab] c'd'} \equiv \frac{1}{2}  (\eta^{ac'}\eta_{a'}^{\, b} - \eta^{bc'}\eta_{a'}^{\, a}) \eta_{c}^{\, d'} + \frac{1}{4} (\eta^{ac'}\eta^{bd'} - \eta^{bc'}\eta^{ad'} ) \eta_{ca'}  .
% & & \etah_{[cd]b}^{c'}  \equiv  \eta_{c}^{\, c'}\eta_{db} - \eta_{d}^{\, c'}\eta_{cb} = \etah_{[cd]b}^{c'} \mOm_{c'}^{ab}, 
\ee
Here, the spin connection $\mOm_{\mu}^{ab}$ has the following explicit form:
\be  \label{GGFS2}
& & \mOm_{\mu}^{ab}(x) = \frac{1}{2}\left( \chih^{a\nu} \sF_{\mu\nu}^{b} - \chih^{b\nu} \sF_{\mu\nu}^{a} -  \chih^{a\rho}  \chih^{b\sigma} \chi_{\mu c }  \sF_{\rho\sigma}^{c} \right) .
\ee

It can be verified that $\mOm_{\mu}^{ab}(x)$ has the same transformation property as the spin gauge field $\cA_{\mu}^{ab}$ under the transformation of the spin gauge symmetry SP(1,3). 

The local group structure factor is interpreted as an Abelian gauge-type field strength, denoted as $\sF_{\mu\nu}^{a}$, for the bi-covariant vector field $\chi_{\mu}^{\; a}(x)$ through its tensor structure, as illustrated in Eq.(\ref{LGSF}). This spin-associated gauge-type bi-covariant vector field, along with its associated field strength, plays an essential role in characterizing gravitational interactions among the spinor fields of leptons and quarks, which are regarded as the fundamental constituents of matter in the SM. This represents a crucial distinction from the metric field, which serves as the fundamental gravitational field in GR.

For clarity, the spin-associated gauge-type field $\chi_{\mu}^{\; a}(x)$ is referred to as {\it gravigauge field } and its corresponding gauge-type field strength $\sF_{\mu\nu}^{a}$ (or local group structure factor $\hsF_{cd}^a\equiv \hmOm_{[cd]}^{a}$) is termed the {\it gravigauge field strength}. 

Notably, the gravigauge field strength $\sF_{\mu\nu}^{a}$ arising from the local group structure factor is not spin gauge covariant. A spin gauge covariant field strength of the gravigauge field is defined as follows:
\be \label{GCGFS}
\cG_{\mu\nu}^{a}  \equiv  D_{\mu} \chi_{\nu}^{\; a}  - D_{\nu} \chi_{\mu}^{\; a} =  \p_{\mu} \chi_{\nu}^{\; a}  - \p_{\nu} \chi_{\mu}^{\; a} + \cA_{\mu b}^{a} \chi_{\nu}^{\; b}  - \cA_{\nu b}^{a} \chi_{\mu}^{\; b} ,
\ee
which is referred to as spin gauge-covariant gravigauge field strength. 

We may refer the spin connection $\mOm_{\mu}^{ab}(x)$ as spin gravigauge field, its spin gauge covariant field strength is defined as:
\be
R_{\mu\nu}^{ab} \equiv \p_{\mu} \mOm_{\nu}^{ab} - \p_{\nu} \mOm_{\mu}^{ab} + \mOm_{\mu}^{ac} \mOm_{\nu c}^{b} - \mOm_{\nu}^{ac} \mOm_{\mu c}^{b},
\ee
which is called as spin gravigauge field strength.

The gravigauge field $\chi_{\mu}^{\; a}(x)$ can be understood as a bi-covariant vector field defined in Minkowski spacetime and valued in spin-associated intrinsic spacetime. This field establishes a connection between two types of spacetime, collectively referred to as bi-frame spacetime, denoted as $T_4\times G_4$ or $T_4^{\ast}\times G_4^{\ast}$ over the coordinate spacetime $M_4$. 

Mathematically, both globally and locally flat vector spacetimes allow for a canonical identification of vectors in tangent Minkowski spacetime $T_4$ at points with vectors in Minkowski spacetime itself $M_4$, and also for a canonical identification of vectors at points with their dual vectors at the same points. Physically, either tangent or dual tangent Minkowski spacetime over globally flat Minkowski spacetime is viewed as a free-motion spacetime $\fM_4$. Meanwhile, the local orthogonal spin-associated intrinsic spacetime characterized by the gravigauge field is regarded as an emergent spacetime $\fG_4$. The canonical identification for the vector spacetimes leads to a simple structure of bi-frame spacetime as:
\be
& & \fB_4 =  \fM_4 \times \bf{G}_4 , \nn \\
 & & \fM_4 \equiv T_4 \cong T_4^{\ast} \cong  M_4 , \quad \fG_4 \equiv G_4 \cong G_4^{\ast} .
\ee

In general, this bi-frame spacetime structure exhibits a spin-fiber bundle structure $\fE_4$, where the globally flat Minkowski spacetime serves as a base spacetime $\fM_4$, and the local orthogonal spin-associated intrinsic spacetime acts as a fiber $\fG_4$. The intrinsic spacetime associated with the spin property, characterized by the gravigauge field strength $\hsF_{cd}^a$, is designated as a spin-fiber gravigauge spacetime. The spin-associated intrinsic derivative operator $\eth_{a}$ and the displacement vector $\dbar \zeta^{a}$ are identified as the gravigauge derivative and the gravigauge displacement, respectively. The spin connection $\mOm_{\mu}^{ab}$ (or $\hmOm_{c}^{ab}$), determined by the gravigauge field, is referred to as the spin gravigauge field. 

Geometrically, the spin-fiber bundle $\fE_4$ is related to the product bi-frame spacetime $\fM_4\times \fG_4$ via a continuous surjective map $\Pi_{\chi}$, which projects the bundle $\fE_4$ onto the base spacetime $\fM_4$. Thus, the spin-fiber bundle $\fE_4$ equipped with this projection $\Pi_{\chi}$ is expressed as: 
\be
& & \Pi_{\chi}: \; \; \fE_4 \to \fM_4 .
\ee
More general, the spin-fiber bundle structure of bi-frame spacetime is represented as  $(\fE_4, \fM_4, \Pi_{\chi}, \fG_4) $. In the trivial case, this reduces to:
\be
& & \fE_4 \sim  \fB_4 = \fM_4 \times \fG_4 .
\ee

Notably, the spin-fiber gravigauge spacetime, spanned by the gravigauge bases $\{\eth_a\}$ and $\{\dbar \zeta^{a}\}$, exhibits a non-commutative spacetime, distinguishing it from the commutative Minkowski spacetime of coordinates (the external spacetime). To investigate this non-commutative property in the local orthogonal gravigauge spacetime $\fG_4$, it is essential to analyze the dynamics of the gravigauge field strength $\hsF_{cd}^a$ or equivalently the spin gravigauge field $\hmOm_{[cd]}^{a}\equiv \hsF_{cd}^a$. 

By employing the gravigauge derivative $\eth_{a}$, the covariant derivative $\bshD_{c}$ of gauge field $\bsA_{c}$ in spin-fiber gravigauge spacetime can be defined as follows: 
\be
& & i\bshD_{c} \equiv \chih_{c}^{\; \mu} i\bsD_{\mu} =  i \eth_{c}+ \bshA_{c} ,
\ee
where $\bshA_{c}$ represents any gauge field arising from gauge symmetries. The corresponding gauge covariant field strength takes the following general form:
\be
& & \bshF_{cd} = \cD_{c}\bshA_{d} - \cD_{d}\bshA_{c} - i [\bshA_{c}, \bshA_{d}] \equiv \bsF_{cd} + \hsF_{cd}^{a} \bshA_{a} , \nn \\
 & & \cD_{c}\bshA_{d} \equiv \eth_{c} \bshA_{d} + \hmOm_{cd}^{a} \bshA_{a} , \; \; \hsF_{cd}^{a} \bshA_{a} \equiv (\hmOm_{cd}^{a} - \hmOm_{dc}^{a}) \bshA_{a}, \nn \\ 
& & \bsF_{cd} \equiv \eth_{c}\bshA_{d} - \eth_{d}\bshA_{c} - i [\bshA_{c}, \bshA_{d}] . 
\ee
The gauge covariant field strength $\bshF_{cd}$, defined in spin-fiber gravigauge spacetime, includes an additional term characterized by the gravigauge field strength, $\hsF_{cd}^{a} \bshA_{a}$. This term arises from the non-commutative nature of gravigauge derivative operator, as demonstrated in Eqs.(\ref{NCR}) and (\ref{EDO}). This additional term ensures that the gauge field strength $\bshF_{cd}$ remains an antisymmetric tensor under the transformation of spin gauge symmetry SP(1,3).

To provide an alternative clearer explanation, let us express the gauge covariant field strength by using the exterior gravigauge derivative formalism, shown in Eq.(\ref{EDO}), as follows:
\be
\bshF & = & \dbar \bshA + \bshA\wedge\bshA = \dbar (\bshA_{d}\, \dbar \zeta^{d}) 
+ \bshA_{c} \bshA_{d}\, \dbar\zeta^{c}\wedge\dbar \zeta^{d} \nn \\ 
& = & (\eth_{c}\bshA_{d} + \bshA_{c} \bshA_{d})\, \dbar\zeta^{c}\wedge\dbar \zeta^{d} + \bshA_{a}\, \dbar (\dbar \zeta^{a}) \nn \\
& = &  \frac{1}{2} ( \bsF_{cd}  + \hsF_{cd}^{a} \bshA_{a})\, \dbar\zeta^{c}\wedge\dbar \zeta^{d} \equiv \frac{1}{2} \bshF_{cd} \, \dbar\zeta^{c}\wedge\dbar \zeta^{d} .
\ee

In our current framework, it involves the following gauge fields:
\be
& & \bshA_{c}  \equiv \chih_{c}^{\; \mu} \bsA_{\mu} \equiv (\hB_{c}^{(l_{\mp})}, \hW_{c}^{(L_{\mp})}, \hA_{c}^{(C)}, \hcA_{c}, \hcW_{c}^{(\mp)}, \hcB_{c}^{(\mp)}, \hcW_{c}) , \nn \\
& & \bsA_{\mu}  \equiv ( B_{\mu}^{(l_{\mp})}, W_{\mu}^{(L_{\mp})}, A_{\mu}^{(C)}, \cA_{\mu}, \cW_{\mu}^{(\mp)}, \cB_{\mu}^{(\mp)}, \cW_{\mu}), 
\ee
where all gauge fields presented in $\bsA_{\mu}$ are introduced from the gauge symmetries U$_Y$(1)$\times$SU$_L$(2)$\times$SU$_C$(3)$\times$WS$_c$(1,3)$\times$SG(1), as detailed in Eqs.(\ref{GFSM}) and (\ref{WSGF}). The corresponding gauge field strengths are expressed as follows:
\be 
 \bshF_{cd}  & = & ( \hF_{cd}^{(l_{\mp})}, \hF_{cd}^{(L_{\mp})}, \hF_{cd}^{(C)}, \hcF_{cd},\hcF_{cd}^{(W\mp)}, \hcF_{cd}^{(\mp)},\hcW_{cd}) \equiv \chih_{c}^{\; \mu} \chih_{d}^{\; \nu} \bsF_{\mu\nu} ,\nn \\
 \bsF_{\mu\nu} & = &  ( F_{\mu\nu}^{(l_{\mp})}, F_{\mu\nu}^{(L_{\mp})}, F_{\mu\nu}^{(C)}, \cF_{\mu\nu},\cF_{\mu\nu}^{(W\mp)}, \cF_{\mu\nu}^{(\mp)},\cW_{\mu\nu}) ,
 \ee
where the relevant gauge field strengths in $\bsF_{\mu\nu}$ are explicitly defined in Eqs.(\ref{SMGFS}) and (\ref{GSMGFS})-(\ref{WSGFS}).  

In summary, by applying the gauge principle, we promote the global WS$_c$(1,3)$\times$GS(1) lepton and quark symmetries to local gauge symmetries. This promotion correspondingly necessitates the introduction of new gauge fields as force carriers. Significantly, the gravigauge field $\chi_{\mu}^{\; a}$ emerges as the fundamental mediator of gravity. From this field, we construct a locally orthogonal spin-fiber gravigauge spacetime. The intrinsic non-commutativity of this structure reveals a profound connection between spin dynamics and the fundamental nature of spacetime and gravity.

In the following section, we will explicitly demonstrate that introducing the field $\chih_{a}^{\; \mu}(x)$ and its dual counterpart $\chi_{\mu}^{\; a}(x)$ is both necessary and unavoidable for constructing a gauge invariant action under conformal inhomogeneous spin gauge transformations in spin-fiber gravigauge spacetime. This introduction plays a crucial role in distinguishing the global Poincar\'e (inhomogeneous Lorentz) symmetry, $\text{PO}(1,3)$, in coordinate Minkowski spacetime from the local conformal inhomogeneous spin gauge symmetry, $\text{WS}_c(1,3)$, in spin-fiber gravigauge spacetime.

\section{The General Standard Model and the Gravitization Equation in Gravigauge Spacetime}

The gravigauge field $\chi_{\mu}^{\; a}(x)$ and its dual $\chih_{a}^{\; \mu}(x)$ are essential for distinguishing between the internal spin symmetry of leptons and quarks and the external symmetry of coordinates, thereby enabling the construction of a gauge-invariant action under conformal inhomogeneous spin gauge transformations. These fields allow the introduction of the gravigauge derivative operator $\eth_{a}$ and the displacement vector $\dbar \zeta^{a}$, which facilitate the formulation of the General Standard Model (GSM) action in spin-fiber gravigauge spacetime.

By extending the gauge symmetries of the SM to the following enlarged internal gauge symmetries: 
\be \label{GSMGS}
\hat{\fG}_{GSM} =  \mbox{U}_Y(1)\times \mbox{SU}_L(2) \times \mbox{SU}_C(3) \times \mbox{WS}_c(1,3)\times \mbox{SG}(1), 
\ee
we can construct the generalized SM action based on its chiral duality-invariant representation, as given in Eq. (\ref{SMaction}). Retaining terms up to second order in derivatives of the fundamental fields, we obtain the explicit form of this action in spin-fiber gravigauge spacetime: 
\be \label{GSMaction}
& & \cS_{GSM} =  \int [\dbar \zeta ]  \cL_{GSM}( \Psi_{\mp}^i,  \hB_{c}, \hW_{c}^{i}, \hA_{c}^{\alpha}, \hat{H}, \hcB_{c}, \hcW_{c}^{a}, \nn \\
& &\;\; \hcA_{c}^{ab},  \hcW_{c}, \hat{\Phi}_{\kappa}, \hat{\phi}_e,\hat{\phi}_w, \bs{\zeta}^{a}, \bs{\zeta} ) = \int [\dbar \zeta]  \sum_{s=\mp}  \frac{1}{2} \{  \sum_{\Psi=l,q}  \bs{\zeta}  \nn \\
& & [ \frac{1}{2} \bar{\Psi}_{s}^{i} \Sigma_{s}^{c}  i \hcD_{c}^{(\Psi s)}\, \Psi_{s}^{i}  + H.c. + \bar{\Psi}_{s}^{i} \hPhi_{s} ( \tilde{\Gamma}_{s} \tilde{\lambda}^{\Psi}_{ij}  +  \hGa_{s}  \hat{\lambda}^{\Psi}_{ij} ) \, \Psi_{s}^{j} ]\nn \\
& & -  \frac{1}{4} \eta^{cc'}\eta^{dd'} [ \frac{1}{2g_1^2} \Tr \hF_{cd}^{(l_{s})} \hF_{c'd'}^{(l_{s})}  + \frac{1}{2g_2^2} \Tr \hF_{cd}^{(L_{s})} \hF_{c'd'}^{(L_{s})} \nn \\
& & + \frac{1}{2g_4^2} \Tr \bshcF_{cd}^{s} \bshcF_{c'd'}^{s} ]  + \frac{1}{8} \eta^{cd} \Tr (\hcD_{c}\hPhi_{s} )^{\dagger} \hcD_{d}\hPhi_{s}      \} \nn \\
& & -\frac{1}{4}\eta^{cc'}\eta^{dd'} [ \frac{4}{g_3^2} \Tr \hF_{cd}^{(C)} \hF_{c'd'}^{(C)} +\hPhi_{\kappa}^2  \Tr (\bs{\hR}_{cd}\Sigma_{c'd'} )  \nn \\
& & + \frac{1}{g_4^2}( \frac{1}{\gamma_c^2} \Tr \hcF_{cd}^{(-)} \hcF_{c'd'}^{(+)}  - \frac{1}{\bs{\zeta}^{2}} \Tr \bshcF_{cd}^{(W-)} \bshcF_{c'd'}^{(W+)} )  \nn \\
& & -  ( \alpha_{\kappa} \hPhi_{\kappa}^2 + \alpha_e\hphi_{e}^2)  \Tr \bshcG_{cd}^{(-)} \bshcG_{c'd'}^{(+)} + \frac{1}{g_w^2} \hcW_{cd}\hcW_{c'd'} \, ]  \nn \\
& & +  \frac{1}{2}  \eta^{cd} [ ( \gamma_{\kappa} \hPhi_{\kappa}^2 + \beta_e\hphi_{e}^2)   \Tr \bshcA_{c}\bshcA_{d}   -  \beta_w^2 \hphi_w^2   \Tr \bshcW_{c}^{(-)}  \bshcW_{d}^{(+)}  ] \nn \\
& &    +   \frac{1}{2} \eta^{cd} [\beta_{\kappa}^2 \hcD_{c}\hPhi_{\kappa} \hcD_{d}\hPhi_{\kappa}   +  \gamma_{w}^2  \hphi_w^2\hcD_{c}\bs{\zeta} \hcD_{d}\bs{\zeta} + \hcD_{c}\hphi_{e} \hcD_{d}\hphi_{e} \nn \\
& & + \hcD_{c}(\bs{\zeta}\hphi_{w})\hcD_{d}(\bs{\zeta}\hphi_{w})   ]  - \hPhi_{\kappa}^4\, \hcV_{S}( \hPhi_{\mp}/\hPhi_{\kappa}, \bs{\zeta} \hphi_w/\hPhi_{\kappa}, \hphi_{e}/\hPhi_{\kappa}) .
\ee
Here, $\hB_{c}$, $\hW_{c}^{i}$, and $\hA_{c}^{\alpha}$ represents electroweak and strong gauge fields associated with the gauge symmetries U$_Y$(1)$\times$SU$_L$(2)$\times$SU$_C$(3), while $\hat{H}$ denotes Higgs field in the SM. $\hcB_{c}$, $\hcW_{c}^{a}$ and $\hcA_{c}^{ab}$ correspond to the chiral conformal-spin, chirality boost-spin and spin gauge fields, respectively, related to the conformal inhomogeneous spin gauge symmetry WS$_c$(1,3). Additionally, $\hcW_{c}$ represents the scaling gauge field (or Weyl gauge field) arising from the scaling gauge symmetry. 

$\hPhi_{\kappa}$, $\hphi_{e}$ and $\hphi_w$ are three singlet scalar fields introduced to preserve scaling symmetries. $\bs{\zeta}$ and $\bs{\zeta}^{a}$ are  the scaling and vector fields, respectively, in gravigauge spacetime, which are essential for constructing both chiral conformal-spin and chirality boost-spin gauge-invariant action. The action concerns in general the additional gauge coupling constants $g_4$, $\gamma_cg_4$ and $g_w$, and the scalar coupling constants  $\alpha_{\kappa}$, $\beta_{\kappa}$, $\gamma_{\kappa}$, $\alpha_e$, $\beta_{e}$, $\beta_{w}$ and $\gamma_{w}$ beyond the SM. 

The covariant derivative of leptons and quarks in the action is given as follows:
\be
 i\hcD_{c}^{(\Psi_{\mp})} & \equiv & \chih_{c}^{\; \mu} i\cD_{\mu}^{(\Psi_{\mp})} \equiv  i\hD_{c}^{(\Psi_{\mp})} + \hcA_{c}  \nn \\
 & \equiv & i \eth_{c}+ \hA_{c}^{(\Psi_{\mp})} + \hcA_{c}  , \nn \\
\hA_{c}^{(\Psi_{\mp})} & \equiv & \chih_{c}^{\; \mu} A_{\mu}^{(\Psi_{\mp})} , \;\; \hcA_{c}\equiv \chih_{c}^{\; \mu} \cA_{\mu}^{ab}\frac{1}{2}\Sigma_{ab} , 
\ee
where $\hA_{c}^{(\Psi_{\mp})} = (A_{\mu}^{(l_{\mp})}, A_{\mu}^{(q_{\mp})}$) are the gauge fields in the SM, as defined in Eq.(\ref{GFSM}), and $\cA_{\mu}^{ab}$ is the spin gauge field. Notably, the covariant derivative $ \hcD_{c}^{(\Psi_{\mp})}$ in the action differs from $\bscD_{\mu}^{(\Psi_{\mp})}$ derived from the gauge symmetries, as shown in Eq.(\ref{WSGF}). In $\hcD_{c}^{(\Psi_{\mp})}$, the chirality boost-spin gauge field $\hcW_{c}^{\; a}\equiv \chih_{c}^{\; \mu} \cW_{\mu}^{\; a}$, the chiral conformal-spin gauge field $\hcB_{c}\equiv \chih_{c}^{\; \mu} \cB_{\mu}$ and also the scaling gauge field $\cW_{c}\equiv \chih_{c}^{\; \mu} \cW_{\mu}$ are no longer involved. This exclusion arises from the chirality property of the leptons and quarks in the chiral spinor representation and the hermiticity requirement of the action, which prevent these gauge fields from coupling to the leptons and quarks.  

In the above action, $\bshcF_{cd}^{(s)}$ ($s=\mp$) represents the gauge field strength of the conformal inhomogeneous spin gauge symmetry WS$_c$(1,3),
\be
\bshcF_{cd}^{(\mp)} \equiv \hcF_{cd} + \hcF_{cd}^{(W\mp)} + \hcF_{cd}^{(\mp)} \equiv \chih_{c}^{\; \mu}\chih_{d}^{\; \nu}\bscF_{\mu\nu}^{(\mp)}, 
\ee
with $\bscF_{\mu\nu}^{(\mp)} \equiv \cF_{\mu\nu} + \cF_{\mu\nu}^{(W\mp)} + \cF_{\mu\nu}^{(\mp)}$ explicitly defined in Eqs.(\ref{GSMGFS}) and (\ref{WSGFS}).

The field strength $\bs{\hR}_{cd}$ in the action represents a local scaling gauge-covariant spin gravigauge field strength, 
\be
& & \bs{\hR}_{cd} \equiv \bs{\hR}_{cd}^{ab} \frac{1}{2}\Sigma_{ab} \equiv \chih_{c}^{\mu}\chih_{d}^{\nu} \bs{R}_{\mu\nu}^{ab}\frac{1}{2}\Sigma_{ab} , \nn \\
& & \bs{\hR}_{cd}^{ab} \equiv \eth_c \bshmOm_{d}^{ab} - \eth_d \bshmOm_{c}^{ab} + \bshmOm_{c}^{aa'} \bshmOm_{d a'}^{b} - \bshmOm_{d}^{aa'} \bshmOm_{c a'}^{b} +  \bshsF_{cd}^{a'} \bshmOm_{a'}^{ab}, 
\ee
with $\bshmOm_{c}^{ab}$  and $\bshsF_{cd}^{a}$  representing local scaling gauge-covariant field strengths:
\be
& & \bshmOm_{c}^{ab}\equiv  \hmOm_{c}^{ab}   - \hcS_{c}^{[ab]} \equiv  \chih_{c}^{\; \mu}\bs{\mOm}_{\mu}^{ab} , \nn \\
& & \bshsF_{cd}^{a} \equiv  \bshmOm_{cd}^{a} - \bshmOm_{dc}^{a} \equiv \hsF_{cd}^{a} + \hcS_{[cd]}^{a}\equiv  - \chih_{c}^{\; \mu} \chih_{d}^{\; \nu} \bs{\sF}_{\mu\nu}^{a} .
\ee
Here, $\bs{\mOm}_{\mu}^{ab}$ and $\bs{\sF}_{\mu\nu}^{a}$ are explicitly defined as:
\be
& & \bs{\mOm}_{\mu}^{ab} = \frac{1}{2}\left( \chih^{a\nu} \bs{\sF}_{\mu\nu}^{b} - \chih^{b\nu} \bs{\sF}_{\mu\nu}^{a} -  \chih^{a\rho}  \chih^{b\sigma} \chi_{\mu c } \bs{\sF}_{\rho\sigma}^{c} \right)  , \nn \\
 & & \bs{\sF}_{\mu\nu}^{a} \equiv d_{\mu}\chi_{\nu}^{\; a}(x) - d_{\nu}\chi_{\mu}^{\; a}(x) , \;\; d_{\mu} \equiv \p_{\mu} + \cS_{\mu} , 
 \ee
and the other tensors are given by,
 \be
& & \hcS_{[cd]}^{a} \equiv \eta_{[cd]}^{[ab]}\hcS_b, \quad \hcS_{c}^{[ab]} \equiv \eta_{[cd]}^{[ab]}\hcS^d, \nn \\
& & \hcS_c \equiv \eth_{c}\ln \hPhi_{\kappa} \equiv \chih_{c}^{\mu} \cS_{\mu}, \;\;  \cS_{\mu} \equiv \p_{\mu}\ln \hPhi_{\kappa},
\ee
where $\hcS_{c}$ is regarded as a pure gauge field characterized by the real scalar field $\hPhi_{\kappa}$. The explicit definitions for the gravigauge field strength $\hsF_{cd}^{a}$ and the spin gravigauge field $\hmOm_{c}^{ab}$ are presented in Eqs.(\ref{GGFS1}) and (\ref{GGFS2}).

The vector field $\bshcW_{c}^{(\mp)}$ in the action associated with the term $\Tr \bshcW_{c}^{(-)}  \bshcW_{d}^{(+)}$ is reconstructed from the gauge field $\hcW_{c}^{\; a}$ to ensure gauge covariance under the transformation of the chirality boost-spin gauge symmetry W$^{1,3}$. Specifically, it takes the following form: 
\be \label{GCGF}
\bshcW_{c}^{(\mp)}  \equiv (\hcW_{c}^{\; a}  - \hcD_{c}\bs{\zeta}^{a} )  \frac{1}{2} \Sigma_{\mp a}, 
\ee
with the covariant derivative acting on $\bs{\zeta}^{a}(x)$ given by:
\be
\hcD_{c}\bs{\zeta}^{a} = (\eth_{c} +  \hcB_{c}) \bs{\zeta}^{a}  + \hcA_{c b}^{a}\bs{\zeta}^{b} .
\ee
Here, $\bs{\zeta}^{a}(x)$ is a spin-associated vector field introduced to ensure gauge covariance for the gauge field $\bshcW_{c}^{(\mp)}$. The vector field $\bs{\zeta}^{a}(x)$ transforms as follows:
\be
\bs{\zeta}^{a}(x) \to \bs{\zeta}^{a}(x) + \varpi^{a}(x), 
\ee
under the translation-like transformation associated with the chirality boost-spin gauge symmetry W$^{1,3}$.

The field strength $\bshcF_{cd}^{(W\mp)}$ in the action defines a gauge covariant field strength corresponding to the gauge covariant field $\bshcW_{c}^{(\mp)}$, with the following explicit form:
\be
& & \bshcF_{cd}^{(W\mp)}  \equiv \hcD_{c} \bshcW_{d}^{(\mp)} - \hcD_{d} \bshcW_{c}^{(\mp)} \equiv (\hcF_{cd}^{a} - \hcF_{cd}\bs{\zeta}^{a}  - \hcF_{cd}^{ab} \bs{\zeta}_{b} )  \frac{1}{2}\Sigma_{\mp a}, \nn \\
& & \hcF_{cd}^{a} \equiv \hcD_{c} \hcW_{d}^{a} - \hcD_{d} \hcW_{c}^{a} ,\; \hcD_{c} \hcW_{d}^{a}\equiv (\hat{\eth}_{c} + \hcB_{c})\hcW_{d}^{a} + \bshmOm_{cd}^{b} \hcW_{b}^{a} + \hcA_{c b}^{a}\hcW_{d}^{b} .
\ee
Notably, a gravigauge derivative $\hat{\eth}_{c}$ is introduced as follows:
\be
\hat{\eth}_{c} \equiv \eth_{c} - \hcS_{c} ,
\ee
which is necessary to maintain scaling gauge covariance for all gauge field strengths defined in gravigauge spacetime.

To describe a more general spin gauge-invariant interaction of the gravigauge field, we extend the action presented in ref. \cite{GSM} by adding a term involving the spin gauge-covariant gravigauge field strength $\bshcG_{cd}^{(\mp)}$, defined such that it is also local scaling gauge-covariant: 
\be
\bshcG_{cd}^{(\mp)} \equiv  (\hcG_{cd}^{a} + \hcS_{[cd]}^{a} )  \frac{1}{2}\Sigma_{\mp a}, \quad  \hcG_{cd}^{a} \equiv \chih_c^{\; \mu} \chih_d^{\; \nu} \cG_{\mu\nu}^{a} , 
\ee
where $ \cG_{\mu\nu}^{a}$ is given in Eq.(\ref{GCGFS}).

The vector field $\bshcA_{c}$ in the action linked to the term $\Tr \bshcA_{c}\bshcA_{d}$ is a gauge-covariant spin gauge field defined as follows:
\be
\bshcA_{c} \equiv (\hcA_{c}^{ab}  -  \bshmOm_{c}^{ab} )\Sigma_{ab}/2 , 
\ee
where the spin gravigauge field $\bshmOm_{c}^{ab}$ has the same transformation property as the spin gauge field $\hcA_{c}^{ab}$ under the spin gauge transformation.

The covariant derivative over $\hPhi_{\mp}$ is defined as follows:
\be
& & \hcD_{c}\hPhi_{\mp} \equiv (\hat{\eth}_{c} - i \hB_{c} \tGa_{\pm}/2  - i \hW_{c}^{i}\Sigma_{L \mp}^{i} ) \hPhi_{\mp}  ,  \nn \\
& &  \eta^{cd} \Tr (\hcD_{c}\hPhi_{\mp} )^{\dagger} \hcD_{d}\hPhi_{\mp} = 8\eta^{cd} (\hcD_{c}\hH)^{\dagger} \hcD_{d}\hH ,
\ee
where the scalar field $\hPhi_{\mp}= \hphi_{I} \Sigma_{h \mp}^{I}$ (or $\hH$ ) corresponding to the Higgs doublet in the SM carries a non-zero local scaling charge, $\hC_s =1$, in gravigauge spacetime.

The real scalar fields, $\bs{\zeta}$, $\hphi_w$, $\hPhi_{\kappa}$, $\hphi_{e}$, and the real vector field, $\bs{\zeta}^a$, are introduced to preserve the chiral conformal-spin gauge symmetry SP$_c$(1,1), the local scaling gauge symmetry SG(1), and the chirality boost-spin gauge symmetry W$^{1,3}$ in gravigauge spacetime. Under the transformations of the gauge symmetries, W$^{1,3}$$\rtimes$SP$_c$(1,1)$\times$SG(1), these fields transform as follows: 
\be
& & \bs{\zeta} \to \bs{\zeta}' = e^{-\varpi} \bs{\zeta},  \nn \\
& &  \hphi_w \to \hphi'_w = e^{\varpi} \hphi_w ,  \;\;  \hphi_w \to \hphi'_w = \xi(x) \hphi_w, \nn \\
& & \bs{\zeta}^a \to \bs{\zeta}^{' a} = e^{-\varpi} \bs{\zeta}^a,, \;\; \bs{\zeta}^a \to \bs{\zeta}^{' a} = \bs{\zeta}^a + \varpi^a ,  \nn \\
& &   \hPhi_{\kappa} \to  \hPhi'_{\kappa} =  \xi(x) \hPhi_{\kappa}, \;\; \hphi_{e} \to  \hphi'_{e} =  \xi(x) \hphi_{e}  ,
\ee
which indicates that $\bs{\zeta}$ and $\bs{\zeta}^a$ carry negative conformal-spin charges $C_c=-1$, while $\hphi_w$ possesses both a conformal-spin charge $C_c=1$ and a local scaling charge $\hC_s=1$. On the other hand, $\hPhi_{\kappa}$ and $\hphi_{e}$ have only a local scaling charge $\hC_s= 1$. 

The covariant derivatives of scalar fields with non-zero conformal-spin charge and non-zero local scaling charge are given as follows:
\be
& & \hcD_{c}\hPhi_{\kappa} \equiv (\eth_{c} + \hcW_{c} ) \hPhi_{\kappa} , \quad \hcD_{c}\bs{\zeta} \equiv (\eth_{c} + \hcB_c ) \bs{\zeta} ,  \nn \\
& & \hcD_{c}\hphi_{e} \equiv (\eth_{c} + \hcW_{c} ) \hphi_{e} , \quad \hcD_{c}(\bs{\zeta}\hphi_w) \equiv (\eth_{c} + \hcW_{c}) (\bs{\zeta}\hphi_w) .
\ee  

The last term in the above action, expressed as $\hPhi_{\kappa}^4\, \hcV_{S}( \hPhi_{\mp}/\hPhi_{\kappa}, \bs{\zeta} \hphi_w/\hPhi_{\kappa}, \hphi_{e}/\hPhi_{\kappa})$, represents local scaling gauge and chiral conformal-spin gauge invariant potentials associated with the scalar fields $\hphi_{I}$, $\bs{\zeta}$, $\hphi_w$ and $\hphi_{e}$. In general, 
the gauge symmetries alone cannot determine a specific form for the scalar potentials.

Recognizing the SM as an effective theory derived from a more fundamental framework, the effective Higgs potential $\hcV_{h}(\hPhi/\Phi_{\kappa})$, which ensures local scaling gauge invariance and reproduces the SM potential when the scaling gauge is fixed, can be expressed as follows:
\be \label{HP}
\hcV_{h}(\hPhi/\hPhi_{\kappa})\equiv \frac{1}{4}\lambda_h [ \frac{1}{16\hPhi_{\kappa}^2} \Tr (\hPhi_{-}^{\dagger}\hPhi_{-} + \hPhi_{+}^{\dagger}\hPhi_{+}) - \delta_{h}^2 ]^2 
= \frac{1}{4} \lambda_h ( \hH^{\dagger}\hH/\hPhi_{\kappa}^2 - \delta_{h}^2 )^2 ,
\ee
where the ordinary Higgs boson $H$ (or $\Phi$) in Minkowski spacetime is replaced by $\hH$ (or $\hPhi$), which carries a local scaling charge in gravigauge spacetime, and  $\delta_h$ is a dimensionless constant parameter characterizing the minimal condition at a stable point of the potential. For the potentials of other scalar fields, a similar property is expected, with minimal conditions at their respective stable points. Their specific forms will be examined in detail later.

The action presented in Eq.(\ref{GSMaction}) describes the GSM in spin-fiber gravigauge spacetime, governed by the enlarged gauge symmetries: $\hat{\fG}_{GSM}$ = U$_Y$(1)$\times$SU$_L$(2)$\times$SU$_C$(3) $\times$WS$_c$(1,3)$\times$SG(1) as shown in Eq.(\ref{GSMGS}) with the additional Z$_2$ discrete symmetry.

The gauge invariance of the action under the transformations of the chirality boost-spin gauge symmetry W$^{1,3}$ and the chiral conformal-spin gauge symmetry SP$_c$(1,1) allows us to choose the following specific gauge fixing conditions:
\be \label{CIGFB}
& & \bs{\zeta}^a \to \bs{\zeta}^{' a} = \bs{\zeta}^a + \varpi^a = 0 , \nn \\
& & \bs{\zeta} \to \bs{\zeta}' = e^{-\varpi} \bs{\zeta} = 1. 
\ee 
Under this gauge prescription, the gauge-covariant chirality boost-spin gauge field and its field strength are simply given by:
\be
& & \bshcW_{c}^{(\mp)} \to \bshcW_{c}^{'(\mp)} = \hcW_{c}^{'(\mp)} \equiv \hcW_{c}^{'\, a} \frac{1}{2}\Sigma_{a \mp } , \nn \\
& & \bshcF_{cd}^{(W\mp)} \to \bshcF_{cd}^{'(W\mp)}  = \hcF_{cd}^{'(W\mp)} \equiv \hcF_{cd}^{' a} \frac{1}{2}\Sigma_{a \mp} .
\ee

For convenience, we may refer the conformal inhomogeneous spin gauge fixing conditions ($\bs{\zeta}=1$ and $\bs{\zeta}^a = 0$) as the unitary conformal-boost gauge. In this gauge basis, the following replacements in the action can be directly applied:
\be
& & \bshcW_{c}^{(\mp)} \to \hcW_{c}^{(\mp)} , \quad \bshcF_{cd}^{(W\mp)}  \to \hcF_{cd}^{(W\mp)},\quad  \hcD_{c}\bs{\zeta} \to \hcB_c , \nn \\
& & \hcV_{S}( \hPhi_{\mp}/\hPhi_{\kappa}, \bs{\zeta} \hphi_w/\hPhi_{\kappa}, \hphi_{e}/\hPhi_{\kappa})  \to  \hcV_{S}( \hPhi_{\mp}/\hPhi_{\kappa}, \hphi_w/\hPhi_{\kappa}, \hphi_{e}/\hPhi_{\kappa})  .
\ee

Furthermore, the local scaling gauge invariance enables the imposition of specific gauge-fixing conditions. Two typical choices are commonly adopted. The first is given by:

Two typical gauge-fixing conditions are usually chosen. A particular one is taken as:
\be
& & \chi_{\mu}^{\;\; a}(x) \to \chi_{\mu}^{'\; a}(x) = \xi_u^{-1}(x) \chi_{\mu}^{\;\; a}(x), \nn \\
& & \det \chi_{\mu}^{\; a}\equiv \chi(x) \to \det \chi_{\mu}^{'\; a}\equiv \chi'(x) = \xi_u^{-4}(x) \chi(x) =1 ,
\ee
which is referred to as the flowing unitary scaling gauge. This gauge convention ($\chi(x)=1$) is frequently used to simplify the Einstein equations, provided the theory exhibits local scaling gauge invariance as such a condition is not achievable merely by selecting a specific coordinate system or reference frame in GR.

A second conventional gauge-fixing condition is: 
\be
\Phi_{\kappa}(x) \to \Phi'_{\kappa}(x) = \xi_e(x) \Phi_{\kappa}(x) = \bM_{\kappa}, 
\ee
where $\bM_{\kappa}$ represents a fundamental mass scale in four-dimensional spacetime. This choice defines the fundamental mass scaling gauge. Under this prescription ($\Phi_{\kappa} = \bM_{\kappa}$), the action further simplifies via the following replacements:
\be
 & & \cS_{c} \to 0, \;\; \hat{\eth}_{c} \to \eth_{c} , \; \; \bshsF_{cd}^{a} \to  \hsF_{cd}^{a}, \;\;   \bshmOm_{c}^{ab} \to  \hmOm_{c}^{ab}, \nn \\
 & &  \cD_{c}\Phi_{\kappa} \to \bM_{\kappa} g_w\hcW_{c}, \;\; \hphi_w \to \phi_w, \;\; \hphi_{e} \to \phi_{e} ,\nn \\
 & &  \hPhi\, (\mbox{or}\, \hH) \to \Phi\, (\mbox{or}\, H), \;\; \delta_h \to v_h/\bM_{\kappa}, \nn \\
 & & \hPhi_{\kappa}^4\, \hcV_{S}(\hPhi/\hPhi_{\kappa}, \hphi_w/\hPhi_{\kappa}, \hphi_{e}/\hPhi_{\kappa})\to \cV_{S}(H, \phi_w, \phi_{e}), 
 \ee
where $v_h$ denotes the vacuum expectation value (VEV) of the Higgs boson in the SM. 

For simplicity, let us adopt both the unitary conformal-boost gauge (where $\bs{\zeta}=1$ and $\bs{\zeta}^a = 0$) and the fundamental mass scaling gauge ($\Phi_{\kappa} = \bM_{\kappa}$). This allows us to simplify the action given in Eq. (\ref{GSMaction}) as follows:
\be \label{GSMaction1}
\cS_{GSM}  & = & \int [\dbar \zeta] \cL_{GSM}(  l_{L,R}^i,  q_{L,R}^i,  \hB_{c}, \hW_{c}^{i}, \hA_{c}^{\alpha}, H, \nn \\
& &\;\; \hcB_{c}, \hcW_{c}^{a}, \hcA_{c}^{ab},  \hcW_{c},  \phi_e, \phi_w ) \nn \\
& = & \int [\dbar \zeta] \lbrace \frac{1}{2} [ \bar{l}_{L}^{i} \gamma^{c}  i \hcD_{c}^{(l_L)} l_{L}^{i}  +\bar{l}_{R}^{i} \gamma^{c}  i  \hcD_{c}^{(l_R)} l_{R}^{i}  \nn \\
& + & \bar{q}_{L}^{i} \gamma^{c}  i \hcD_{c}^{(q_L)} q_{L}^{i}  +\bar{q}_{R}^{i} \gamma^{c}  i\hcD_{c}^{(q_R)} q_{R}^{i}  + \bar{l}_{L}^{i} H \lambda^{e}_{ij} e_{R}^{j} \nn \\
& + &   \bar{l}_{L}^{i} \tilde{H} \lambda^{\nu}_{ij} \nu_{R}^{j}  + \bar{q}_{L}^{i} H \lambda^{d}_{ij} d_{R}^{j} +  \bar{q}_{L}^{i} \tilde{H} \lambda^{u}_{ij} u_{R}^{j}  + H.c. ] \nn \\
& - & \frac{1}{4} \eta^{c c'}\eta^{d d'} ( \hF_{cd} \hF_{c'd'} 
+ \hF_{cd}^{i} \hF_{c'd'}^{i} + \hF_{cd}^{\alpha} \hF_{c'd'}^{\alpha}  \nn \\
& + &  \hcF_{cd}\hcF_{c'd'} -  \hcF_{cd}^{a} \hcF_{c'd' a} +   \hcF_{cd}^{ab} \hcF_{c'd' ab}  + \hcW_{cd}\hcW_{c'd'} ) \nn \\
& + & \eta^{cd} (\hD_{c}H)^{\dagger} \hD_{d}H + \frac{1}{4} \bM_{\kappa}^2\etat^{cd c'd'}_{aa'}\hsF_{cd}^{a}\hsF_{c'd'}^{a'}  \nn \\
 & + & \frac{1}{2} M_{\cA}^2 \chi_{\cA}^2  (\cA_{[cda]} - g_4^{-1}\mOm_{[cda]})(\cA^{[cda]} - g_4^{-1} \mOm^{[cda]})   \nn \\
 & + & \frac{1}{2} M_{\cG}^2\chi_{\cG}^2 (\cA_{[cda)} - g_4^{-1}\mOm_{[cda)})(\cA^{[cda)} - g_4^{-1}\mOm^{[cda)})   \nn \\
& + &  \frac{1}{2}  \eta^{cd} [  \phi_w^2 (g_c^2\gamma_w^2 \hcB_{c} \hcB_{d} - g_4^2\beta_{w}^2 \hcW_{c}^{a}  \hcW_{d a}  )  + \beta_{\kappa}^2 g_w^2 \bM_{\kappa}^2 \hcW_{c}\hcW_{d} ] \nn \\
& + & \frac{1}{2} \eta^{cd}  (\hcD_{c}\phi_{w} \hcD_{d}\phi_{w} + \hcD_{c}\phi_{e} \hcD_{d}\phi_{e} )  - \cV_{S}(H, \phi_w, \phi_{e})  \rbrace .
\ee

In the above action, the gauge fields have been rescaled as: 
\be
 (\hB_{c}, \hW_{c}^i , \hA_{c}^{\alpha}, \hcA_{c}^{ab},  \hcW_{c}^{a}, \hcB_{c}, \hcW_{c}) \to (g' \hB_{c}, g \hW_{c}^i , g_3 \hA_{c}^{\alpha}, g_4\hcA_{c}^{ab}, g_4  \hcW_{c}^{a},, g_c\hcB_{c}, g_w\hcW_{c}) ,
\ee
in order to normalize their kinetic terms. Here, the gauge coupling $g_c$ is given by, 
\be
g_c \equiv g_4\gamma_c/\sqrt{1-\gamma_c^2}.
\ee
We introduced the following definition:
\be
 i\hcD_{c}^{(\Psi_{L,R})}  \equiv i \hD_{c}^{(\Psi_{L,R})} + g_4 \hcA_{c}^{ab}\Sigma_{ab}/2 \equiv \chih_{c}^{\; \mu} (  i D_{\mu}^{(\Psi_{L,R})} + g_4 \cA_{\mu}^{ab}\Sigma_{ab}/2 ) , 
\ee
with $D_{\mu}^{(\Psi_{L,R})}$ ($\Psi =l, q$) representing the covariant derivatives acting on leptons and quarks in the SM.

To derive the action, we used the following relation:
\be \label{GGGR}
& &  R_{cd}^{ab} \eta_a^c\eta_b^d \equiv - \frac{1}{4} \etat_{aa'}^{cdc'd'} \hsF_{cd}^{a} \hsF_{c'd'}^{a'} 
+ 2 \hD_{b} (\eta^{bc}\hsF_{cd}^{a}\eta_a^d ) , 
\ee
where $\etat^{cdc'd'}_{aa'}$ is a constant tensor defined as:
\be \label{STensor1}
\tilde{\eta}^{cd c'd'}_{a a'}  & \equiv &    \eta^{c c'} \eta^{d d'} \eta_{a a'}  
+  \eta^{c c'} ( \eta_{a'}^{d} \eta_{a}^{d'}  -  2\eta_{a}^{d} \eta_{a'}^{d'}  ) +  \eta^{d d'} ( \eta_{a'}^{c} \eta_{a}^{c'} -2 \eta_{a}^{c} \eta_{a'}^{c'} ) ,
\ee
which ensures the spin gauge invariance up to a total derivative. 

Additionally, the following identity is employed:
\be
\frac{1}{4} \eta^{cc'}\eta^{dd'}\eta_{aa'}\cG_{cd}^{a}\cG_{c'd'}^{a'}  & \equiv &  \frac{1}{4} ( g_4 \hcA_{[cd]a} + \hsF_{cd a})  ( g_4 \hcA^{[cd]a} + \hsF^{cd a})  \nn \\
& = &  ( g_4 \hcA_{[cda]}-\hmOm_{[cda]})  ( g_4 \hcA^{[cda]}-\hmOm^{[cda]}) \nn \\
& + &  \frac{1}{4}  ( g_4 \hcA_{[cda)}-\hmOm_{[cda)})  ( g_4 \hcA^{[cda)}-\hmOm^{[cda)}) .
\ee
with the definitions:
\be \label{SGF}
& & \hcA_{cd a} \equiv \hcA_{[cda]} + \hcA_{[cda)} , \quad \hcA_{[cd]a} \equiv \hcA_{cd a} - \hcA_{dc a},  \nn \\
& &  \hcA_{[cda]} \equiv \frac{1}{3} (  \hcA_{cda} +  \hcA_{dac} +  \hcA_{acd} ) , \nn \\
& &  \hcA_{[cda)} \equiv \frac{1}{3} (  \hcA_{cda} + \hcA_{dca} -  \hcA_{cad} -  \hcA_{acd} ) ,
\ee
where $\hcA_{[cda]}$ is the totally antisymmetric spin gauge field (4 independent degrees of freedom), and $\hcA_{[cda)}$ is the symmetric-antisymmetric mixed component (20 degrees of freedom), so $\hcA_{[cda]}$ and $\hcA_{[cda)}$ are orthogonal, $\hcA_{[cda]} \hcA^{[cda)} =0$. Analogous definitions and relations apply to the spin gravigauge field $\hmOm_{cda}$. 

We also introduce the following mass parameters and field-dependent terms: 
\be \label{MP}
& & M_{\cA}^2 = (\gamma_{\kappa} + 2\alpha_{\kappa} ) g_4^2\bM_{\kappa}^2 + (\beta_e+ 2\alpha_e) g_4^2 v_e^2, \nn \\
& & M_{\cG}^2 = ( \gamma_{\kappa} + \alpha_{\kappa}/2 ) g_4^2 \bM_{\kappa}^2 + (\beta_e + \alpha_e/2 )  g_4^2 v_e^2, 
\ee
and
\be
& & \chi_{\cA}^{2} \equiv 1  + (\beta_e + 2\alpha_e )g_4^{2}(\phi_{e}^2-v_e^2)/M_{\cA}^2 , \nn \\
& &  \chi_{\cG}^{2} \equiv 1  +  (\beta_e + \alpha_e/2 )g_4^{2}(\phi_{e}^2-v_e^2)/M_{\cG}^2 .
\ee
Here, $M_{\cA}^2$ and $M_{\cG}^2$ correspond to the masses of the spin gauge field components $\hcA_{[cda]}$  and $\hcA_{[cda)}$, respectively, in spin-fiber gravigauge spacetime. 
The constant $v_e$ is a proposed VEV of the scalar field $\phi_e$.

Notably, the introduction of the dual bi-covariant vector fields $\chih_{a}^{\; \mu}(x)$ and $\chi_{\mu}^{\; a}(x)$ is essential to preserve the principle of gauge invariance, which is fundamentally linked to the spin gauge symmetry SP(1,3). The spin-fiber gravigauge spacetime, emerging from the gravigauge bases constructed via these dual bi-covariant vector fields, exhibits the properties of a local orthogonal and non-commutative spacetime. Its non-commutative nature is characterized by the gravigauge field strength, $\hsF_{cd}^a$, which serves as a local group structure factor. This structural framework makes it clear that the gravigauge field strength should not be regarded as an independent dynamical field within the spin-fiber gravigauge spacetime.

Indeed, the absence of a dynamic term for the gravigauge field strength $\hsF_{cd}^a$ in the action enables us to derive, through the least action principle, the following relation:
\be \label{GE}
\bscM_{cda}^{\; c'd'a'} \hsF_{c'd' a'}  = \widehat{\bscF}_{cda} ,  \quad \mbox{or}\quad \hsF_{cd a}  = \widehat{\bscM}_{cda}^{\; c'd'a'} \widehat{\bscF}_{c'd'a'}
\ee
which provides a constraint equation for the gravigauge field strength $\hsF_{cd}^a$ in spin-fiber gravigauge spacetime. In the unitary conformal-boost gauge basis, $\bscM_{cda}^{\; c'd'a'}$ has the following explicit form:
\be
& & \bscM_{cda}^{\; c'd'a'}  \equiv \frac{1}{2} (\eta_{c}^{\, c'} \eta_{d}^{\, d'} -\eta_{d}^{\, c'} \eta_{c}^{\, d'} ) ( \eta_{a}^{\, a'} \hat{\cM}_{+}^2   - \widehat{\cV}_{a b} \eta^{b a'} ) \nn \\
& & + \frac{1}{2} [ ( \eta_{d}^{\, c'} \eta_{c}^{\, a'}- \eta_{c}^{\, c'}  \eta_{d}^{\, a'}   ) \eta_{a}^{\, d'} -  (\eta_{d}^{\, d'} \eta_{c}^{\, a'} - \eta_{c}^{\, d'} \eta_{d}^{\, a'} )\eta_{a}^{\, c'} ] \hat{\cM}_{-}^{2} \nn \\
& & -  [ \eta_{ca} (\eta_{d}^{\, d'}  \eta^{c'a'} - \eta_{d}^{\, c'}  \eta^{d'a'}) -  \eta_{da}( \eta_{c}^{\, d'}  \eta^{c'a'} -\eta_{c}^{\, c'}  \eta^{d'a'} ) ] \bM^2_{\kappa} , 
\ee
with
\be
& & \widehat{\cV}_{a b}  \equiv  \hcB_{a} \hcB_{b} - \hcW_{a c} \hcW_{b}^{c} + \hcA_{a a'b'}\hcA_{b}^{a'b'}  +\hcW_{a} \hcW_{b} \nn \\
& & \qquad + \hB_{a}\hB_{b}  +  \hW_{a}^{i} \hW_{b}^{i} + \hA_{a}^{\alpha} \hA_{b}^{\alpha } = \widehat{\cV}_{ba} , \nn \\
& & \hat{\cM}_{+}^2  =   \bM_{\kappa}^2 + \frac{1}{2} [(\gamma_{\kappa} + \alpha_{\kappa})\bM^2_{\kappa} + (\beta_{e} + \alpha_e) \hphi_{e}^2 ]  +  (\gamma_{\kappa}\bM^2_{\kappa} + \beta_{e}\hphi_{e}^2), \nn \\
& & \hat{\cM}_{-}^2  =   \bM_{\kappa}^2 + \frac{1}{2} (\gamma_{\kappa}\bM^2_{\kappa} + \beta_{e}\hphi_{e}^2) -  (\gamma_{\kappa}\bM^2_{\kappa} + \beta_{e}\hphi_{e}^2), 
\ee
and $\widehat{\bscF}_{cda}$ is given by,
\be
\widehat{\bscF}_{cda}  & \equiv &  \cF_{cd}\hcB_{a} -  \cF_{cd}^{b} \hcW_{a b}  + \cF_{cd}^{a'b'}\hcA_{a a'b'}  \nn \\
& + & \cW_{cd} \hcW_{a}  + F_{cd} \hB_{a} +  F_{cd}^{i} \hW_{a}^{i} + F_{cd}^{\alpha}\hA_{a}^{\alpha} \nn \\
& - &  (\gamma_{\kappa}\bM^2_{\kappa} + \beta_{e}\hphi_{e}^2)g_4  ( \hcA_{acd}  + 2\hcA_{[cd]a} )  \nn \\
& -&  (\alpha_{\kappa}\bM^2_{\kappa} + \alpha_{e}\hphi_{e}^2)  g_4 \hcA_{[cd]a} =  - \widehat{\bscF}_{dc a} .
\ee
Here, $\bscM_{cda}^{\; c'd'a'}$ may be regarded as a $24\times 24$ matrix, antisymmetric under exchanges of $c, d$ and $c', d'$. While $\widehat{\bscM}_{cda}^{\; c'd'a'} $ represents the inverse matrix and satisfies the following condition:
\be
& & \widehat{\bscM}_{cda}^{\; c'd'a'} \bscM_{c'd'a'}^{\; \tc \td \ta}  = \frac{1}{2} (\eta_{c}^{\, \tc} \eta_{d}^{\, \td} - \eta_{c}^{\, \tc} \eta_{d}^{\, \td} ) \eta_{a}^{\, \ta}  . 
\ee

In constructing the action for the General Standard Model (GSM) within spin-fiber gravigauge spacetime, as shown in Eq.(\ref{GSMaction}), it becomes evident that the gravigauge field strength, $\hsF_{cd}^a$, or the spin gravigauge field $\hmOm_{c}^{ab}$, emerges as an auxiliary field. The resulting constraint equation presented  in Eq. (\ref{GE}) indicates that $\hsF_{cd}^a$ is intricately governed by the collective dynamics of the conformal inhomogeneous spin gauge field, along with all other fundamental gauge fields and their corresponding field strengths. For convenience, we refer to Eq. (\ref{GE}) as the {\it gravitization equation}.

The gravitization equation enables the determination of the gravigauge field strength $\hsF_{cd}^a$, demonstrating that the spin-fiber gravigauge spacetime manifests as an emergent spacetime characterized by non-commutative geometry. Consequently, the gravitational interaction revealed through $\hsF_{cd}^a$ arises due to the non-commutative nature of the spin-fiber gravigauge spacetime.

In a word, we formulate a complete and gauge-invariant action for the GSM. A key result is the derivation of a constraint equation for the gravigauge field strength. It is interpreted as the gravitization equation, which shows how the gravitational effect emerges as a collective manifestation of the dynamics of all other gauge fields.

\section{Gravidynamics of the General Standard Model in GQFT}

The gravigauge field $\chi_{\mu}^{\;\; a}$, as a bi-covariant vector field in bi-frame spacetime, acts as a Goldstone-type boson. It facilitates the projection of the action constructed in spin-fiber gravigauge spacetime into an action formulated within the framework of gravitational quantum field theory (GQFT) based on bi-frame spacetime. 

By reformulating the action in spin-fiber gravigauge spacetime presented in Eq.(\ref{GSMaction1}), we arrive at the GSM within the framework of GQFT as follows: 
\be \label{GSMactionGQFT}
\cS_{GSM}  & = & \int [d x]\chi \cL_{GSM}(  l_{L,R}^i,  q_{L,R}^i, B_{\mu}, W_{\mu}^{i}, A_{\mu}^{\alpha}, H,
 \nn \\
& & \qquad \qquad \qquad \cB_{\mu}, \cW_{\mu}^{a}, \cA_{\mu}^{ab}, \cW_{\mu}, \phi_e,\phi_w )  \nn \\
& = & \int [d x]\chi \lbrace \frac{1}{2} [ \bar{l}_{L}^{i} \gamma^{a}  i \chih_{a}^{\; \mu} \cD_{\mu}^{(l_L)} l_{L}^{i}  +\bar{l}_{R}^{i} \gamma^{a}  i \chih_{a}^{\; \mu} \cD_{\mu}^{(l_R)} l_{R}^{i}  \nn \\
& + & \bar{q}_{L}^{i} \gamma^{a}  i \chih_{a}^{\; \mu} \cD_{\mu}^{(q_L)} q_{L}^{i}  +\bar{q}_{R}^{i} \gamma^{a}  i \chih_{a}^{\; \mu} \cD_{\mu}^{(q_R)} q_{R}^{i}  + \bar{l}_{L}^{i} H \lambda^{e}_{ij} e_{R}^{j} \nn \\
& + &   \bar{l}_{L}^{i} \tilde{H} \lambda^{\nu}_{ij} \nu_{R}^{j}  + \bar{q}_{L}^{i} H \lambda^{d}_{ij} d_{R}^{j} +  \bar{q}_{L}^{i} \tilde{H} \lambda^{u}_{ij} u_{R}^{j}  + H.c. ] \nn \\
& - & \frac{1}{4} \chih^{\mu\mu'}\chih^{\nu\nu'} ( F_{\mu\nu} F_{\mu'\nu'} 
+ F_{\mu\nu}^{i} F_{\mu'\nu'}^{i} + F_{\mu\nu}^{\alpha} F_{\mu'\nu'}^{\alpha}  \nn \\
& + &  \cF_{\mu\nu}\cF_{\mu'\nu'} -  \cF_{\mu\nu}^{a} \cF_{\mu'\nu' a} +   \cF_{\mu\nu}^{ab} \cF_{\mu'\nu' ab}  + \cW_{\mu\nu}\cW_{\mu'\nu'} ) \nn \\
& + &\frac{1}{4} \bM_{\kappa}^2\tchi^{\mu\nu\mu'\nu'}_{aa'}\sF_{\mu\nu}^{a}\sF_{\mu'\nu'}^{a'} + \frac{1}{4} m_G^2 \chih^{\mu\mu' \nu\nu'}_{aa'}  \cG_{\mu\nu}^{a}\cG_{\mu'\nu' }^{a'} \nn \\
 & + & \chih^{\mu\nu}  [ (D_{\mu}H)^{\dagger} D_{\nu}H + \frac{1}{2}\beta_{\kappa}^2 g_w^2 \bM_{\kappa}^2 \cW_{\mu}\cW_{\nu} ] \nn \\
& + &  \frac{1}{2}  \chih^{\mu\nu} \phi_w^2 (g_c^2\gamma_w^2 \cB_{\mu} \cB_{\nu} - g_4^2\beta_{w}^2 \cW_{\mu}^{a}  \cW_{\nu a}  )   \nn \\
& + & \frac{1}{2} \chih^{\mu\nu}  (\cD_{\mu}\phi_{w} \cD_{\nu}\phi_{w} + \cD_{\mu}\phi_{e} \cD_{\nu}\phi_{e} )  - \cV_{S}(H, \phi_w, \phi_{e})  \rbrace ,
\ee
where we have adopted the relation between the two integral measures, $[\dbar \zeta]\equiv [dx]\, \chi$, with $\chi = \det \chi_{\mu}^{\; a} $, and introduced the following definitions:
\be
& & \cF_{\mu\nu}^{ab} =  \p_{\mu} \cA_{\nu}^{ab} - \p_{\nu} \cA_{\mu}^{ab} + g_4(\cA_{\mu c}^{a} \cA_{\nu}^{cb} - \cA_{\nu c}^{a} \cA_{\mu}^{cb}), \nn \\
& & \cF_{\mu\nu}^{a}  =  (\p_{\mu} + g_c\cB_{\mu})\cW_{\nu}^{\; a} - (\p_{\nu} + g_c\cB_{\nu})\cW_{\mu}^{\; a} \nn \\
& & \quad \quad + g_4(\cA_{\mu c}^{a} \cW_{\nu}^{\; c} - \cA_{\nu c}^{a} \cW_{\mu}^{\; c} ), 
\ee
and
\be
& & D_{\mu}H \equiv (\p_{\mu} - i g' B_{\mu} - i \frac{1}{2}g W_{\mu}^{i}\sigma^i ) H, \nn \\
& & \cD_{\mu}\phi_{e} \equiv (\p_{\mu} + g_w \cW_{\mu} ) \phi_{e}, \; \cD_{\mu}\phi_{w} \equiv (\p_{\mu} + g_w \cW_{\mu} ) \phi_w .
\ee

The relevant tensors in the above action are defined as follows: 
\be \label{STensor2}
& & \tchi_{aa'}^{\mu\nu \mu'\nu'} \equiv \chih_{c}^{\;\, \mu}\chih_{d}^{\;\, \nu} \chih_{c'}^{\;\, \mu'} \chih_{d'}^{\;\, \nu'}  \etat^{c d c' d'}_{a a'} , \nn \\
& &  \chih^{\mu\nu} \equiv \chih_{a}^{\; \mu} \chih_{b}^{\; \nu} \eta^{ab} , \; \; \chi_{\mu\nu} \equiv \chi^{\;a}_{\mu} \chi^{\; b}_{\nu} \eta_{ab} , \nn \\
& & \chi = \det \chi_{\mu}^{\; a} = \sqrt{-\det \chi_{\mu\nu}}, \;\; \chih = 1/\chi ,
\ee  
with $\etat^{c d c' d'}_{a a'}$ given in Eq.(\ref{STensor1}). Here, $\chi_{\mu\nu}$ is regarded as a symmetric composite field of the gravigauge field, defining a {\it gravimetric field}, and $\chih^{\mu\nu}$ is the dual tensor of the gravimetric tensor $\chi_{\mu\nu}$.

In deriving the above action, we have expressed the spin gauge invariant mass-like term $\eta^{cd}\Tr \bshcA_{c}\bshcA_{d}$ into a spin gauge invariant field strength in the fundamental mass scaling gauge, which has the following explicit relation:
\be \label{MSG}
& & \frac{1}{2} \eta^{cd}\Tr \bshcA_{c}\bshcA_{d}  \equiv  \frac{1}{2} \eta^{cd} ( g_4 \hcA_{c}^{ab}-\hmOm_{c}^{ab})  ( g_4\hcA_{d ab}-\hmOm_{d ab} ) \nn \\
& & = \frac{1}{2} \chih^{\mu\nu}  ( g_4 \cA_{\mu}^{ab}-\mOm_{\mu}^{ab} )  ( g_4\cA_{\nu ab}-\mOm_{\nu ab}) \equiv  \frac{1}{4} \bchi^{\mu\nu \mu' \nu'}_{aa'}\cG_{\mu\nu}^{a}\cG_{\mu'\nu'}^{a'}  , 
 \ee
where the spin gauge covariant field strength $\cG_{\mu\nu}^{a}$ and the tensor $\bchi^{\mu\nu \mu' \nu'}$ are given as follows:
\be \label{STensor3}
& & \cG_{\mu\nu}^{a}  \equiv  \p_{\mu} \chi_{\nu}^{\; a}  - \p_{\nu} \chi_{\mu}^{\; a} + g_4(\cA_{\mu b}^{a} \chi_{\nu}^{\; b}  - \cA_{\nu b}^{a} \chi_{\mu}^{\; b} ), \nn \\
& & \chib_{aa'}^{\mu\nu \mu'\nu'} \equiv \chih_{c}^{\;\, \mu}\chih_{d}^{\;\, \nu} \chih_{c'}^{\;\, \mu'} \chih_{d'}^{\;\, \nu'}  \etab^{c d c' d'}_{a a'} , \nn \\
& & \etab^{c d c' d'}_{a a'} \equiv \frac{3}{2}  \eta^{c c'} \eta^{d d'} \eta_{a a'}  
+ \frac{1}{2} ( \eta^{c c'} \eta_{a'}^{d} \eta_{a}^{d'}  +  \eta^{d d'} \eta_{a'}^{c} \eta_{a}^{c'} ) .
\ee

We also defined the following tensors in the above action: 
\be \label{STensor3}
& & \chih_{aa'}^{\mu\nu \mu'\nu'} \equiv \chih_{c}^{\;\, \mu}\chih_{d}^{\;\, \nu} \chih_{c'}^{\;\, \mu'} \chih_{d'}^{\;\, \nu'} [ \hat{\eta}^{c d c' d'}_{a a'} +  \check{\eta}^{c d c' d'}_{a a'} (\phi_e^2-v_e^2)/m_G^2 ] , \nn \\
& & \hat{\eta}^{c d c' d'}_{a a'} \equiv \alpha_G \eta^{c c'} \eta^{d d'} \eta_{a a'}  
- \frac{1}{2} \alpha_W ( \eta^{c c'} \eta_{a'}^{d} \eta_{a}^{d'}  
+  \eta^{d d'} \eta_{a'}^{c} \eta_{a}^{c'} ) , \nn \\
& &  \check{\eta}^{c d c' d'}_{a a'} \equiv (\frac{3}{2} \beta_e + \alpha_e) \eta^{c c'} \eta^{d d'} \eta_{a a'}  +\frac{1}{2}\beta_e  ( \eta^{c c'} \eta_{a'}^{d} \eta_{a}^{d'}  
+  \eta^{d d'} \eta_{a'}^{c} \eta_{a}^{c'} ) .
\ee  
The following constant parameters are introduced:
\be \label{CPA}
& & \alpha_G \equiv  \frac{3}{2} + \frac{\beta_G}{\gamma_G }  , \quad  \alpha_W \equiv -1 , \nn \\
& & \gamma_G \equiv \gamma_{\kappa} + \beta_e \frac{v_e^2}{\bM_{\kappa}^2},\;\;  \beta_G \equiv \alpha_{\kappa}+ \alpha_e \frac{v_e^2}{\bM_{\kappa}^2 } ,  \nn \\
& & m_G^2\equiv \gamma_G \bM_{\kappa}^2 , \quad 16\pi G_{\kappa} = 1/\bM_{\kappa}^2 .
\ee

In the unitary conformal-boost gauge ($\bs{\zeta}=1$ and $\bs{\zeta}^a = 0$) and the fundamental mass scaling gauge ($\Phi_{\kappa} = \bM_{\kappa}$), the action in Eq.(\ref{GSMactionGQFT}) for the GSM within the framework of GQFT exhibits the following joint symmetry:
\be
\fG_{GSM} = \mbox{SC}(1)\times \mbox{PO}(1,3) \Join \mbox{U}_Y(1) \times \mbox{SU}_L(2) \times \mbox{SU}_C(3)\times \mbox{SP}(1,3),
\ee
with PO(1,3) and SC(1) representing the Poincar\'e group and global scaling symmetries. 

By applying the principle of least action to the gravigauge field $\chi_{\mu}^{\;\; a}(x)$, we derive the following general gauge-type gravitational equation: 
\be  \label{GaGE}
 \p_{\nu} \whsF^{\mu\nu }_{a}  = \whfJ_{a}^{\; \mu}   , 
\ee
 with an associated conserved current:
 \be
 \p_{\mu} \whfJ_{a}^{\; \mu} = 0 ,
 \ee
which arises from the antisymmetric property of the field strength $\whsF^{\mu\nu }_{a} = -\whsF^{\nu\mu }_{a}$,  leading to $\p_{\mu} \whfJ_{a}^{\; \mu} =  \p_{\mu} \p_{\nu} \whsF^{\mu\nu }_{a} = - \p_{\mu} \p_{\nu} \whsF^{\mu\nu }_{a} = 0$. Here, the field strength $\whsF^{\mu\nu }_{a}$ and the current $\whfJ_{a}^{\; \mu}$ are defined as follows:
\be \label{FSCC}
\whsF^{\mu\nu }_{a} & \equiv & \chi \1 \tchi^{[\mu\nu]\rho\sigma}_{a b}  \sF_{\rho\sigma }^{b}  , \nn \\
\whfJ_{a}^{\;\mu} & \equiv & \gamma_G \hcG_{a}^{\;\mu}  + \hsF_{a}^{\;\mu} -16\pi G_{\kappa} \hmJ_{a}^{\;\mu} ,
\ee
where the current $\whfJ_{a}^{\; \mu}$ has been decomposed into three parts with the following explicit forms:
\be \label{GCC}
& & \hcG_{a}^{\;\mu} = - \cD_{\nu}(\widehat{\cG}_{a}^{\mu\nu} ) + (\eta_{\sigma}^{\; \mu}\chih_{a}^{\; \rho} -  \frac{1}{4} \chih_{a}^{\; \mu} \eta_{\sigma}^{\; \rho} ) \cG_{\rho\nu}^{b} \widehat{\cG}^{\sigma\nu}_{b} ] , \nn \\
& & \hsF_{a}^{\;\mu} =  (\eta_{\sigma}^{\; \mu}\chih_{a}^{\; \rho} -  \frac{1}{4} \chih_{a}^{\; \mu} \eta_{\sigma}^{\; \rho} ) \sF_{\rho\nu}^{b} \whsF^{\sigma\nu}_{b},
\ee
and
\be \label{GCC1}
\hmJ_{a}^{\;\mu} & = & \chi \1 \{ \frac{1}{2} [  ( \chih_{a}^{\; \; \mu} \chih_{b}^{\; \rho}   - \chih_{b}^{\; \mu}  \chih_{a}^{\;\rho}) ( \bar{l}_{L}^{i} \gamma^{b} i\cD_{\rho}^{(l_L)} l_{L}^{i}  + \bar{l}_{R}^{i} \gamma^{b} i\cD_{\rho}^{(l_R)} l_{R}^{i}  \nn \\
& + & \bar{q}_{L}^{i} \gamma^{b} i\cD_{\rho}^{(q_L)} q_{L}^{i} + \bar{q}_{R}^{i} \gamma^{b} i\cD_{\rho}^{(q_R)} q_{R}^{i}  ) +  \chih_{a}^{\;\;\mu}( \bar{l}_{L}^{i} H \lambda^{e}_{ij} e_{R}^{j}   \nn \\
& + &  \bar{l}_{L}^{i} \tilde{H} \lambda^{\nu}_{ij} \nu_{R}^{j} +  \bar{q}_{L}^{i} H \lambda^{d}_{ij} d_{R}^{j} +  \bar{q}_{L}^{i} \tilde{H} \lambda^{u}_{ij} u_{R}^{j} )  + H.c.  ]  \nn \\
%& & =  \chi \1 \{  \sum_{\Psi=l,q} \frac{1}{2} [  (   \chih_{a}^{\; \; \mu} \chih_{b}^{\; \nu} -\chih_{b}^{\; \mu}  \chih_{a}^{\;\nu} )  \bar{\Psi}_{L}^{i} \gamma^{b} i\cD_{\nu}^{(\Psi_L)} \Psi_{L}^{i}  \nn \\
%& &\, + \chih_{a}^{\;\;\mu}( \bar{\Psi}_{L}^{i} H \lambda^{\Psi^{(-)}}_{ij} \Psi_{R}^{(-)j} +  \bar{\Psi}_{L}^{i} \tilde{H} \lambda^{\Psi^{(+)}}_{ij} \Psi_{R}^{(+)j} + H.c. ) ] \nn \\
& + & ( \chih^{\mu\sigma} \chih_{a}^{\; \; \rho} - \frac{1}{4} \chih_{a}^{\; \; \mu}  \chih^{\rho\sigma} )  \chih^{\nu\nu'} 
 [ F_{\mu\nu} F_{\mu'\nu'} + F_{\mu\nu}^{i} F_{\mu'\nu'}^{i} \nn \\
 & + &  F_{\mu\nu}^{\alpha} F_{\mu'\nu'}^{\alpha}  + \cF_{\mu\nu}\cF_{\mu'\nu'} + \cF_{\mu\nu}^{ab} \cF_{\mu'\nu' ab}     -  \cF_{\mu\nu}^{a} \cF_{\mu'\nu' a} \nn \\ 
 & + & \cW_{\mu\nu}\cW_{\mu'\nu'}  ]  - ( \chih^{\mu\sigma} \chih_{a}^{\; \; \rho} - \frac{1}{2} \chih_{a}^{\; \; \mu}  \chih^{\rho\sigma} ) [ 2(\cD_{\rho}H)^{\dagger} \cD_{\sigma}H \nn \\
 &  +& \cD_{\rho}\phi_{w} \cD_{\sigma}\phi_{w} + \cD_{\rho}\phi_{e} \cD_{\sigma}\phi_{e} + \beta_{\kappa}^2 g_w^2 \bM_{\kappa}^2 \cW_{\rho}\cW_{\sigma}  \nn \\
& + & \phi_w^2 ( g_c^2\gamma_w^2 \cB_{\rho} \cB_{\sigma} - g_4^2\beta_{w}^2 \cW_{\rho}^{a}  \cW_{\sigma a}) ] -  \chih_{a}^{\; \; \mu} \cV_{S}(H, \phi_w, \phi_{e}) \}.
\ee
The following definitions have been utilized:
\be
& & \tchi^{[\mu\nu]\mu'\nu'}_{a a'} \equiv \frac{1}{2}( \tchi^{\mu\nu\mu'\nu'}_{a a'} - \tchi^{\nu\mu\mu'\nu'}_{a a'} ) , \nn \\
& & \whcG_a^{\mu\nu} \equiv \chi\, \chih^{[\mu\nu]\mu'\nu'}_{a a'}  \cG_{\mu'\nu' }^{a'}, \nn \\
& & \chih^{[\mu\nu]\mu'\nu'}_{a a'} \equiv \frac{1}{2}( \chih^{\mu\nu\mu'\nu'}_{a a'} - \chih^{\nu\mu\mu'\nu'}_{a a'} ) .
\ee 

It is evident that $\hsF_{a}^{\;\mu}$ and $\hcG_{a}^{\; \mu}$ represent the source currents generated from the dynamics of the gravigauge field itself, while $\hmJ_{a}^{\;\mu}$ corresponds to the source current arising from the dynamics of all other basic fields.

The equation of motion for the gravigauge field presented in Eq.(\ref{GaGE}) offers a general description of gravidynamics within the framework of GQFT, referred to as the general {\it gauge-type gravitational equations}.

It is interesting to demonstrate that the quadratic term of the gravigauge field strength $\sF_{\mu\nu}^{a}$ in the action of the GSM shown in Eq.(\ref{GSMactionGQFT}) is equivalent to the Einstein-Hilbert action, up to a total derivative. The explicit relationship is expressed as follows:
\be \label{GGGR}
& & \frac{1}{4} \chi\, \tchi_{aa'}^{\mu\nu \mu'\nu'} \sF_{\mu\nu}^{a} \sF_{\mu'\nu'}^{a'} = \chi\, R
- 2 \p_{\mu} (\chi \chih^{\mu\rho} \chih_{a}^{\;\sigma} \sF_{\rho\sigma}^{a} ) , 
\ee
with the identity:
\be \label{GGGI}
 R \equiv  \chih_{b}^{\; \mu} \chih_{a}^{\; \nu} R_{\mu\nu}^{ab}  \equiv \chih^{\mu\sigma} \chih^{\nu\rho} R_{\mu\nu\rho\sigma} \equiv \chih^{\mu\sigma} R_{\mu\sigma} .
\ee
This relation and identity illustrate the gauge-gravity-geometry correspondence. Here, $R_{\mu\nu}^{ab}$ is defined as the field strength of the spin gravigauge field $\mOm_{\mu}^{ab}$,
\be
R_{\mu\nu}^{ab} = \p_{\mu}\mOm_{\nu}^{ab} - \p_{\nu}\mOm_{\mu}^{ab} + \mOm_{\mu c}^{a} \mOm_{\nu}^{cb} - \mOm_{\nu c}^{a} \mOm_{\mu}^{cb} .
\ee
Geometrically, $R_{\mu\nu\rho\sigma}$ is referred to as the Riemann curvature tensor, and $R_{\mu\sigma}$ and $R$ denote the Ricci curvature tensor and scalar, respectively. The explicit form of $R_{\mu\nu\sigma}^{\rho}$ is given as follows:  
\be \label{RMC3}
& & R_{\mu\nu\sigma}^{\;\rho}(x)  = \p_{\mu} \Gamma_{\nu\sigma}^{\rho} - \p_{\nu} \Gamma_{\mu\sigma}^{\rho}  + \Gamma_{\mu\lambda}^{\rho} \Gamma_{\nu\sigma}^{\lambda}  - \Gamma_{\nu\lambda}^{\rho} \Gamma_{\mu\sigma}^{\lambda} .
\ee
Here, $\Gamma_{\mu\sigma}^{\rho}(x)$ is the affine connection or the Christoffel symbol in geometry, which is determined by the gravigauge field, $\chi_{\mu}^{\;\; a}$, or the gravimetric field, $\chi_{\mu\nu}$, as follows:
\be \label{SGMF}
\Gamma_{\mu\sigma}^{\rho}(x)  & \equiv & \chih_{a}^{\;\; \rho} \cD_{\mu} \chi_{\sigma}^{\;\; a}  \equiv  \chih_{a}^{\;\; \rho} \p_{\mu} \chi_{\sigma}^{\;\; a} +  \chih_{a}^{\;\; \rho}   \mOm_{\mu\1 b}^{a} \chi_{\sigma}^{\;\;b} , \nn \\
& = & \frac{1}{2}\chih^{\rho\lambda} (\p_{\mu} \chi_{\lambda\sigma} + \p_{\sigma} \chi_{\lambda\mu} - \p_{\lambda}\chi_{\mu\sigma} ) =\Gamma_{\sigma\mu}^{\rho}.
 \ee

The action presented in Eq. (\ref{GSMactionGQFT}) is found to exhibit an emergent hidden GL(1,3,R) symmetry in coordinate spacetime, providing a local extension of the Poincar\'e group symmetry:
\be
PO(1,3) \to GL(1,3, R) .
\ee

This GL(1,3,R) symmetry, which forms the basis of Einstein's GR and governs gravity in curved spacetime, demonstrates the background-independent nature of the theory. 

When projecting the gauge-type gravitational equation given in Eq.(\ref{GaGE}) into coordinate spacetime through the Goldstone-type gravigauge field $\chi_{\mu}^{\; a}$ (or its inverse $\chih_{a}^{\; \mu}$), we obtain the following general gravitational equation:
\be \label{GGE}
& & R_{\mu\nu} -  \frac{1}{2}\chi_{\mu\nu} R  + \gamma_G \cG_{\mu\nu} = 8\pi G_{\kappa} \mT_{\mu\nu} . 
\ee
In this general gravitational equation, there are ten symmetric and six antisymmetric components:  
 \be \label{GGE2}
& & R_{\mu\nu} -  \frac{1}{2}\chi_{\mu\nu} R  + \gamma_G \cG_{(\mu\nu)} = 8\pi G_{\kappa} \mT_{(\mu\nu)} , \nn \\
& &  \gamma_G\cG_{[\mu\nu]}  = 8\pi G_{\kappa} \mT_{[\mu\nu]} .
\ee
Here, the symmetric equations yield a generalized form of the Einstein equations, and the antisymmetric equations provide an additional set of equations that extend beyond the GR.
 
The symmetric tensors $\mT_{(\mu\nu)}$ and $\cG_{(\mu\nu)}$, as well as the antisymmetric tensors $\mT_{[\mu\nu]}$ and $\cG_{[\mu\nu]}$ are derived from the currents $\hmJ_{a}^{\;\mu}$ and $\hcG_{a}^{\; \mu}$, presented in Eqs.(\ref{GCC1}) and (\ref{GCC}). They are given as follows:
\be
 \mT_{\mu\nu}  & \equiv & \chih \chi_{\mu\rho} \hmJ_{a}^{\; \rho} \chi_{\nu}^{\; a}  \equiv \mT_{(\mu\nu)} + \mT_{[\mu\nu]}, \nn \\
\mT_{(\mu\nu)}  & \equiv &  \frac{1}{2} (\mT_{\mu\nu} + \mT_{\nu\mu} ) , \quad \mT_{[\mu\nu]}  \equiv  \frac{1}{2} (\mT_{\mu\nu} - \mT_{\nu\mu} ) , \nn \\
\mT_{(\mu\nu)} & = & \frac{1}{2} [ \chi_{\mu\nu} \chih_{a}^{\; \rho}   - \frac{1}{2}(\eta_{\mu}^{\; \rho}  \chi_{\nu a} + \eta_{\nu}^{\; \rho}  \chi_{\mu a}) ]\nn \\
&\cdot & [ \bar{l}_{L}^{i} \gamma^{a} i\cD_{\rho}^{(l_L)} l_{L}^{i}  + \bar{l}_{R}^{i} \gamma^{a} i\cD_{\rho}^{(l_R)} l_{R}^{i}  + \bar{q}_{L}^{i} \gamma^{a} i\cD_{\rho}^{(q_L)} q_{L}^{i} \nn \\
& + &  \bar{q}_{R}^{i} \gamma^{a} i\cD_{\rho}^{(q_R)} q_{R}^{i}   + H.c.  ] + \frac{1}{2}\chi_{\mu\nu}[ \bar{l}_{L}^{i} H \lambda^{e}_{ij} e_{R}^{j}  \nn \\
& + &  \bar{l}_{L}^{i} \tilde{H} \lambda^{\nu}_{ij} \nu_{R}^{j} +  \bar{q}_{L}^{i} H \lambda^{d}_{ij} d_{R}^{j} +  \bar{q}_{L}^{i} \tilde{H} \lambda^{u}_{ij} u_{R}^{j}  + H.c.  ]  \nn \\
& + &( \eta_{\mu}^{\; \rho}\eta_{\nu}^{\; \sigma} - \frac{1}{4} \chi_{\mu\nu}  \chih^{\rho\sigma} )\chih^{\rho'\sigma'}   
 [ F_{\rho\rho'} F_{\sigma\sigma'} + F_{\rho\rho'}^{i}  F_{\sigma\sigma'}^{i} \nn \\
 & + &  F_{\rho\rho'}^{\alpha} F_{\sigma\sigma'}^{\alpha}  F_{\mu\nu}^{\alpha} + \cF_{\rho\rho'} \cF_{\sigma\sigma'} + \cF_{\rho\rho'}^{ab} F_{\sigma\sigma' ab} - \cF_{\rho\rho'}^{a} F_{\sigma\sigma' a}  \nn \\ 
 & + & \cW_{\rho\rho'} \cW_{\sigma\sigma'}  ]  - ( \eta_{\mu}^{\; \rho}\eta_{\nu}^{\; \sigma} - \frac{1}{2} \chi_{\mu\nu}  \chih^{\rho\sigma} ) [ 2(\cD_{\rho}H)^{\dagger} \cD_{\sigma}H \nn \\
 &  +& \cD_{\rho}\phi_{w} \cD_{\sigma}\phi_{w} + \cD_{\rho}\phi_{e} \cD_{\sigma}\phi_{e} + \beta_{\kappa}^2 g_w^2 \bM_{\kappa}^2 \cW_{\rho}\cW_{\sigma}  \nn \\
& + & \phi_w^2 ( g_c^2\gamma_w^2 \cB_{\rho} \cB_{\sigma} - g_4^2\beta_{w}^2 \cW_{\rho}^{a}  \cW_{\sigma a}) ] -  \chi_{\mu\nu} \cV_{S}(H, \phi_w, \phi_{e}) , \nn \\
\mT_{[\mu\nu]}  & \equiv &  - \frac{1}{4}( \eta_{\mu}^{\; \rho}  \chi_{\nu a} - \eta_{\nu}^{\; \rho}  \chi_{\mu a} ) ( \bar{l}_{L}^{i} \gamma^{a} i\cD_{\rho}^{(l_L)} l_{L}^{i}  + \bar{l}_{R}^{i} \gamma^{a} i\cD_{\rho}^{(l_R)} l_{R}^{i} \nn \\
&  + & \bar{q}_{L}^{i} \gamma^{a} i\cD_{\rho}^{(q_L)} q_{L}^{i} +   \bar{q}_{R}^{i} \gamma^{a} i\cD_{\rho}^{(q_R)} q_{R}^{i}   + H.c.  ) ,
\ee 
and 
\be
\cG_{\mu\nu}  &  \equiv & \frac{1}{2} \chih \chi_{\mu\rho} \hcG_{a}^{\; \rho} \chi_{\nu}^{\; a}  \equiv \cG_{(\mu\nu)} + \cG_{[\mu\nu]}, \nn \\
\cG_{(\mu\nu)} & \equiv & \frac{1}{2} (\cG_{\mu\nu}  + \cG_{\nu\mu}  )  ,\quad \cG_{[\mu\nu]} \equiv   \frac{1}{2} (\cG_{\mu\nu}  - \cG_{\nu\mu}  ) , \nn \\
\cG_{(\mu\nu)} & = & \frac{1}{4}  \{ \chi_{\mu\rho} \bs{\nabla}_{\sigma}\whcG_{\nu}^{\sigma\rho} + \chi_{\nu\rho}\bs{\nabla}_{\sigma}\whcG_{\mu}^{\sigma\rho}  \nn \\
& + & ( \chi_{\mu\rho} \eta_{\nu}^{\; \sigma'} + \chi_{\nu\rho} \eta_{\mu}^{\; \sigma'} - \frac{1}{2} \chi_{\mu\nu} \eta_{\rho}^{\sigma'} ) \cG_{\sigma'\sigma}^{\rho'} \whcG_{\rho'}^{\rho\sigma}  \} , \nn \\
  \cG_{[\mu\nu]} &= &  \frac{1}{4} \{ \chi_{\mu\rho} \bs{\nabla}_{\sigma}\whcG_{\nu}^{\sigma\rho}- \chi_{\nu\rho} \bs{\nabla}_{\sigma}\whcG_{\mu}^{\sigma\rho}  \nn  \\
& + & ( \chi_{\mu\rho} \eta_{\nu}^{\; \sigma'} - \chi_{\nu\rho} \eta_{\mu}^{\; \sigma'}  ) \cG_{\sigma'\sigma}^{\rho'} \whcG_{\rho'}^{\rho\sigma}  \} ,
\ee
where we have introduced the following definitions:
\be
& &   \cG_{\mu\nu}^{\rho} \equiv \chih_{a}^{\; \rho} \cG_{\mu\nu}^{a} ,  \;\; \whcG_{\mu}^{\rho\sigma} \equiv \chih \chi_{\mu}^{\; a} \whcG_{a}^{\rho\sigma} = \chi_{\mu}^{\; a} \chih^{[\rho\sigma]\mu'\nu'}_{a a'}  \cG_{\mu'\nu' }^{a'}, \nn \\
& &  \bs{\nabla}_{\sigma}\whcG_{\mu}^{\rho\sigma} \equiv \p_{\sigma}\whcG_{\mu}^{\rho\sigma} + \Gamma_{\sigma\nu}^{\sigma} \whcG_{\mu}^{\rho\nu} - \cA_{\sigma\mu}^{\nu} \whcG_{\nu}^{\rho\sigma} ,  \nn \\
& & \cA_{\sigma\mu}^{\nu}\equiv  \chih_{a}^{\; \nu} D_{\sigma}\chi_{\mu}^{\; a}  =  \chih_{a}^{\; \nu}\p_{\sigma}\chi_{\mu}^{\; a} + \chih_{a}^{\; \nu} \cA_{\sigma b}^{a}\chi_{\mu}^{\; b}.
\ee

The equations of motion presented in Eqs.(\ref{GGE}) and Eq.(\ref{GGE2}) are referred to as general geometric-type gravitational equations beyond Einstein's GR.  Unlike in GR, where the GL(1,3,R) symmetry governs gravity through the dynamics of Riemannian geometry, characterized solely by the gravimetric field $\chi_{\mu\nu}$, the current framework reveals GL(1,3,R) as a hidden group symmetry. This symmetry emerges naturally as a consequence of the action constructed based on the gauge invariance principle within the spin-fiber graviguage spacetime.

Despite the presence of the hidden group symmetry GL(1,3,R) and the equivalence to the Riemannian geometry of the Einstein-Hilbert action term, the gravitational interactions involving spinor fermions (leptons and quarks) and spin gauge bosons associated with the field strength $\cG_{\mu\nu}^{a}$ in the action occur predominantly through the spin-associated gravigauge field $\chi_{\mu}^{\; a}$, rather than the composite gravimetric field, $\chi_{\mu\nu} = \chi_{\mu}^{\; a} \chi_{\nu}^{\; b} \eta_{ab}$. Consequently, it is the gravigauge field, $\chi_{\mu}^{\; a}$, that acts as the fundamental gravitational field within the framework of GQFT.

It has been demonstrated \cite{GQFT4,GQFT6} that the gravitational equations governed by the antisymmetric energy-momentum tensor predict three novel gravitational wave polarizations: one spin-0 scalar-transverse mode and two spin-1 vector-transverse modes. Furthermore, refs. \cite{GQFT4,CXW,GN} establish a direct constraint on the combined parameter,
\be
 \gamma_W \equiv  \gamma_G( \alpha_G - \alpha_W/2) = \frac{2M_{\cG}^2}{g_4^2\bM_{\kappa}^2}, 
 \ee
for which the current experimental upper bound is approximately $\gamma_W \lesssim 10^{-6}$. This implies that the mass $M_{\cG}$ of the spin gauge field component $\cA_{[cab)}$ defined in Eq.(\ref{SGF}) must be less than the fundamental mass scale $\bM_{\kappa} \sim \bM_P/\sqrt{2}$ (where $\bM_P$ is the reduced Planck mass) by roughly three orders of magnitude, i.e., 
\be
M_{\cG} \lesssim 10^{-3} \bM_P, \quad M_{\cG}  = g_4 m_G \sqrt{1 + 3\beta_G/2\gamma_G } .
\ee

A lower bound on the mass $M_{\cA}$ of the spin gauge field component $\cA_{[cab]}$ defined in Eq.(\ref{SGF}) is set directly by collider experiments. The most stringent bound is derived from analyses of $\mu^{+}\mu^{-}$ scattering amplitudes across all initial and final helicity configurations\cite{HTW}. Since the spin gauge boson couples universally to all leptons and quarks, it can be produced at hadronic colliders such as the LHC via the gluon fusion channel. Comprehensive searches for its decay channels, including dijets, $t\bar{t}$, dileptons, diphoton, and diboson final states, conducted by the ATLAS \cite{A1,A2,A3,A4,A5} and CMS \cite{C1,C2,C3,C4,C5,C6} collaborations, provide strong constraints. These results already exclude the existence of a light vector gauge boson within the energy scales currently probed\cite{PDG}, i.e., 
\be
M_{\cA} \gtrsim (1\sim 3)\, \mbox{TeV}, \quad M_{\cA} = g_4 m_G \sqrt{ 1 + 2\beta_G/\gamma_G } .
\ee

If $\beta_G =0$, or equivalently, $\alpha_{\kappa}=0$ and $\alpha_e=0$, the spin gauge components $\cA_{[cab]}$ and $\cA_{(cab]}$ become degenerate in mass, with $M_{\cA} = M_{\cG}$. The current experimental bounds on this mass are:
\be
 (1\sim 3)\times 10^3\, \text{GeV}  \lesssim M_{\cA} = M_{\cG} \lesssim 10^{16}\, \text{GeV} .
\ee

In summary, a central result of this section is the derivation of the field equations for the gravigauge field from the unified action. This establishes the gravidynamics of the GSM, demonstrating how the gravity emerges naturally from gauge-theoretic principles within the GQFT framework.

%%%%%%%%%%%

\section{ The Energy-momentum Cancellation Law from Translational Symmetry and  the Zero Energy-momentum Tensor Theorem in GSM}

Within the framework of GQFT, a profound implication of translational symmetry in Minkowski spacetime was explored in \cite{GQFT5}, establishing a direct connection between the energy-momentum cancellation law and the zero energy-momentum tensor theorem. This theorem states that the energy-momentum tensor vanishes identically when the equations of motion for all fundamental fields in GQFT are utilized, in contrast to the conventional conservation law derived from translation symmetry in QFT. Through rigorous mathematical analysis and physical interpretation, it has been demonstrated how the energy-momentum cancellation law, together with the zero energy-momentum tensor theorem, forms the foundational structure of GQFT. This provides deep insights into the interplay between the energy-momentum conservation law in QFT and the energy-momentum cancellation law in GQFT under translational symmetry, as well as their influence on the dynamics of quantum fields.

The GSM constructed within spin-fiber graviguage spacetime reveals that the laws of nature are independent of the choice of coordinate systems. Its alternative formulation, reformulated within the framework of GQFT based on the concept of bi-frame spacetime, demonstrates that gravitational interaction is described by the spin-associated gravigauge $\chi_{\mu}^{\; a}$ and its field strength $\sF_{\mu\nu}^{a}$. Although the action possesses a hidden group symmetry GL(1,3,R), it always allows to choose a globally flat Minkowski spacetime as a base spacetime, which is fundamentally different from GR. This distinction arises because the fundamental gravitational field is the gravigauge field $\chi_{\mu}^{\; a}$, rather than the composite gravimetric field $\chi_{\mu\nu}$. It is the Goldstone-type gravigauge field $\chi_{\mu}^{\; a}$ that is identified to the massless graviton. Consequently, this framework enables a meaningful definition of the energy-momentum tensor for the GSM through the translational invariance of coordinates in globally flat Minkowski spacetime within the GQFT framework.

According to Noether’s theorem, every differentiable symmetry of an action corresponds to a conservation law. Consider the translational transformation of coordinates, $x^{\mu} \to x^{'\mu} = x^{\mu} + a^{\mu}$,  for an arbitrary $a^{\mu}$. The invariance of the action under such a transformation leads to the conservation law of the energy-momentum tensor,
\be  \label{EMTC}
\delta S_{GSM} = \int d^4x\, \p_{\nu} ( \hcT_{\mu}^{\;\, \nu}) a^{\mu} =0  \quad \to \;\;  \p_{\nu}\hcT_{\mu}^{\;\, \nu}= 0 ,
\ee
where the surface terms have been ignored. The energy-momentum tensor is found to have the following explicit form:
\be \label{EMTensor}
\hcT_{\mu}^{\; \, \nu} & = & \chi \{ \eta_{\mu}^{\; \nu}  \cL_{GSM} -  \chih_{a}^{\; \nu} \frac{1}{2} [ \bar{l}_{L}^{i} \gamma^{a} i\cD_{\mu}^{(l_L)} l_{L}^{i}    \nn \\
& + & \bar{l}_{R}^{i} \gamma^{a} i\cD_{\mu}^{(l_R)} l_{R}^{i} + \bar{q}_{L}^{i} \gamma^{a} i\cD_{\mu}^{(q_L)} q_{L}^{i} + \bar{q}_{R}^{i} \gamma^{a} i\cD_{\mu}^{(q_R)} q_{R}^{i}  + H.c.  ]  \nn \\
& + &  \eta_{\mu}^{\; \; \rho}  \chih^{\nu\sigma}[  ( \sum \bscF_{\rho\rho'}^{A} \bscF_{\sigma\sigma'}^{A} - \cF_{\rho\rho'}^{a} \cF_{\sigma\sigma' a}) \chih^{\rho'\sigma'} \nn \\
& - & 2(\cD_{\rho}H)^{\dagger} \cD_{\sigma}H - \cD_{\rho}\phi_{w} \cD_{\sigma}\phi_{w} - \cD_{\rho}\phi_{e} \cD_{\sigma}\phi_{e}  \nn \\
& - &  \beta_{\kappa}^2 g_w^2 \bM_{\kappa}^2 \cW_{\rho}\cW_{\sigma}  - g_4^2 \phi_w^2 ( \gamma_w^2 \cB_{\rho} \cB_{\sigma} - \beta_{w}^2  \cW_{\rho}^{a}  \cW_{\sigma a}) ]  \} \nn \\
& - &  \bM_{\kappa}^2  [ \chi_{\mu}^{\; a} ( \p_{\rho} \whsF^{\rho\nu }_{a}    + \cD_{\rho}\widehat{\cG}_{a}^{\rho\nu}) +  \sF_{\mu\sigma}^{a}  \whsF^{\nu\sigma}_{a}  +\cG_{\mu\sigma}^{a} \widehat{\cG}^{\nu\sigma}_{a} ) ] ,
\ee
where we have employed the equations of motion for all gauge fields except the gravigauge field $\chi_{\mu}^{\; a}$ to obtain a manifestly gauge-invariant energy-momentum tensor in both the unitary conformal-boost gauge basis and the fundamental mass-scaling gauge basis. Note that the gauge-dependent term, which appears as a total derivative in the energy-momentum tensor, $\hcT_{\mu}^{\; \, \nu}  \sim \p_{\nu'} \tilde{\cT}_{\mu}^{\nu\nu'}$, with $\tilde{\cT}_{\mu}^{\nu\nu'} = \bM_{\kappa}^2  \chi_{\mu}^{\; a}( \whsF^{\nu\nu' }_{a} -\widehat{\cG}_{a}^{\nu\nu'})  - \chi  (\sum \bscF_{\rho\sigma}^{A} \bscA_{\mu} ^{A} - \cF_{\rho\sigma}^{a} \cW_{\mu a}) \chih^{\nu\rho} \chih^{\nu'\sigma}$, has been removed. This term vanishes automatically in the derivative of the energy-momentum tensor in Eq.(\ref{EMTC}), $\p_{\nu} \p_{\nu'} \tilde{\cT}_{\mu}^{\nu\nu'} = 0 $, due to the antisymmetric property, $\tilde{\cT}_{\mu}^{\nu\nu'} = - \tilde{\cT}_{\mu}^{\nu'\nu}$. 

It is noteworthy that the energy-momentum tensor in Eq. (\ref{EMTensor}) can be rewritten as follows:
\be
\hcT_{\mu}^{\; \, \nu} & \equiv &  \bM_{\kappa}^2 \chi_{\mu}^{\; a} ( \p_{\rho} \whsF^{\nu\rho }_{a} -\whfJ_{a}^{\; \, \nu}  ) ,
\ee
with the current $\whfJ_{a}^{\; \, \nu}$ presented explicitly in Eqs.(\ref{FSCC})-(\ref{GCC1}). The energy-momentum conservation law, $\p_{\nu} ( \hcT_{\mu}^{\;\, \nu})=0$, leads to the following relation:
\be
\p_{\nu} \whfJ_{a}^{\; \, \nu}  = (\chih_{a}^{\; \mu} \p_{\nu} \chi_{\mu}^{\; b} ) (\p_{\rho} \whsF^{\nu\rho }_{b} - \whfJ_{b}^{\; \, \nu} )  = -(\p_{\nu}\chih_{a}^{\; \mu} ) \hcT_{\mu}^{\; \, \nu}/\bM_{\kappa}^2 .
\ee
This indicates that by utilizing the equation of motion for the gravigauge field $\chi_{\mu}^{\; a}$, as given in Eq.(\ref{GaGE}), which acts as the general gauge-type gravitational equation, the energy-momentum conservation law aligns with the conserved current, $\p_{\nu} \whfJ_{a}^{\; \, \nu}=0$. 

Consequently, when the equations of motion for all gauge fields, including the gravigauge field, are applied, not only is the current conserved, but the entire energy-momentum tensor also vanishes:
\be
 \p_{\nu} \whfJ_{a}^{\; \, \nu}  = 0, \quad \hcT_{\mu}^{\; \, \nu} =  0 , 
\ee
which introduces an additional constraint that goes beyond the conventional energy-momentum conservation law in classical and quantum field theories.

Based on the above analysis within the framework of GQFT, we arrive at a conclusion that strengthens Noether's theorem in the context of translational invariance. This enables us to formulate a more general theorem regarding the energy-momentum tensor, which states the following: a fundamental theory describing the basic constituents of matter and their fundamental interactions must exhibit a vanishing energy-momentum tensor under translational invariance within the framework of GQFT, with globally flat Minkowski spacetime serving as the base spacetime. This general theorem is termed the zero energy-momentum tensor theorem in GQFT, which extends beyond the conventional energy-momentum conservation law in QFT. It may also be referred to as the energy-momentum cancellation law of translational invariance in GQFT.

The energy-momentum cancellation law implies that a fundamental theory constructed within the framework of GQFT should be capable of describing the entire observable universe through all fundamental interactions occurring across the entirety of Minkowski spacetime. It suggests that globally flat Minkowski spacetime, as a complete spacetime, is infinitely large, and no energy-momentum flows out of the spacetime. Consequently, the total energy-momentum tensor across the entire spacetime vanishes due to the cancellation of contributions from the fundamental gravigauge field and all other fields.

To comprehensively understand the energy-momentum cancellation law, let us further analyze the intrinsic features of gravitational interactions as indicated in the GSM. One fundamental feature is revealed in the gravitization equation presented in Eq. (\ref{GE}), where gravitational effects are characterized by the gravigauge field strength $\hsF_{cd}^{a}$. This field strength is associated with the group structure factor of the non-commutative gravigauge derivative operator $\eth_{c}$ in spin-fiber gravigauge spacetime and appears as an auxiliary field strength in the action. This observation implies that gravitational effects arise not only from the spin-associated kinematics of all basic fields but also from the collective dynamics of all fundamental gauge fields.

Another essential feature is demonstrated from the sources of gravitational interactions, as revealed in the gauge-type gravitational equation (Eq. (\ref{GaGE})), all fundamental fields with kinetic motion and all gauge fields, as vector fields in Minkowski spacetime, couple to gravitational interactions through the inverse (dual) gravigauge field $\chih_{a}^{\; \mu}$. Specifically, the inverse gravigauge field $\chih_{a}^{\; \mu}$ is associated with all kinetic terms characterized by the derivative operator vector and with all gauge fields introduced as covariant vectors in Minkowski spacetime. Since gravitational interactions of all fundamental fields occur through coupling to the inverse gravigauge field, they provide contributions to the energy-momentum tensor that are opposite to those of the gravigauge field $\chi_{\mu}^{\; a}$. This results in a complete cancellation, leading to a zero energy-momentum tensor theorem for a fundamental theory grounded in the entirety of Minkowski spacetime.

Alternatively, in light of the energy-momentum cancellation law under the translational invariance of the theory grounded in the entirety of Minkowski spacetime within the framework of GQFT, it becomes possible to derive both gauge-type and geometric-type gravitational equations. Explicitly, the general gauge-type gravitational equation can be simply inferred from the following relation:
\be
\hcT_{a}^{\; \nu} & \equiv & \chih_{a}^{\; \mu} \hcT_{\mu}^{\; \, \nu} = \bM_{\kappa}^2 ( \p_{\rho} \whsF^{\nu\rho }_{a} -\whfJ_{a}^{\; \, \nu}  ) , \nn \\
& = & \bM_{\kappa}^2  [ \p_{\rho} \whsF^{\nu\rho }_{a} + 16\pi G_{\kappa} \hmJ_{a}^{\; \, \nu} -  \hsF_{a}^{\; \nu}  -\gamma_G \hcG_{a}^{\; \nu}  ] = 0 ,
\ee
and the general geometric-type gravitational equation is obtained directly as follows:
\be
\cT_{\mu\nu}  & \equiv & \hcT_{\mu}^{\; \, \rho} \chi_{\rho\nu} =  2\bM_{\kappa}^2  [ 8\pi G_{\kappa} \hT_{\mu}^{\; \, \rho} - ( \hR_{\mu}^{\; \, \rho} - \frac{1}{2} \eta_{\mu}^{\; \, \rho} R + \gamma_G\hcG_{\mu}^{\; \, \rho}) ]   \chi_{\rho\nu}  \nn \\
& = & 2\bM_{\kappa}^2  [  8\pi G_{\kappa} \mT_{\mu\nu} - (R_{\mu\nu} - \frac{1}{2} \chi_{\mu\nu} R  +  \gamma_G\cG_{\mu\nu} )  ]  = 0 ,
\ee
with the following relations and definitions:
\be
& & \mT_{\mu\nu} \equiv \hT_{\mu}^{\; \, \rho} \chi_{\rho\nu}  , 
\;\;  \hT_{\mu}^{\; \, \rho} \equiv \chi_{\mu}^{\; a} \hmJ_{a}^{\; \, \rho},  \nn \\
& & \cG_{\mu\nu}  \equiv \hcG_{\mu}^{\; \, \rho}  \chi_{\rho\nu}, 
\;\; \hcG_{\mu}^{\; \, \rho} \equiv \chi_{\mu}^{\; a} \hcG_{a}^{\; \rho} /2 , \nn \\
& & R_{\mu\nu} \equiv \hR_{\mu}^{\; \, \rho} \chi_{\rho\nu}, \; \; \hR_{\mu}^{\; \, \rho} - \frac{1}{2} \eta_{\mu}^{\; \, \rho} R  \equiv \chi_{\mu}^{\; a}  (\hsF_{a}^{\; \rho} - \p_{\sigma} \whsF^{\rho\sigma}_{a})/2 .
\ee

Thus, within the GQFT framework, we demonstrate the equivalence of the energy-momentum cancellation law and the gravigauge field equation of motion (i.e., the general gravitational equation).

In summary, a central achievement is the reformulation of energy-momentum dynamics. Imposing consistency with the fundamental Minkowski spacetime necessitates energy-momentum cancellation laws, which replace the standard conservation law. This reformulation enables a proof of the zero energy-momentum tensor theorem for the full GSM, thereby establishing a novel and coherent mechanism through which quantum fields generate gravitational sources.

%%%%%%%%%%%%%

\section{ Gauge Conditions for WS$_{c}$(1,3)$\times$GS(1), Mass Generation and Spontaneous Breaking of Scalar Potentials in GSM}

In this section, we are going to examine the gauge conditions associated with the WS$_c$(1,3)$\times$GS(1) gauge symmetries and discuss the spontaneous breakdown of scalar potentials as a mechanism for mass generation in GSM. The analysis focuses on the interplay between gauge symmetries and scalar fields, and show how the gauge fixing conditions imposed on the WS$_c$(1,3)$\times$GS(1) symmetry groups influence the behavior of scalar potentials and their role in generating masses for gauge and scalar bosons. 

It is well known that gauge symmetry introduces redundant degrees of freedom, which must be eliminated by imposing an appropriate gauge-fixing condition. Let us first discuss the gauge-fixing condition for the spin gauge symmetry SP(1,3) of leptons and quarks. Unlike conventional internal gauge symmetries, the spin gauge symmetry SP(1,3) not only requires the introduction of a corresponding spin gauge field $\cA_{\mu}^{ab}$ but also necessarily entails an invertible bi-covariant vector field $\chih_{a}^{\;\; \mu}$ along with its dual field $\chi_{\mu}^{\; a}$, serving as the gravigauge field. 

The gravigauge field $\chi_{\mu}^{\; a}$ contains sixteen degrees of freedom, six more than the composite symmetric gravimetric field $\chi_{\mu\nu}= \chi_{\mu}^{\; a} \chi_{\mu}^{\; b} \eta_{ab}$ in GR. These additional degrees of freedom are associated with the equivalence classes of the spin gauge symmetry SP(1,3), allowing us to eliminate the redundancy by imposing a gauge-fixing condition.

A useful gauge fixing condition can be implemented by performing a special spin gauge transformation, $S(\tilde{\Lambda})$, in the spinor representation. Under this transformation, the $\gamma$-matrix and the gravigauge field $\chi_{\mu}^{\; a}$ transform in the following way:
\be
& & S^{-1}(\tilde{\Lambda}) \gamma^a S(\tilde{\Lambda}) = \tilde{\Lambda}^{a}_{\;\, b}(x) \gamma^b, \nn \\
& & \chi_{\mu a }(x) \to  \tchi_{\mu a}(x) = \tilde{\Lambda}_{a}^{\; \, b}(x) \chi_{\mu b}(x) = \tchi_{a\mu}(x)  ,
\ee
so that the gravigauge field $\tilde{\chi}_{\mu a }(x)$, behaving as a Goldstone boson, becomes symmetric.

Such a spin gauge transformation naturally fixes the gauge for the SP(1,3) spin gauge symmetry. This gauge choice is called the flowing unitary spin gauge. In the flowing unitary gauge, the symmetric gravigauge field $\chi_{\mu a}(x)= \chi_{a\mu}(x)$ has exactly the same degrees of freedom as the composite gravimetric field with $\chi_{\mu\nu}(x) \equiv \chi_{\mu a}\eta^{ab}\chi_{b\nu}\equiv \chi_{\nu a}\eta^{ab}\chi_{b\mu}$. 
However, since the total independent degrees of freedom must remain unchanged, six degrees of freedom are transferred from the gravigauge field $\chi_{\mu a}(x)$ to the spin gauge field $\cA_{\mu}^{ab}(x)$. As a result, the spin gauge field acquires a mass-like term (as shown in Eqs.(\ref{GSMaction1}) and (\ref{MP}) ), causing it to behave as a massive gauge boson with:
\be \label{MP1}
& & M_{\cA}^2 \chi_{\cA}^{2} = (\gamma_{\kappa} + 2\alpha_{\kappa} ) g_4^2\bM_{\kappa}^2 + (\beta_e+ 2\alpha_e) g_4^2 \phi_e^2, \nn \\
& & M_{\cG}^2  \chi_{\cG}^{2} = ( \gamma_{\kappa} + \alpha_{\kappa}/2 ) g_4^2 \bM_{\kappa}^2 + (\beta_e + \alpha_e/2 )  g_4^2 \phi_e^2 .
\ee

To eliminate redundant degrees of freedom caused from the chirality boost-spin gauge symmetry $W^{1,3}$,  the chiral conformal-spin gauge symmetry SP$_c$(1,1) and the scaling gauge symmetry GS(1), we can adopt the unitary conformal-boost gauge basis, and the fundamental mass scaling gauge basis or unitary scaling gauge basis. This gauge fixing is implemented by transforming the vector field $\bs{\zeta}^a(x)$, the singlet scalar field $\bs{\zeta}(x)$, the determinant $\chi(x)$ of the gravigauge field, or the single scaling field $\Phi_{\kappa}(x)$, into the specific fixing conditions: 
\be
& & \bs{\zeta}^a(x) = 0, \quad \bs{\zeta}(x)=1, \nn \\
& & \chi(x) = 1 \;\; \mbox{or}\;\;  \Phi_{\kappa}(x) = \bM_{\kappa},
\ee
which remove six redundant degrees of freedom, four from the chirality boost-spin gauge symmetry, one from the chiral conformal-spin gauge symmetry and one from the scaling gauge symmetry. Consequently, the corresponding gauge fields $\cW_{\mu}^{a}$ and $\cB_{\mu}$ develop mass-like terms as follows:
\be
m_{\cW} \sim  g_4^2 \beta_{w}^2 \phi_w^2  , \quad m_{\cB} \sim  g_c^2\gamma_w^2 \phi_w^2. 
\ee

To extract the mass term for the scaling gauge boson $\cW_{\mu}$, it is useful to make the following redefinition in the fundamental mass scaling gauge: 
\be \label{SGB1}
\mW_{\mu} \equiv \cW_{\mu} + \frac{1}{2g_w}\p_{\mu} \ln(\beta_{\kappa}^2\bM_{\kappa}^2 + \phi_w^2 + \phi_e^2 ),
\ee
and its kinetic term and interactions with scalar fields can be rewritten as follows:
\be \label{SGB2}
\cL_{\mW} & = & -\frac{1}{4} \chih^{\mu\mu'}\chih^{\nu\nu'}  \mW_{\mu\nu}\mW_{\mu'\nu'}  + \frac{1}{2} g_w^2(\beta_{\kappa}^2  \bM_{\kappa}^2 + \phi_e^2 +\phi_w^2)  \chih^{\mu\nu} \mW_{\mu}\mW_{\nu} .
\ee
The mass-like term for the redefined scaling gauge boson $\mW_{\mu}$ is given by,
\be
m_{\mW}^2 \sim g_w^2(\beta_{\kappa}^2  \bM_{\kappa}^2 + \phi_e^2 +\phi_w^2).
\ee

Notably, the flowing unitary spin gauge and scaling gauge are only valid locally at fixed points in coordinate spacetime. This is because, under transformations governed by the emergent hidden group symmetry GL(1,3,R), the gravigauge field $\chi_{\mu}^{\; a}$ loses its symmetric property. 
To restore symmetry in the gravigauge field after a general coordinate transformation, a corresponding spin gauge transformation must be applied at a fixed point in the transformed new coordinate system.

To eliminate the flowing nature of the unitary spin and scaling gauges, an additional gauge prescription for the emergent hidden group symmetry GL(1,3,R) is required. A simple gauge-fixing condition can be achieved by imposing a vanishing spin and scaling gauge covariant derivative of the gravigauge field,
\be
\cD^{\mu} \chi_{\mu}^{\; a} \equiv \chih^{\mu\nu}(\p_{\nu}\chi_{\mu}^{\; a} + \cS_{\nu}\chi_{\mu}^{\; a} + \cA_{\nu b}^{a} \chi_{\mu}^{\; b} )  =0 ,
\ee 
with $\cS_{\nu}\equiv \p_{\nu}\ln \Phi_{\kappa}$. This prescription fixes the group symmetry GL(1,3,R) while allowing specific spin and scaling gauge transformations to preserve the symmetric gravigauge field. As a result, the flowing unitary spin gauge is extended to a full unitary gauge.

When maintaining the symmetric property of the gravigauge field in the full unitary gauge, the action retains an associated global symmetry SO(1,3)$\adjoin$SP(1, 3). Under this symmetry transformation, the gravigauge field explicitly keeps its symmetric form,
\be
& & \chi_{\mu a}(x) \to \chi'_{\mu a}(x') = L_{\mu}^{\; \nu}\chi_{\nu b}(x) \Lambda^{b}_{\; a} =  \Lambda_{a}^{\; b}\chi_{b \nu }(x) L^{\nu}_{\; \mu}=\chi'_{a \mu}(x'), \nn \\
& &  x^{\mu} \to x^{'\mu} =  L^{\mu}_{\; \nu} x^{\nu} , \;\;  L_{\mu}^{\; \nu} = \Lambda_{a}^{\; b} \in SO(1,3) \cong SP(1,3) .
\ee

Therefore, in the full unitary gauge for the spin gauge symmetry SP(1,3) and the emergent hidden group symmetry GL(1,3,R), along with the unitary conformal-boost gauge, $\bs{\zeta}^a = 0$ and $\bs{\zeta}=1$, for the gauge symmetries $W^{1,3}\times SP_c(1,1)$, as well as  the fundamental mass scaling gauge, $\Phi_{\kappa}(x) = \bM_{\kappa}$, for the scaling gauge symmetry SG(1), the action of GSM within the framework of GQFT possesses the following associated global symmetry and internal gauge symmetry $G_{SM}$ of the SM:
\be
\mG_{GSM} = \mbox{SC(1)}\ltimes \mbox{P}^{1,3} \ltimes \mbox{SO(1,3)} \wtjoin \mbox{SP(1,3)} \times G_{SM} .
\ee
This is recognized as the fundamental symmetry group in GSM after applying the appropriate gauge-fixing conditions for the gauge symmetry WS$_c$(1,3)$\times$GS(1). Notably, the global Poincar\'e symmetry PO(1,3) appears as a basic symmetry in this framework. 

It is observed that all mass-like terms for the gauge fields associated with the gauge symmetry WS$_c$(1,3)$\times$GS(1) are characterized by the scalar fields. Typically, these gauge fields acquire their masses when the corresponding scalar fields develop vacuum expectation values (VEVs) through their potentials. However, since the form of the scalar potential is not uniquely determined by gauge symmetries alone, a definitive scalar potential cannot be constructed purely from symmetry principles.

To facilitate mass generation, the scalar potential is constructed such that the scalar field enables to stabilize at a minimum with a non-zero VEV. For simplicity in the analysis, we assume the following effective forms for the scalar potentials at low energies:
\be \label{EHP}
& & \cV_h \sim \frac{1}{4} \lambda_h^2 ( H^{\dagger}H - v_{h}^2 )^2 , \nn \\
& & \cV_w \sim \frac{1}{4} \lambda_w^2 ( \phi_w^2 - v_{w}^2 )^2, \quad  \cV_e \sim \frac{1}{4} \lambda_e^2 ( \phi_e^2 - v_{e}^2 )^2, 
\ee
which induce VEVs around the minimum points of the scalar fields:
\be
& & H^0 = (v_h + h^0)/\sqrt{2}, \nn \\
& & \phi_w = v_w + \varphi_w, \quad \phi_e = v_e + \varphi_e . 
\ee
Here, $\cV_h$ denotes the Higgs potential with a VEV $v_h$ in the SM, responsible for generating masses for leptons, quarks, and weak gauge bosons.

Upon spontaneous symmetry breaking, where the scalar potentials reach their minimum in the unitary conformal-boost gauge basis, $\bs{\zeta}=1$ and $\bs{\zeta}^a = 0$, and the fundamental mass scaling gauge, $\Phi_{\kappa}= \bM_{\kappa}$, the following particles acquire masses: the spin gauge boson $\cA_{\mu}^{ab}$ with mass $M_{\cA}$, the chirality boost-spin gauge boson $\cW_{\mu}^{\; a}$ with mass $m_{\cW}$, the chiral conformal-spin gauge boson $\cB_{\mu}$ with mass $m_{\cB}$, the scaling gauge boson $\mW_{\mu}$ with mass $M_{\mW}$, the scalar bosons $\varphi_w$ and $\varphi_e$ with masses $m_{\varphi_w}$ and $m_{\varphi_e}$, respectively, all arising from the VEVs $v_w$ and $v_e$ as well as the fundamental mass scale $\bM_{\kappa}$. Explicitly, the masses of these gauge and scalar bosons are given by,
\be \label{Masses}
& & M_{\cA}  = g_4 \gamma_G \bM_{\kappa}  \sqrt{1 + 2\beta_G/\gamma_G } , \nn \\
& & M_{\cG} = g_4 \gamma_G \bM_{\kappa} \sqrt{ 1 + 3\beta_G/2\gamma_G } , \nn \\
& & m_{\cW} = g_4\beta_w v_w, \quad  m_{\cB} = g_c\gamma_w v_w, \nn \\
& &  M_{\mW} = g_{w}\beta_{\kappa}\bM_{\kappa} \sqrt{1 + (v_w^2 + v_e^2)/(\beta_{\kappa}^2\bM_{\kappa}^2) }, \nn \\
& &  m_{\varphi_w} = \lambda_{w} v_w, \quad  m_{\varphi_e} = \lambda_{e} v_e ,
\ee
with $\gamma_G$ and $\beta_G$ defined in Eq.(\ref{CPA}). These masses involve additional parameters beyond the SM, including the gauge couplings ($g_4$, $g_c$ and $g_w$), the scalar couplings ($\beta_{\kappa}$, $\beta_w$, $\beta_e$, $\gamma_{\kappa}$ and $\gamma_w$), the VEVs ($v_w$ and $v_e$) and the potential coupling constants ($\lambda_w$ and $\lambda_e$). 

Symmetry breaking renders all particles massive, except for the gravigauge field $\chi_{\mu}^{\; a}$, which behaves as a Goldstone boson. Notably, the scaling gauge invariance implies that all VEVs arise from dimensionless parameters. As shown in Eq.(\ref{HP}) that the Higgs boson attains the minimum at $\langle H^2/\Phi_{\kappa}^2 \rangle = \epsilon_h$. In the fundamental mass scaling gauge basis, $\Phi_{\kappa}=\bM_{\kappa}$, the VEV is given by, $v_h/\sqrt{2} = \sqrt{\epsilon_h}\bM_{\kappa}$ with $\epsilon_h$ being the dimensionless parameter. 

It is postulated that there exists a fundamental dimensionless parameter $\epsilon_{\kappa}$, determined by the ratio of two fundamental mass scales: 
\be
\epsilon_{\kappa} \equiv \Lambda_{\kappa}/\bM_{\kappa} \sim 10^{-30}, 
\ee
where $\Lambda_{\kappa}$ represents the basic cosmological energy scale, $\Lambda_{\kappa}\sim 10^{-3}$ eV. This parameter $\epsilon_{\kappa}$ provides an extremely small ratio. All relevant VEVs are hypothesized to be governed by this tiny number:
\be
v_i = \sqrt{\epsilon_i}\bM_{\kappa}, \quad \epsilon_i \equiv \epsilon_{\kappa}^{\gamma_i}
\ee

The smallness of the VEV $v_h$ is linked to this ratio via the relation, $\epsilon_h\equiv \epsilon_{\kappa}^{\gamma_h}$, where $\gamma_h \sim 1$. Nevertheless, the naturalness of such a small VEV and the stability of the scalar potential remain unresolved issues. Further exploration of mechanisms such as dynamical spontaneous symmetry breaking\cite{DSSB} may shed light on these questions.

In summary, we present a detailed analysis of the spontaneous symmetry breaking that reduces the WS$_c$(1,3) gauge symmetry to SP(1,3). A central component of this study is a comprehensive examination of the mass generation mechanisms for all particles. This process accounts for the masses of matter and gauge fields, while the Goldstone-type gravigauge field is identified as the massless graviton.

%%%%%%%%%%%%%

\section{Dark Graviton as a Dark Matter Candidate in GSM}

In GQFT, the gravigauge field $\chi_{\mu}^{\; a}$ is always massless as it behaves as a Goldstone boson, 
corresponding to the massless graviton. We now examine several interesting properties of the massive bosons in this framework. Notably, the chirality boost-spin gauge field $\cW_{\mu}^{\; a}$, the chiral conformal-spin gauge field $\cB_{\mu}$ and the scaling gauge field $\cW_{\mu}$ (or $\mW_{\mu}$) do not  couple directly to the leptons and quarks in GSM due to the chirality property and hermiticity requirement of the action. 

Particularly significant is the behavior of the chirality boost-spin gauge boson $\cW_{\mu}^{\; a}$ and the redefined scaling gauge boson $\mW_{\mu}$ (Eq.(\ref{SGB2})), which exhibit the following $Z_2$ discrete symmetry:
\be
\cW_{\mu}^{\; a}\to -\cW_{\mu}^{\; a} , \quad \mW_{\mu}\to - \mW_{\mu} .
\ee
This symmetry implies that both $\cW_{\mu}^{\; a}$ and $\mW_{\mu}$ are stable particles, making them natural candidates for dark matter. 
 
 As shown from Eq. (\ref{Masses}), the gauge boson $\cW_{\mu}^{a}$ acquires its mass through the vacuum expectation value (VEV) of the scalar field $\phi_w$. The mass depends on the parameters $\beta_w$ and $v_w$, which must ultimately be determined experimentally. Interestingly, if the VEV $v_w$ follows the same pattern as the Higgs VEV $v_h$ in the SM, being governed by the small parameter, $\epsilon_{\kappa}\equiv \Lambda_{\kappa}/\bM_{\kappa}\sim 10^{-30}$, through the relation, 
 \be
 v_w= \sqrt{\epsilon_w}\bM_{\kappa}, \quad \epsilon_w\equiv \epsilon_{\kappa}^{\gamma_w}, 
 \ee
with $\gamma_w = 1.0\sim 1.2$, the predicted mass range falls within current and future experimental sensitivity windows, spanning from the TeV scale down to MeV energies, i.e., $m_{\cW}\sim$ TeV$\sim$MeV. For large values, $\gamma_w > 2$,  the mass could even become smaller than neutrino masses, $m_{\cW} < 10^{-3}$ eV.

In GSM, the chirality boost-spin gauge boson $\cW_{\mu}^{a}$ interacts with all leptons and quarks through the spin gauge boson $\cA_{\mu}^{ab}$. The interaction strength is determined by the heavy spin gauge boson mass $M_{\cA}$, which depends on the free parameters $\gamma_{\kappa}$ and $g_4$, scaled by the fundamental mass scale $\bM_{\kappa}$. Through the heavy spin gauge boson, the dominant interactions of $\cW_{\mu}^{a}$ with leptons and quarks can be approximated by the following effective dimension-6 operator:
\be
\cO_{\cW}^{(6)} \sim \frac{g_4^2}{4M_{\cA}^2} \chih_{c}^{\; \mu} \chih^{\nu\sigma} \cW_{\sigma}^{\; b} (\p_{\mu} \cW_{\nu}^{\; a}  - \p_{\nu} \cW_{\mu}^{\; a})  \sum_{\psi_i} \bar{\psi}_i \{ \Sigma_{ab}, \gamma^c \} \psi_i ,
\ee
where $\psi_i$ represents all leptons and quarks in the SM, $\psi_i \equiv \nu_i, e_i, u_i, d_i$ for the three families $i=1,2,3$. 

Notably, after fixing the chirality boost-spin gauge, the boson $\cW_{\mu}^{a}$ behaves as a bi-covariant vector field. In analogy to the massless graviton associated with the gravigauge field $\chi_{\mu}^{\; a}$, we may characterize this dark matter candidate $\cW_{\mu}^{a}$ as a massive dark graviton. 

The massive dark graviton $\cW_{\mu}^{a}$ exhibits self-interactions mediated by both the spin gauge boson $\cA_{\mu}^{ab}$ and the chiral conformal-spin gauge boson $\cB_{\mu}$. These interactions generate an effective dimension-6 operator for dark graviton self-coupling. In the low-momentum regime, $q^2 \ll M_{\cA}^2$, the spin gauge boson mediation yields:
\be
\cO_{\cW}^{(\cG)} & \sim & \frac{g_4^2}{M_{\cG}^2} \frac{1}{3} \chih^{\rho\sigma} (2\cW_{\mu\rho}^{a} \cW_{\sigma}^{\; b} \chih^{\mu\; c}  - \cW_{\mu\rho}^{b} \cW_{\sigma}^{\; c} \chih^{\mu\; a} - \cW_{\mu\rho}^{c} \cW_{\sigma}^{\; a} \chih^{\mu\; b} ) \cW_{\nu\rho' a } \cW_{\sigma' b } \chih_{c}^{\; \nu} \chih^{\rho'\sigma'} , \nn \\
\cO_{\cW}^{(\cA)} & \sim & \frac{g_4^2}{M_{\cA}^2} \frac{1}{3} \chih^{\rho\sigma} (\cW_{\mu\rho}^{a} \cW_{\sigma}^{\; b} \chih^{\mu\; c}  + \cW_{\mu\rho}^{b} \cW_{\sigma}^{\; c} \chih^{\mu\; a} + \cW_{\mu\rho}^{c} \cW_{\sigma}^{\; a} \chih^{\mu\; b} ) \cW_{\nu\rho' a } \cW_{\sigma' b } \chih_{c}^{\; \nu} \chih^{\rho'\sigma'} ,
\ee
with $\cW_{\mu\nu}^{a} \equiv \p_{\mu}\cW_{\nu}^{\; a} - \p_{\nu}\cW_{\mu}^{\; a}$. Similarly, the chiral conformal-spin gauge boson mediates an effective interaction described by: 
\be
\cO_{\cW}^{(\cB)} & \sim & g_c^2 \eta^{\mu\rho} \int d^4q  \frac{e^{i q\cdot (x-y)}}{q^2 - m_{\cB}^2} \chih^{\nu\nu'}(x) (\p_{\mu} \cW_{\nu}^{\; a}(x)  - \p_{\nu} \cW_{\mu}^{\; a}(x)) \nn \\ 
& & \quad \quad  \cW_{\nu' a}(x)  \chih^{\sigma\sigma'}(y)  (\p_{\rho} \cW_{\sigma}^{\; b}(y)  - \p_{\sigma} \cW_{\rho}^{\; b}(y))  \cW_{\sigma' b} (y),
\ee
where the interaction strength depends on the chiral conformal-spin gauge boson mass relative to the exchanged momentum.

The massive scaling gauge boson $\mW_{\mu}$ appears as a promising heavy dark matter candidate. Its mass is primarily determined by the free parameters $g_w$ and $\beta_{\kappa}$, multiplied by the fundamental mass scale $\bM_{\kappa}$ for the relatively small VEVs $v_w$ and $v_e$. The properties of this heavy scaling gauge boson as a dark matter candidate have been comprehensively analyzed in references\cite{TW1,TW2}.

The gauge boson $\cB_{\mu}$ and the scalar boson $\varphi_w$ may also serve as potential dark matter candidates, depending on their mass spectrum and decay properties. When their masses are lighter than both the dark graviton and neutrinos, being determined by the VEV $v_{w}$ along with the parameters $\gamma_w$, $\lambda_w$ and $g_c$, these particles become stable and viable as dark matter. Alternatively, even with relatively larger masses, these particles could still act as dark matter candidates if their decay rates (mediated through dark graviton and spin gauge boson interactions) are sufficiently suppressed to yield lifetimes exceeding the age of the universe.
 
In conclusion, a key prediction of the GSM is the emergence of a stable, massive bi-covariant vector gauge boson from the chirality boost-spin sector, which we term the dark graviton. We advance this particle as a compelling candidate for dark matter. Its feeble interactions with the SM sector are mediated by the heavy spin gauge field, while its self-interactions are facilitated by both  spin-related gauge bosons and scalar fields.

%%%%%%%%

\section{Primordial Potential Energy, Dynamical Dark Energy, and Cosmic Evolution in GSM}

We now turn to a discussion of the scalar fields $\phi_e$ and $\phi_w$ and examine their properties. With the redefinition of the scaling gauge boson $\mW_{\mu}$ given in Eqs.(\ref{SGB1}) and (\ref{SGB2}), the kinetic terms of the scalar fields can be expressed as:
\be \label{SG}
\cL_{\phi}^{(1)} & = & \frac{1}{2} \chih^{\mu\nu}  ( \p_{\mu}\phi_{e} \p_{\nu}\phi_{e}   +   \p_{\mu}\phi_{w} \p_{\nu}\phi_{w}  ) \nn \\
& - & \frac{1}{8} \chih^{\mu\nu}   \frac{\p_{\mu}(\phi^2_{e} + \phi_w^2) \p_{\nu}(\phi_e^2 +\phi_{w}^2) }{\beta_{\kappa}^2  \bM_{\kappa}^2 + \phi_e^2 +\phi_w^2}  .
\ee
The relevant interactions of the scalar fields are given by:
\be
\cL_{\phi}^{(2)} & = & \frac{1}{4}( \gamma_{\kappa}^2\bM_{\kappa}^2 + \beta_e\phi_{e}^2) \bchi^{\mu\nu\mu' \nu'}\cG_{\mu\nu}^{a}\cG_{\mu'\nu' a} \nn \\
& + & \frac{1}{2} g_w^2(\beta_{\kappa}^2  \bM_{\kappa}^2 + \phi_e^2 +\phi_w^2)  \chih^{\mu\nu} \mW_{\mu}\mW_{\nu} \nn \\
& + &  \frac{1}{2}  \chih^{\mu\nu}  \phi_w^2 (g_c^2\gamma_w^2 \cB_{\mu} \cB_{\nu} - g_4^2\beta_{w}^2 \cW_{\mu}^{a}  \cW_{\nu a}  ).
\ee

To better analyze the scalar properties and their interactions, it is useful to introduce the following nonlinear representations:
\be
 & & \phi_e \equiv M_S \sinh(\frac{\phi_s}{M_S}) \sin (\chi_c), \nn \\
 & & \phi_w \equiv M_S \sinh(\frac{\phi_s}{M_S} ) \cos (\chi_c), 
 \ee
with the relations:
\be
& & \phi_w^2 + \phi_e^2 \equiv M_S^2 \sinh^2 \frac{\phi_s}{M_S}, \nn \\
& & \frac{\phi_e}{\phi_w} \equiv \tan \chi_c , \quad M_S \equiv \beta_{\kappa}\bM_{\kappa} .
\ee 
Here, $\phi_s$ and $\chi_c$ correspond to the redefined scalar fields in the non-linear representation. The Lagrangian for the non-linear scalar fields can then be rewritten into the following form:
\be \label{TSL}
\cL_{\phi} & \equiv & \cL_{\phi}^{(1)} + \cL_{\phi}^{(2)}  \nn \\
& = & \frac{1}{2} \chih^{\mu\nu}  ( \p_{\mu}\phi_{s} \p_{\nu}\phi_{s}   + M_S^2\sinh^2 \frac{\phi_s}{M_S}\,  \p_{\mu}\chi_{c} \p_{\nu}\chi_{c} ) \nn \\
& + &  \frac{1}{4}(\gamma_{\kappa}^2\bM_{\kappa}^2 + \beta_e M_S^2 \sinh^2\frac{\phi_s}{M_S} \sin^2 \chi_c )\, \bchi^{\mu\nu\mu' \nu'}\cG_{\mu\nu}^{a}\cG_{\mu'\nu' a} \nn \\\
& + &  \frac{1}{2}  M_S^2 \sinh^2\frac{\phi_s}{M_S} \cos^2\chi_c\, \chih^{\mu\nu}  (g_c^2\gamma_w^2 \cB_{\mu} \cB_{\nu} - g_4^2\beta_{w}^2 \cW_{\mu}^{a}  \cW_{\nu a}  ) \nn \\
& + &  \frac{1}{2} g_w^2 M_S^2(1 + \sinh^2 \frac{\phi_s}{M_S}) \chih^{\mu\nu} \mW_{\mu}\mW_{\nu}.
\ee 

We now examine the possible form of the scalar potential $\cV_S(\phi_w, \phi_e)$, which is not fully constrained by gauge symmetries alone. Beyond symmetry requirements, the scalar potential must satisfy an additional physical criterion: the fundamental scalar fields $\phi_w$ and $\phi_e$ should stabilize at a minimum with a non-zero VEV. Moreover, these fields must serve a dual cosmological purpose: providing both the primordial energy source for early universe inflation and the dynamical dark energy responsible for the current epoch's accelerated expansion.

The fundamental scalar fields $\phi_w$ and $\phi_{e}$ are postulated to simultaneously generate the initial potential energy driving inflation in the high-energy early universe, and produce the dynamical dark energy underlying today's cosmic acceleration. To incorporate these dual cosmological roles while enabling mass generation, we propose that the scalar potential comprises two distinct components:
\be
 \cV_S(\phi_w, \phi_e) \equiv \cV_S(\phi_s, \chi_c)=\cV_p(\phi_s) + \cV_d(\phi_s, \chi_c) ,
\ee
where $\cV_p(\phi_s)$ and $\cV_d(\phi_s, \chi_c)$ have the following specific forms:
\be
& & \cV_{p}(\phi_s) = \frac{1}{8}  \lambda_{P}^{4} M_{S}^4 \frac{ (\sinh^{2} \frac{\phi_s}{M_S} - \epsilon_p^2)^{2} }{( 1  + \lambda_p^{-2}\sinh^{2} \frac{\phi_s}{M_S})^{2}} ,   \nn \\
& & \cV_{d}(\phi_s, \chi_c)  =  \frac{1}{8}  \lambda_{D}^{4} M_{S}^4 \frac{ \sinh^{4} \frac{\phi_s}{M_S} \sin^{4}\chi_c + \epsilon_{c}^{4} }{( 1  + \lambda_d^{-2} \sinh^{2} \frac{\phi_s}{M_S} \sin^{2} \chi_c )^2 } , \nn \\
& &  \epsilon_p \equiv \frac{v_{p}}{M_S} , \quad  \epsilon_c \equiv \frac{\Lambda_{\kappa}}{M_S} ,
\ee
with $\lambda_P$, $\lambda_p$ and $\lambda_D$, $\lambda_d$ the coupling constants. Where  $v_p$ is expected to be a VEV with $\epsilon_p=v_p/M_{S}\ll 1$ and $\Lambda_{\kappa}$ represents the basic cosmological constant. 

We now illustrate how this scalar potential acts as the driving force behind inflationary expansion in the early universe. Consider the case where the nonlinear scalar field $\phi_s$ is significantly larger than the mass scale $M_S= \beta_{\kappa}\bM_{\kappa}$. When the following condition is satisfied:
\be
& & \sinh^2 \frac{\phi_s}{M_S} \gg  \lambda_p^2  \gg \epsilon_p^2, 
\ee
the scalar potential $\cV_p(\phi_s)$ asymptotically approaches to a constant value,
\be
& & \cV_p(\phi_s) \to \frac{1}{8}  \lambda_{P}^{4} \lambda_{p}^{4} M_{S}^4 = \frac{1}{8}  \lambda_{P}^{4} \lambda_{p}^{4} \beta_{\kappa}^4 \bM_{\kappa}^4 .
\ee
Similarly, for the scaling field $\chi_c$, if it satisfies:
\be
& &  \sinh^2 \frac{\phi_s}{M_S}\sin^2\chi_c \gg \lambda_c^2 \gg \epsilon_{c}^2 ,
\ee
its potential $\cV_d(\phi_s,\chi_c)$ also approaches a constant:
\be
& & \cV_{d}(\phi_s, \chi_c)  \to  \frac{1}{8}  \lambda_{D}^{4} \lambda_d^4 M_{S}^4 = \frac{1}{8}  \lambda_{D}^{4} \lambda_{d}^{4} \beta_{\kappa}^4 \bM_{\kappa}^4. 
\ee
In this regime, the scalar potential provides a nearly constant potential energy.

For the scalar field to act as a source of primordial energy driving early-universe inflation, its potential must satisfy the slow-roll conditions. We now examine how the evolution of the nonlinear scalar field $\phi_s$ fulfills these conditions. The slow-roll parameters are generally defined as:
\be \label{SRP1}
\epsilon_{V} \simeq  \bM_{\kappa}^2 \left( \frac{\cV_p'(\phi_s)}{\cV_p(\phi_s)} \right)^2  , \quad \eta_{V}  \simeq  \bM_{\kappa}^2 \frac{\cV_p''(\phi_s)}{\cV(\phi)} ,
\ee
where $\cV_p'(\phi_s)$ and $\cV_p''(\phi_s)$ denote the first and second derivatives of $\cV_p(\phi_s)$ with respect to $\phi_s$. Explicitly, these parameters are given by:
\be \label{SRP2}
\epsilon_V & \simeq & \frac{1}{\beta_{\kappa}^2} \frac{8 \sinh^2 \frac{2\phi_s}{M_S} (1+ \epsilon_p^2/\lambda_p^2)^2}{ (1+ \lambda_p^{-2}\sinh^{2} \frac{\phi_s}{M_S})^{2} (\sinh^{2} \frac{\phi_s}{M_S} - \epsilon_p^2)^{2} } , \nn \\ 
\eta_V & \simeq & \frac{1+ \epsilon_p^2/\lambda_p^2 }{\beta_{\kappa}^2} [ \frac{ 4 \cosh \frac{2\phi_s}{M_S}(\sinh^{2} \frac{\phi_s}{M_S} - \epsilon_p^2) + 2 \sinh^2 \frac{2\phi_s}{M_S} } { (1+ \lambda_p^{-2}\sinh^{2} \frac{\phi_s}{M_S}) (\sinh^{2} \frac{\phi_s}{M_S} - \epsilon_p^2)^{2}  } \nn \\
& & - \frac{ 6 \lambda_p^{-2} \sinh^2 \frac{2\phi_s}{M_S}}{ (1+ \lambda_p^{-2}\sinh^{2} \frac{\phi_s}{M_S})^{2} (\sinh^{2} \frac{\phi_s}{M_S} - \epsilon_p^2) }  ] .
\ee
Now, consider the case where the initial values of $\phi_s$ are of the order of the fundamental mass scale $\bM_{\kappa}$ and the parameter $\beta_{\kappa}$ is much smaller than the unity: 
\be
& & |\phi_s| \sim \bM_{\kappa}, \;\; \beta_{\kappa} \ll 1, \quad  |\phi_s|/M_S = 1/\beta_{\kappa} \gg 1, \nn \\
& & \sinh^2 \frac{\phi_s}{M_S} \sim  \frac{1}{4} e^{2|\phi_s|/M_S} =  \frac{1}{4}e^{2/\beta_{\kappa}}\gg 1 .
\ee
Under these conditions, the slow-roll parameters simplify to:
\be \label{SRP3}
& & \epsilon_V \simeq   \frac{2^8\lambda_p^4}{\beta_{\kappa}^2}  e^{-4/\beta_{\kappa}}  , \quad 
\eta_V \simeq    \frac{2^5\lambda_p^2}{\beta_{\kappa}^2}  e^{-2/\beta_{\kappa}}, \nn \\ 
& & \epsilon_V/\eta_V \simeq 8\lambda_p^2  e^{-2/\beta_{\kappa}} .
\ee
 
 The slow-roll process occurs when $\epsilon_V \ll 1$ and $\eta_V \ll 1$, leading to the following conditions:
\be \label{SRC1}
& & \sqrt{\beta_{\kappa}} e^{1/\beta_{\kappa}} \gg  4\lambda_p,  \quad  \epsilon_V \ll 1 ,  \nn \\
& &  \beta_{\kappa} e^{1/\beta_{\kappa}} \gg  4\sqrt{2} \lambda_p , \quad \eta_V \ll 1 , 
\ee
which indicates that the slow-roll conditions can be realized as long as $\beta_{\kappa}$ is sufficiently small, provided that $\lambda_p$ is of order unity or smaller. Notably, since $\epsilon_V$ is always smaller than $\eta_V$ (as indicated by eq.(\ref{SRP3})), once $\eta_V$ satisfies the slow-roll condition, $\epsilon_V$ automatically meets the requirement as well. 
 
As the scalar field $\phi_s$ decreases to small values, the slow-roll parameters grow to $\epsilon_V \sim 1$ and $\eta_V\sim 1$, violating the slow-roll conditions and causing the inflationary expansion of the early universe to terminate. The scalar boson $\phi_s$ may be referred to as an inflaton. 

Eventually, the potentials  $\cV_p(\phi_s)$ and $\cV_d(\phi_s, \chi_c)$ reach their minima at the following points:
\be
& & \langle \phi_s \rangle = v_s, \quad \langle \chi_c \rangle \equiv v_c/v_s, \nn \\
& & \sinh^{2} \frac{v_s}{M_S} \sin^{2} \frac{v_c}{v_s} = \frac{\epsilon_{c}^4}{\lambda_d^{2}} , \nn \\
& & \sinh^{2} \frac{v_s}{M_S} = \epsilon_p^2 , \;\; \sin^{2} \frac{v_c}{v_s}=  \frac{\epsilon_{c}^4}{ \lambda_d^{2}\epsilon_p^2} ,
\ee 
which implies that the initial scalar fields $\phi_e$ and $\phi_w$ acquire vacuum expectation values (VEVs):
\be
& & \langle \phi_e\rangle  \equiv v_e = M_S \sinh\frac{v_s}{M_S} \sin\frac{v_c}{v_s} = M_S  \frac{\epsilon_{c}^2}{\lambda_d} =  \frac{\Lambda_{\kappa}^2}{ \lambda_d\beta_{\kappa} \bM_{\kappa}}, \nn \\
& & \langle\phi_w \rangle \equiv v_w = \sqrt{ M_S^2 \epsilon_p^2 - v_e^2 } = v_p \sqrt{ 1 - 
v_e^2/v_p^2 } \simeq v_p - v_e^2/2v_p.
\ee

As $\epsilon_{c}$ and $\epsilon_p$ are very small parameters with $\Lambda_{\kappa}\ll v_p \ll M_S$, the following approximations hold:
\be
v_w \simeq v_s\simeq v_p, \quad v_e \simeq v_c \simeq \frac{ \Lambda_{\kappa}}{ \lambda_d\beta_{\kappa} \bM_{\kappa}} \Lambda_{\kappa}, 
\ee 
Consequently, the potential energy of the scaling field $\chi_c$ at the minimum point yields a cosmic energy density $\Lambda_D^4$:
\be \label{CED}
 \Lambda_D^4 \equiv \frac{1}{8}  \lambda_{D}^{4} M_{S}^4 \frac{v_e^4/M_S^4 + \epsilon_{c}^{4} }{( 1  + v_e^2/(\lambda_dM_S)^2)^2 } \simeq \frac{1}{8}  \lambda_{D}^{4} M_{S}^4 \epsilon_{c}^{4} \simeq \frac{1}{8}  \lambda_{D}^{4} \Lambda_{\kappa}^{4}  .
\ee
This energy density serves as dark energy, providing an explanation for the current accelerated expansion of the universe.

Around the minima of the potentials, the scalar fields $\phi_s$ and $\chi_c$ in the non-linear representation are expressed as:
\be
\phi_s = v_s + \varphi_s, \quad \chi_c = (v_c + \varphi_c)/v_s .
\ee 
To a good approximation, the following relations hold:
\be
& & \phi_w \simeq v_s + \varphi_s  - 2\varphi_c v_c/v_s + O(v_c^2/v_s, \varphi_c^2/v_s) , \nn \\
& & \phi_e \simeq v_c + \varphi_c + \varphi_s v_c/v_s + O(v_c^2/v_s, \varphi_c\varphi_s/v_s) , \nn \\
& & \phi_w \equiv v_w + \varphi_w, \quad  \varphi_w \simeq \varphi_s , \nn \\
& & \phi_e \equiv v_e + \varphi_e , \quad  \varphi_e \simeq \varphi_c , 
\ee
which indicates that the leading kinetic terms of $\varphi_s$ and $\varphi_c$ from Eq.(\ref{TSL}) take the normalized standard form:
\be \label{TSL}
\cL_{\varphi} \simeq \frac{1}{2} \chih^{\mu\nu}  ( \p_{\mu}\varphi_{s} \p_{\nu}\varphi_{s}   +  \p_{\mu}\varphi_{c} \p_{\nu}\varphi_{c} ) .
\ee

From the scalar potentials, the masses of the scalar fields $\varphi_s$ and $\varphi_c$ (or, equivalently,  $\varphi_w$ and $\varphi_e$ to a good approximation) are obtained as:
\be
& & m_{\varphi_s} \simeq \lambda_P^2 v_s \simeq \lambda_P^2 v_p \simeq m_{\varphi_w} , \nn \\
& & m_{\varphi_c} \simeq \sqrt{2} \lambda_D^2 v_c \simeq \sqrt{2} \lambda_D^2 v_e \simeq \frac{\sqrt{2} \lambda_D^2 }{ \lambda_d\beta_{\kappa} }\frac{\Lambda_{\kappa}^2}{\bM_{\kappa}}  \simeq m_{\varphi_e} .
\ee
The scalar boson $\varphi_c$ (or $\varphi_e$) acquires an extremely small cosmic mass $m_{\varphi_c}$ (or $m_{\varphi_e}$), which serves as a canonical field for dark energy. We refer to this scalar boson as the dark cosmino. 

Since the scalar potentials cannot be uniquely determined from symmetry principles alone, a rigorous and general investigation of the dynamical properties of both $\varphi_s$ (as the inflaton) and $\varphi_c$ (as the dark cosmino) would be particularly valuable.

The GSM inherently accounts for key cosmological phenomena. The model features a scalar inflaton field whose potential satisfies the slow-roll conditions, providing a mechanism for early-universe inflation. Additionally, dark energy is explained by the potential of another scalar field with a small cosmological energy, termed the dark cosmino. This field yields a tiny but finite cosmic mass, effectively functioning as a dynamical cosmological constant.

%%%%%%%

\section{Conclusions and Discussion}

We have presented a comprehensive theoretical framework for the GSM by systematically developing its foundations based exclusively on the intrinsic properties of leptons and quarks within GQFT. Utilizing a left-right symmetric chiral representation of leptons and quarks with a chiral duality symmetry $Z_2$, we explicitly demonstrate an extended gauge symmetry structure associated with intrinsic spin properties, namely, the inhomogeneous spin gauge symmetry and the scaling gauge symmetry, WS$_c$(1,3)$\times$GS(1). This structure naturally generalizes the SM's gauge symmetries, U$_Y$(1)$\times$SU$_L$(2)$\times$SU$_C$(3).

By applying the gauge invariance principle and introducing the corresponding gauge fields, including the spin-associated gravigauge field, we have constructed a gauge-invariant and chiral-duality-invariant action. This formulation leads to the emergence of a local orthogonal gravigauge spacetime, which constitutes a spin-associated intrinsic non-commutative spacetime, fundamentally distinct from coordinate-based Minkowski spacetime. The non-commutative structure of $\fG_4$ manifests gravitational effects through the gravigauge field strength $\hsF_{cd}^a$ or, equivalently, the spin gravigauge field $\hmOm_{c}^{ab}$ (with $\hmOm_{[cd]}^{a}\equiv \hsF_{cd}^a$). The derived gravitization equation (via the gravigauge field strength constraints in $\fG_4$) reveals how gravigauge field strength emerges from collective gauge field dynamics. This provides crucial insights into the quantum nature of gravity.

The GSM formalism within GQFT provides a framework that naturally incorporates all four known fundamental interactions, electromagnetic, weak, strong, and gravitational, while predicting novel ones, including a spin gauge force, as well as chirality boost-spin and chiral conformal-spin gauge forces, in addition to scaling gauge and scalar interactions. Within this framework, we derive general gauge-type gravitational equations to describe the gravidynamics of the GSM.

The GQFT-based GSM is background-independent, due to an emergent hidden general linear group symmetry GL(1,3,R), yet it admits globally flat Minkowski spacetime as a base spacetime. Furthermore, translational invariance in Minkowski spacetime leads to an energy-momentum cancellation law in GQFT, replacing the conventional conservation law of QFT, which allows for a proof of the zero energy-momentum tensor theorem in the GSM.

Our analysis reveals special gauge-fixing conditions for the extended gauge symmetries WS$_c$(1,3)$\times$GS(1). Mass generation for fundamental fields occurs via spontaneous symmetry breaking, with the notable exception of the gravigauge field, which behaves as a Goldstone-type boson. The chirality boost-spin gauge field is identified as a stable, massive dark graviton due to a discrete $Z_2$ symmetry, making it a viable dark matter candidate. This dark graviton interacts with all leptons and quarks through spin gauge field mediation and exhibits self-interactions via multiple channels, including mediation by the spin gauge field and interactions through the chiral conformal-spin gauge and scalar fields.

We also investigate how a fundamental scalar field can serve as an inflaton with a slow-roll potential, providing the primordial potential energy necessary for early-universe inflation. Simultaneously, we demonstrate how another scalar field acts as a dark cosmino, generating a cosmological constant while acquiring a tiny cosmic mass that accounts for canonical dark energy. However, the inherent limitation that scalar potentials cannot be uniquely determined from symmetry principles alone remains an unresolved issue requiring further investigation. 

Founded on fundamental gauge symmetries, this framework offers a unified description of gravitational, cosmological, and elementary particle phenomena. Beyond recovering established physics, it yields testable predictions for new interactions beyond the SM and for cosmic effects, including novel gravitational wave polarizations, the nature of dark matter, inflation, and dynamic dark energy. Notably, several predictions are directly testable. For instance, future space-based gravitational wave observatories, such as LISA \cite{LISA}, Taiji \cite{Taiji}, and Tianqin \cite{Tianqin}, will be able to probe the new gravitational wave polarization predicted by GQFT.

The GSM establishes a rigorous theoretical foundation for probing new physics frontiers, including: quantum gravity phenomenology, dark sector dynamics, early-universe cosmology, and the unification of fundamental forces. The present study also highlights several key directions for future research: phenomenological exploration of the predicted novel interactions, quantitative predictions for dark matter detection, cosmological modeling of inflation and dark energy scenarios, and a deeper investigation into the mathematical implications and physical treatment of non-commutative gravigauge spacetime structures within GQFT.

In summary, while the standard model of particle physics has achieved remarkable success, its theoretical framework encompasses only three fundamental interactions: electromagnetic, weak, and strong, without incorporating gravity into a unified description. Furthermore, it struggles to explain key cosmological phenomena such as dark matter, dark energy, and cosmic inflation. GR interprets gravity as a curvature effect of spacetime geometry, while quantum field theory describes the other three interactions through quantum fields propagating in flat spacetime. Within the framework of GQFT, the GSM reformulates gravity as a gauge-like interaction associated with the intrinsic spin gauge symmetry of elementary particles. This theoretical construction places gravity on equal footing with the electromagnetic, weak, and strong interactions, grounding all four fundamental forces in the intrinsic properties of matter's basic constituents and their underlying gauge symmetries. By reinterpreting the nature of spacetime and gravity, GSM reintegrates gravity into a unified framework of particle interactions, transcending the limitations of traditional geometric descriptions and opening new pathways toward a quantum theory of gravity.

The GSM framework integrates the fundamental laws governing microscopic particles and the evolutionary history of the macroscopic universe within a single theoretical architecture. This unification ensures that the laws of microscopic particle behavior and the dynamics of cosmic evolution share a common theoretical origin and maintain internal logical consistency. The theory predicts a series of new particles and interaction forms, among which the dark graviton serves as a candidate particle for dark matter. This prediction provides a first-principles theoretical foundation for dark matter research, advancing beyond traditional phenomenological descriptions. Moreover, the new scalar fields introduced by the theory exhibit unique dynamical properties that can naturally drive both the early inflationary phase of the universe and its current accelerated expansion. This offers a fresh perspective for revealing the microscopic nature of dark energy, moving beyond simple attribution to a cosmological constant.

The GSM represents more than an extension of existing theories, it is a foundational endeavor  to reconstruct physical reality based on deeper symmetry principles. It aims to establish a unified, self-consistent theory from first principles that simultaneously explains the unified description of fundamental interactions and the evolution of the universe. Should its predictions of new interactions and novel gravitational wave polarizations be experimentally confirmed, our understanding of the universe, from the subatomic realm to the cosmic scale, would be fundamentally transformed.

\vspace{1.0cm}

\centerline{{\bf Acknowledgement}}
\vspace{0.2cm}

This work was supported in part by  the National Key Research and Development Program of China under Grant No.2020YFC2201501, the National Science Foundation of China (NSFC) (Grants Nos. 12547104, 12441504, 12147103 and 11821505), and the Strategic Priority Research Program of the Chinese Academy of Sciences. I would like to thank Q.X. Zhang for carefully reading and checking the manuscript.

\vspace{0.5cm}
\section*{Conflict of interest}

The author declares that he has no conflict of interest.

%%%%%%%%%%%%%%%%%%%
\vspace{1.0cm}

\centerline{\bf Supplementary Materials}
\vspace{1.0cm}

The Supplementary Materials provide a detailed description of the enlarged symmetries within the chiral duality formulation of the Standard Model (SM) as presented in the paper ``Theoretical Foundations of the General Standard Model: A Unified Framework for Particle Physics and Cosmology." It also illustrates how this chiral duality formulation reduces to the standard formalism found in the existing literature. The presentation is designed to assist readers who may be less familiar with relativistic quantum field theory for spinor fields or with the Standard Model of particle physics.

\subsection{Enlarged Symmetries in the Chiral Duality Formulation of the SM}

As indicated in eq.(27) of the content, the chiral duality formalism of the SM action presented in eq.(9) possesses an enlarged global symmetry:
\be \label{GSMAS}
\mG_{GSM} & = &  \mbox{SC}(1)\times \mbox{PO}(1,3) \adjoin \mbox{WS}_c(1,3) \times \mbox{SG}(1) \times Z_2, \nn \\
& \equiv & \mbox{SC}(1)\times \mbox{P}^{1,3}\ltimes \mbox{SO}(1,3) \adjoin \mbox{SP}(1,3)\rtimes \mbox{W}^{1,3} \rtimes \mbox{SP}_c(1,1) \times \mbox{SG}(1) \times Z_2 . \nn
\ee
For clarity, we now list the constituent symmetries explicitly: $SO(1,3) \adjoin SP(1,3)$, $SC(1)\adjoin SG(1)$, $W^{1,3}$ , $SP_c(1,1)$ and $Z_2$. The coordinate translation symmetry $P^{1,3}$ is omitted here, as it is manifest. 

%%%%%%%

\subsubsection{Chiral Duality $Z_2$ Symmetry}

We begin by examining the chiral duality $Z_2$ symmetry. Under the $Z_2$ operation, the equivalent chiral representations of leptons and quarks transform as:
\be
& & \Psi_{\mp}^{c_d} \equiv {\cal C}_{d} \Psi_{\mp} {\cal C}_{d}^{-1} = C_d\Psi_{\mp} = \Psi_{\pm} . \nn
\ee
Consequently, the projection operators transform into their counterparts: 
\[
C_d \Gamma_{\mp} C_d^{-1} = \Gamma_{\pm}, \quad C_d \tilde{\Gamma}_{\mp} C_d^{-1} = \tilde{\Gamma}_{\pm}, \quad \text{and} \quad C_d \hat{\Gamma}_{\mp} C_d^{-1} = \hat{\Gamma}_{\pm}.
\]
This ensures that the covariant derivative and the Higgs field also transform into their chirality-dual parts:
\be
D_{\mu}^{(\Psi_{\mp})c_d}\equiv C_{d}D_{\mu}^{(\Psi_{\mp})} C_{d}^{-1}  \equiv D_{\mu}^{(\Psi_{\pm})}, \quad \Phi_{\mp}^{c_d}\equiv C_{d}\Phi_{\mp} C_{d}^{-1}  \equiv \Phi_{\pm} . \nn
\ee
Since the total action in Eq.(9) is a sum over both $s=\mp$ terms, the entire Lagrangian is invariant under this chiral duality, exhibiting the $Z_2$ symmetry.

Furthermore, the group generators of WS$_c$(1,3) corresponding to the group symmetries, SP(1,3), W$^{1,3}$ and SP$_c$(1,1), transform into their equivalent chiral-dual counterparts:
\be
\Sigma_{+}^{\ta\tb}  \equiv (\Sigma^{a b}, \Sigma_{+}^{a}, \Sigma_{+})  = C_d\Sigma_{-}^{\ta\tb}C_d^{-1} \equiv C_d(\Sigma^{a b}, \Sigma_{-}^{a}, \Sigma_{-}) C_d^{-1} .\nn
\ee
This transformation underlies the enlarged global symmetries present in the chiral duality formulation of the SM action. We will now verify this explicitly.

%%%%%%%

\subsubsection{ $SO(1,3)\adjoin SP(1,3)$ Associated Symmetry}

We now demonstrate the associated symmetry $SO(1,3)\adjoin SP(1,3)$. It is useful to first outline the transformation properties of the field components under this symmetry:
\be
& & \Psi_{s}(x) \to \Psi'_{s}(x') = e^{i\varpi_{ab}\Sigma^{ab}} \Psi_{s}(x) \equiv S(\Lambda) \Psi_{s}(x) , \nn \\
 & & x^{\mu} \to x'^{\mu} = L^{\mu}_{\; \nu} x^{\nu},\;\; S^{-1}(\Lambda) \Gamma^{a} S(\Lambda) = \Lambda_{\; b}^{a}\Gamma^{b}, \nn \\
& & \Lambda_{\; b}^{a}, L^{\mu}_{\; \nu} \in SO(1,3), \;\;  S(\Lambda) \in SP(1,3). 
\ee
Let us explicitly verify the invariance of the chiral spinor kinetic term:
\be
    \mathcal{L'}(x') & = & \bar{\Psi}^\prime_{s}(x')\Sigma_{s}^{a}\delta_{a}^{\mu}D'_{\mu}\Psi'_{s}(x') \nn \\
    &=& (\bar{\Psi}_{s} S^{-1}) \Sigma_{s}^{a} \delta_{a}^{\mu} \left( (L^{-1})^{\; \nu}_{\mu} D_{\nu} \right) (S \Psi_{s})\nn \\
    &= & \bar{\Psi}_{s} (S^{-1} \Sigma_{s}^{a} S) \delta_{a}^{\mu} (L^{-1})^{\; \nu}_{\mu} D_{\nu} \Psi_{s} \nn \\
    &=& \bar{\Psi}_{s} (\Lambda_{\; b}^{a} \Sigma_{s}^{b}) (L^{-1})^{\; \nu}_{a} D_{\nu} \Psi_{s} \nn \\
    & \equiv & \bar{\Psi}_{s} \Sigma_{s}^{b} \delta_{b}^{\nu} D_{\nu} \Psi_{s} = \mathcal{L}(x) .
\ee
The last equality follows from the identity, 
\be
\Lambda_{\; b}^{a} (L^{-1})^{\; \nu}_{a}  \equiv \delta_{b}^{\nu} ,
\ee
which implies the relation:
\be
\Lambda_{\; b}^{a} \equiv L_{\; b}^{a}  = \eta_{\mu}^{a} L^{\mu}_{\; \nu} \eta_{b}^{\nu} .
\ee

This indicates that the Lorentz transformations for the coordinates (governed by SO(1,3)) and the spin transformations for the spinor fields (governed by SP(1,3)) must be applied simultaneously and consistently. In other words, they constitute an associated symmetry.

Given the commutation properties $[\Sigma^{ab},\Gamma_{\mp}]=[\Sigma^{ab},\hat{\Gamma}_{\mp}]=[\Sigma^{ab},\tilde{\Gamma}_{\mp}]=0$, all other terms, namely the Yukawa coupling term, Higgs kinetic term, Higgs potential term and gauge field terms, are all invariant under the $SO(1,3)\adjoin SP(1,3)$ transformation. This is because their respective generators are constructed from these projectors.

%%%%%%

\subsubsection{$SC(1)\adjoin SG(1)$ Associated Symmetry}

The transformation properties for all components under the scaling symmetry are as follows. The coordinates and derivatives scale as:
\be
   x^\mu \to \lambda^{-1}x^\mu, \quad d^4x \to \lambda^{-4}d^4x, \quad D^{(\Psi)}_\mu \to \lambda D^{(\Psi)}_\mu, \quad \mathcal{D}_\mu \to \lambda \mathcal{D}_\mu ,
 \ee
 while the basic fields transform as:
 \be 
  \Psi \to \lambda^{3/2}\Psi, \quad A_{\mu} \to  \lambda A_{\mu} , \quad \Phi \to \lambda \Phi .
  \ee
  Consequently, the scaling dimension of each term in the chiral duality formalism of the SM Lagrangian (Eq. 9) is given by:
 \be
 \mathcal{L}_{\psi,\text{kin}} \sim  \bar{\Psi}_{\mp} D_\mu \Psi_{\mp} \to (\lambda^{3/2}\bar{\Psi}_{\mp}) (\lambda D_\mu) (\lambda^{3/2}\Psi_{\mp}) = \lambda^{4} \mathcal{L}_{\psi,\text{kin}}, \nn
 \ee
 for the kinetic term of spinor fields, and
\be
 \mathcal{L}_{\text{gauge}} \sim  \text{Tr}[F_{\mu\nu}F^{\mu\nu}] \to \text{Tr}[(\lambda^2 F_{\mu\nu})(\lambda^2 F^{\mu\nu})] = \lambda^4 \mathcal{L}_{\text{gauge}}, 
 \ee
for the kinetic term of gauge fields. 

The scaling dimension for Higgs sector terms are given by:
\be
 \mathcal{L}_\Phi \sim    \text{Tr}[(D_\mu \Phi_{\mp})^\dagger (D^\mu \Phi_{\mp})] \to \text{Tr}[(\lambda^2 D_\mu \Phi_{\mp})^\dagger (\lambda^2 D^\mu \Phi_{\mp})] = \lambda^4 \mathcal{L}_\Phi, \nn
 \ee
 for the kinetic term of Higgs field, and  
\be
  \mathcal{L}_Y \sim \bar{\Psi}_{\mp} \Phi_{\mp} \Psi_{\mp} \to (\lambda^{3/2}\bar{\Psi}_{\mp}) (\lambda \Phi_{\mp}) (\lambda^{3/2}\Psi_{\mp}) = \lambda^{4} \mathcal{L}_Y .
 \ee
for the Yukawa coupling term, as well as,
\be
  (\text{Tr}[\Phi_{\mp}^\dagger \Phi_{\mp}])^2 \to (\text{Tr}[(\lambda \Phi_{\mp})^\dagger (\lambda \Phi_{\mp})])^2  = \lambda^4 (\text{Tr}[\Phi_{\mp}^\dagger \Phi_{\mp}])^2, \nn
  \ee 
for the quartic term of Higgs field.

Notably, the Higgs mass term breaks the scaling symmetry for $v_h \neq 0$,  as it transforms with a different weight:
\be
v_h^2 \text{Tr}[\Phi_{\mp}^\dagger \Phi_{\mp}] \to v_h^2 \text{Tr}[(\lambda \Phi_{\mp})^\dagger (\lambda \Phi_{\mp})] = \lambda^2(v_h^2 \text{Tr}[\Phi_{\mp}^\dagger \Phi_{\mp}]) . \nn
\ee
As mentioned in the text, SC(1) and SG(1) represent the associated homogeneous scaling symmetries acting on the coordinates and basic fields, respectively, with the exception of the Higgs mass term.

%%%%%%%%%%%%%%

\subsubsection{$W^{1,3}$ Chirality Boost-Spin Symmetry}

The chirality boost-spin symmetry $W^{1,3}$ transforms the chiral spinor fields $\Psi_s$ as:
\be
    \Psi_s \to \Psi'_s = U_s \Psi_s, \quad U_s \equiv e^{i\omega^a \Sigma_{as}}, 
\ee
where $\omega^a$ is a constant vector. The generators $\Sigma_{as}$ ($s=\pm$) obey the Abelian commutative relations, $[\Sigma_{as}, \Sigma_{bs}]=0$. It is straightforward to verify the commutation relations:
\be
 [\Sigma_{as},\hat{\Gamma}_{s}]=[\Sigma_{as},\tilde{\Gamma}_{s}]=0.
 \ee

The invariance of the Lagrangian under the $W^{1,3}$ transformation can be verified term by term.

For the spinor kinetic term:
\be
\bar{\Psi}_s \Sigma_{as} D_\mu \Psi_s & \to &  (\bar{\Psi}_s U_s^{-1}) \Sigma_{as} (U_s D_\mu \Psi_s) \nn \\
        &= & \bar{\Psi}_s (U_s^{-1} \Sigma_{as} U_s) D_\mu \Psi_s \nn \\
        &= & \bar{\Psi}_s \Sigma_{as} D_\mu \Psi_s , 
 \ee
which holds due to the commutation relation $[\Sigma_{as}, \Sigma_{bs}]=0$. 

For the Yukawa coupling term: 
\be
\bar{\Psi}_s \Phi_s (\tilde{\Gamma}_s \tilde{\lambda}^\Psi + \hat{\Gamma}_s \hat{\lambda}^\Psi) \Psi_s 
& \to & (\bar{\Psi}_s U^{-1}) \Phi_s (\tilde{\Gamma}_s \tilde{\lambda}^\Psi + \hat{\Gamma}_s \hat{\lambda}^\Psi) (U \Psi_s)  \nn \\
& = & \bar{\Psi}_s \Phi_s (\tilde{\Gamma}_s \tilde{\lambda}^\Psi + \hat{\Gamma}_s \hat{\lambda}^\Psi) (U^{-1} U \Psi_s) \nn \\
& = & \bar{\Psi}_s \Phi_s (\tilde{\Gamma}_s \tilde{\lambda}^\Psi + \hat{\Gamma}_s \hat{\lambda}^\Psi) \Psi_s, 
\ee
based on the commutation relations:
\be
[\Sigma_{as}, \Sigma_{h s }^{I}]=[\Sigma_{as}, \tilde{\Gamma}_s]=[\Sigma_{as}, \hat{\Gamma}_s]=0 .
\ee

Notably, the generators of the $W^{1,3}$ group symmetry are nilpotent, $(\Sigma_{as})^2=0$, as a consequence of the chirality property. This nilpotency simplifies the transformation operator to the linear form $U_s = 1 + i\omega^a \Sigma_{as}$.

Another interesting feature is the set of nilpotent-like relations $\Sigma_{as} \Sigma_{h s }^{I} = \Sigma_{h s }^{I} \Sigma_{as} = 0$. These properties streamline the verification of the $W^{1,3}$ symmetry's invariance. The demonstration becomes straightforward upon observing the following relations:
\be
& & \Sigma_{as}U_s = \Sigma_{as}(1 + i\omega^a \Sigma_{as}) = \Sigma_{as} , \nn \\
& & \Sigma_{h s }^{I} U_s = \Sigma_{h s }^{I}(1 + i\omega^a \Sigma_{as}) = \Sigma_{h s }^{I} .
\ee

In conclusion, the entire Lagrangian is invariant under the global $W^{1,3}$ transformation. This symmetry can be interpreted as a translation-like symmetry for chiral spinor fields, analogous to the coordinate translation symmetry $P^{1,3}$ in Minkowski spacetime.

%%%%%%%%

\subsubsection{$SP_c(1,1)$ Chiral Conformal-Spin Symmetry }

The chiral conformal-spin symmetry $SP_c(1,1)$, which acts on chiral spinor fields, requires the coordinates to undergo a simultaneous scaling transformation. This structure is analogous to the associated symmetry $SC(1)\adjoin SG(1)$ discussed previously.

The explicit transformations under $SP_c(1,1)$ are given by:
\be
 \Psi_{\mp}(x) \to \Psi'_{\mp}(x') = S_{\mp}(\varpi) \Psi_{\mp}(x) = e^{\mp \frac{\varpi}{2} \gamma_9} \Psi_{\mp}(x) = e^{\mp \frac{\varpi}{2} (\mp 1)} \Psi_{\mp} = e^{\varpi/2} \Psi_{\mp}(x), 
\ee
for the chiral spinor field, and 
\be
x^{\mu} \to x^{' \mu} = e^{-\varpi/3} x^{\mu} ,  
\ee
for the coordinates. Here, the transformation operator is $S_{\mp}(\varpi) = \exp(i\varpi\Sigma_{\mp}) = \exp(\mp \frac{\varpi}{2} \gamma_9)$, and we have used the property $\gamma_9 \Psi_{\mp} = \mp \Psi_{\mp}$.

The gauge and Higgs fields transform as:
\be
A_{\mu}(x) \to A'_{\mu}(x') = e^{\varpi/3} A_{\mu}(x) , \quad \Phi(x) \to \Phi'(x') = e^{\varpi/3} \Phi(x) , 
\ee
where $A_\mu$ represents all gauge fields.

By performing transformations analogous to those for the symmetry $SC(1)\adjoin SG(1)$ in the previous subsection, one can straightforwardly verify that the action is invariant under the chiral conformal-spin symmetry 
$SP_c(1,1)$, with the exception of the Higgs mass term.

It is important to emphasize that $SP_c(1,1)$ constitutes a chiral scaling transformation. This distinguishes it from the conventional scaling transformation of the symmetry $SC(1)\adjoin SG(1)$, which commutes with all other symmetries. In contrast, $SP_c(1,1)$ is a subgroup of the semi-direct product group WS$_c$(1,3) and does not commute with the chirality boost-spin symmetry W$^{1,3}$. This non-commutativity is explicitly revealed when WS$_c$(1,3) is considered as a local gauge symmetry acting on the chiral spinor field, which has been analyzed detailed in the content of this paper.

%%%%%%%%%%%%%%%%%%%%%%%%%%%%%%%%%%%

\subsection{Reduction to the Standard Formalism of the SM Lagrangian}

This section of the appendix demonstrates how the chiral duality formulation of the SM presented in this paper reduces to the standard SM Lagrangian found in the literature. The only exception is the neutrino Yukawa coupling term, which, consistent with experimental evidence, becomes massive and incorporates a leptonic mixing matrix. The detailed reduction is presented below.

\subsubsection{Equivalence of the Two Chiral Duality Sectors}

The action of the chiral duality formulation in Eq. (9) involves a sum over the chirality index 
$s=\mp$ with a prefactor of $1/2$. This formulation essentially uses two equivalent chiral spinor representations for leptons and quarks to describe the same physical content, thereby exhibiting a chiral duality symmetry $Z_2$. To reduce this to the standard SM formalism, it is sufficient to consider only one of the two sectors. For this purpose, the $s=-$ sector is selected.

\subsubsection{Fermion Kinetic Sector}

In the chiral duality representation, the fermion kinetic term is given by
\be
    \frac{1}{2} \bar{\Psi}_s \Sigma_s^a \delta_a^{\mu} i D_{\mu}^{(\Psi_s)} \Psi_s + \text{h.c.},
\ee
with the covariant derivative and SM group generators defined as:
\be
i D_{\mu}^{(\Psi_{\mp})} &=& i \partial_{\mu} + A_{\mu}^{(\Psi_{\mp})}, 
\quad \Psi_{\mp} = l_{\mp}, \; q_{\mp}, \nn \\
A_{\mu}^{(l_{\mp})} &=& B_{\mu} \Sigma_{Y \mp}^{(l)} + W_{\mu}^i \Sigma_{L \mp}^i, \nn \\
A_{\mu}^{(q_{\mp})} &=& B_{\mu} \Sigma_{Y \mp}^{(q)} + W_{\mu}^i \Sigma_{L \mp}^i + A_{\mu}^a T_a, \nn \\
\Sigma_{Y \mp}^{(\Psi)} &\equiv& \frac{1}{2} N^{(\Psi)} + \frac{1}{2} \hat{\Gamma}_{\mp}, \;\; N^{(l)} = -1, \; N^{(q)} = \frac{1}{3}, \nn \\
\Sigma_{L \mp}^i & \equiv & \Sigma^i \tilde{\Gamma}_{\mp}, \quad \hat{\Gamma}_{\mp} =  \hat{\gamma}_9 \tilde{\Gamma}_{\mp} . \nn
\ee

We now verify two key features to connect with the SM.

(1) Weak Isospin Coupling: The generators for the weak interaction are $\Sigma_{L-}^i = \Sigma^i \tilde{\Gamma}-$. The projector $\tilde{\Gamma}-$ ensures that only the left-handed doublet components participate, eliminating the right-handed parts. Consequently, right-handed fields acquire no $SU(2)_L$ gauge coupling, thereby reproducing the SM structure of weak interactions.

(2) Hypercharge Assignment: The $U_Y(1)$ generator is $\Sigma_{Y-}^{(\Psi)} = \frac{1}{2}N^{(\Psi)} + \frac{1}{2}\hat{\Gamma}_{-}$, where $\hat{\Gamma}_{-}$ is defined as $\hat{\Gamma}_{-}=\hat{\gamma}_9 \tilde{\Gamma}_{-}$.

For left-handed doublets, the term $\frac{1}{2}\hat{\Gamma}_{-}$ vanishes due to the action of $\tilde{\Gamma}_{-}$, yielding the hypercharges:
\be
    Y(l_L) = \frac{1}{2}N^{(l)} = -\frac{1}{2}, \quad Y(q_L) = \frac{1}{2}N^{(q)} = \frac{1}{6} .\nn 
 \ee
 
 For right-handed singlets, $\tilde{\Gamma}_-$ acts as the identity. The chirality matrix $\hat{\gamma}_9 = \sigma_0 \otimes \sigma_3 \otimes I_4$ then assigns $+1$ to the upper isospin component and $-1$ to the lower, resulting in the hypercharges:
\be
    Y(\nu_R) &=& \frac{1}{2}(-1) + \frac{1}{2}(+1) = 0, \quad Y(e_R) = \frac{1}{2}(-1) + \frac{1}{2}(-1) = -1, \nn \\
    Y(u_R) &= &\frac{1}{2}\left(\frac{1}{3}\right) + \frac{1}{2}(+1) = \frac{2}{3}, \quad Y(d_R) = \frac{1}{2}\left(\frac{1}{3}\right) + \frac{1}{2}(-1) = -\frac{1}{3}.
\ee

These results reproduce the correct SM hypercharges for all leptons and quarks. Upon expanding all terms in the $s=-$ chirality sector, the standard fermion kinetic Lagrangian is recovered:
\be
\mathcal{L}_{\text{F}} = \bar{l}_L^i \gamma^\mu i D_\mu^{(l_L)} l_L^i + \bar{l}_R^i \gamma^\mu i D_\mu^{(l_R)} l_R^i + \bar{q}_L^i \gamma^\mu i D_\mu^{(q_L)} q_L^i + \bar{q}_R^i \gamma^\mu i D_\mu^{(q_R)} q_R^i . \nn
\ee

%%%%%%%%%%%%%%%

\subsubsection{Gauge Kinetic Sector}

The kinetic terms for the gauge fields in the $s=-$ sector are:
\be
\cL_{G} = - \eta^{\mu\mu'} \eta^{\nu\nu'} \left( \frac{1}{8g_1^2} \Tr F_{\mu\nu}^{(l-)} F_{\mu'\nu'}^{(l-)} + \frac{1}{8g_2^2} \Tr F_{\mu\nu}^{(L-)} F_{\mu'\nu'}^{(L-)} + \frac{1}{g_3^2} \Tr F_{\mu\nu}^{(C)} F_{\mu'\nu'}^{(C)} \right), 
\ee
where the field strengths and generators are defined as:
\be
F_{\mu\nu}^{(Y)} &= & F_{\mu\nu} \Sigma_{Y\mp}^{(\Psi)}, \quad F_{\mu\nu} = \partial_\mu B_\nu - \partial_\nu B_\mu, \;\; (\Psi = l, q) ,\nn \\
F_{\mu\nu}^{(L)} &=& F_{\mu\nu}^i \Sigma_{L\mp}^i, \quad F_{\mu\nu}^i = \partial_\mu W_\nu^i - \partial_\nu W_\mu^i + \epsilon^{ijk} W_\mu^j W_\nu^k, \nn \\
F_{\mu\nu}^{(C)} &= & F_{\mu\nu}^\alpha T_\alpha, \quad F_{\mu\nu}^\alpha = \partial_\mu A_\nu^\alpha - \partial_\nu A_\mu^\alpha + C^{\alpha\beta\gamma} A_\mu^\beta A_\nu^\gamma, \nn
\ee
with the generators,
\be
& & \Sigma_{Y\mp}^{(\Psi)} \equiv \frac{1}{2} N^{(\Psi)} + \frac{1}{2} \hat{\Gamma}_{\mp}, \quad \Sigma_{L\mp}^i \equiv \Sigma^i \tilde{\Gamma}_{\mp}, \nn \\
& &  T_\alpha = \lambda_\alpha/2, \quad \Tr(\lambda_\alpha \lambda_\beta) = \delta_{\alpha\beta} . \nn
\ee

To recover the standard SM Lagrangian, we compute the trace normalizations for these generators.

(1) $SU_L(2)$ left-handed symmetry. From the explicit representation of the $\Gamma_I$ matrices and the definition $\Sigma_{JK} = \frac{i}{4}[\Gamma_J, \Gamma_K]$, we have:
\be
\Sigma^i = \frac{1}{2}(I_2 \otimes \sigma^i \otimes I_4), \nn
\ee
Since these matrices commute with $\tilde{\Gamma}_\pm = \frac{1}{2}(I \mp \tilde{\gamma}_9)$, we obtain,
\be
\Tr(\Sigma_{L\mp}^i \Sigma_{L\mp}^j) = \Tr(\Sigma^i \Sigma^j \tilde{\Gamma}_\mp) = \frac{1}{8} \Tr(I_2) \Tr(\sigma^i \sigma^j) \Tr(I_4)  = 2 \delta^{ij}, \nn
\ee
a result that follows directly from $\Tr(\sigma^i\sigma^j) = 2\delta^{ij}$ and $\Tr(\tilde{\Gamma}_\mp) = 8$. This leads to:
\be
\Tr[F_{\mu\nu}^{(L-)} F_{\mu'\nu'}^{(L-)}] = 2 F_{\mu\nu}^i F_{\mu'\nu'}^i . \nn
\ee

(2) $U(1)_Y$ hypercharge symmetry. The generator is given by $\Sigma_{Y\mp} = \frac{1}{2}N^{(\Psi)} + \frac{1}{2}\hat{\Gamma}_\mp$, with the properties $\hat{\Gamma}_+^2 = \tilde{\Gamma}_+$ and $\Tr(\hat{\Gamma}_\mp) = 0$. Hence:
\[
\Tr(\Sigma_{Y\mp}^{(\Psi)})^2 = \frac{1}{4}[ (N^{(\Psi)})^2 \Tr(I) + \Tr(\tilde{\Gamma}_+) ] = \frac{1}{4}(16 (N^{(\Psi)})^2 + 8).
\]
For $N^{(l)}=-1$,  this yields:
\be
\Tr[F_{\mu\nu}^{(l-)} F_{\mu'\nu'}^{(l-)}] = \left( \Tr(\Sigma_{Y-}^{(l)})^2 \right)  F_{\mu\nu} F_{\mu'\nu'} = 6 F_{\mu\nu} F_{\mu'\nu'}, \quad \Tr(\Sigma_Y^{(l)})^2 = 6 .
\ee

(3) $SU_C(3)$ color symmetry. Using the definition $T_\alpha = \lambda_\alpha/2$ and the normalization $\Tr(\lambda_\alpha\lambda_\beta) = \delta_{\alpha\beta}$, we have:
\be
\Tr[F_{\mu\nu}^{(C)} F_{\mu'\nu'}^{(C)}] = \frac{1}{4} F_{\mu\nu}^\alpha F_{\mu'\nu'}^\alpha, \quad \Tr[T_\alpha T_\beta] = \frac{1}{4}\delta_{\alpha\beta} . \nn
\ee

Substituting these trace normalizations into the gauge kinetic term gives: 
\be
\mathcal{L}_{G} &=& -\frac{1}{4} \left( \frac{1}{2g_1^2} (6 F_{\mu\nu}F^{\mu\nu}) + \frac{1}{2g_2^2} (2 F_{\mu\nu}^i F^{i\,\mu\nu}) \right) -\frac{1}{g_3^2} \left(\frac{1}{4} F_{\mu\nu}^\alpha F^{\alpha\,\mu\nu}\right) \nn \\
&= & -\frac{1}{4} \left( \frac{3}{g_1^2} F_{\mu\nu}F^{\mu\nu} + \frac{1}{g_2^2} F_{\mu\nu}^i F^{i\,\mu\nu} + \frac{1}{g_3^2} F_{\mu\nu}^\alpha F^{\alpha\,\mu\nu} \right).
\ee
Rescaling the gauge fields as:
\be
B_\mu \rightarrow \frac{g_1}{\sqrt{3}}B_\mu, \quad W^i_\mu \rightarrow g_2 W^i_\mu, \quad A^\alpha_\mu \rightarrow g_3 A^\alpha_\mu, \nn
\ee
we obtain the canonical kinetic terms: 
\be
\mathcal{L}_{G} = - \frac{1}{4} \eta^{\mu\mu'} \eta^{\nu\nu'} ( F_{\mu\nu} F_{\mu'\nu'} + F_{\mu\nu}^i F_{\mu'\nu'}^i + F_{\mu\nu}^\alpha F_{\mu'\nu'}^\alpha ). \nn 
\ee

Focusing on the electroweak sector,
\be
\mathcal{L}_{EW} = - \frac{1}{4} \eta^{\mu\mu'} \eta^{\nu\nu'} ( F_{\mu\nu} F_{\mu'\nu'} + F_{\mu\nu}^i F_{\mu'\nu'}^i ),
\ee
and comparing this to the conventional rescaling in the SM: 
\be
B_\mu \to g' B_\mu, \quad W_\mu^i \to g W_\mu^i , \nn
\ee
with the standard covariant derivative:
\be
iD_\mu = i\partial_\mu + g' Y B_\mu + g \frac{\sigma^i}{2} W_\mu^i + \dots , \nn
\ee
we identify the relations between the coupling constants:
\be
g'^2 = \frac{g_1^2}{3}, \quad g^2 = g_2^2. \nn
\ee

From the definition of the weak mixing angle:
\be
\sin^2 \theta_W = \frac{g'^2}{g^2 + g'^2}, \nn 
\ee
and assuming that at a certain high energy scale the electroweak interactions unify with an equal coupling: $g_1 = g_2$, we predict:
\be
    \sin^2 \theta_W = \frac{g_1^2/3}{g_2^2 + g_1^2/3} = \frac{1}{4}. \nn
\ee

It is important to note that $g_1 = g_2$ is a high-energy boundary condition. The prediction $\sin^2 \theta_W = 1/4$ holds at this unification scale.  As the energy scale $\mu$ runs down, the coupling constants $g_1$ and $g_2$ evolve differently under renormalization group flow. Consequently, $\sin^2 \theta_W(\mu)$ naturally deviates from $1/4$, becoming consistent with the experimental value of $\sin^2 \theta \approx 0.23153$ presented in Particle Data Group$[43]$.

%%%%%%%%%%%%%%%%

\subsubsection{Higgs Sector}

The kinetic term for the Higgs field in the $s=-$ sector is given by
\be
    \frac{1}{8} \eta^{\mu\nu} \Tr ((\mathcal{D}_\mu \Phi_-)^\dagger (\mathcal{D}_\nu \Phi_-) , 
\ee
where the Higgs field $\Phi_-$ is defined in terms of its scalar components $\phi_I$ as:
\[
\Phi_- = \sum_{I=5}^8 \phi_I \Sigma_{h-}^I, \quad \text{with} \quad \Sigma_{h-}^I \equiv \Gamma^I \Gamma_-, \quad \Gamma_- = \frac{1}{2}(1-\gamma_9) ,
\]
The standard Higgs doublet is constructed from these components as:
\[
H = \begin{pmatrix} H^+ \\ H^0 \end{pmatrix} = \begin{pmatrix} \phi_7 + i\phi_6 \\ \phi_5 - i\phi_8 \end{pmatrix},
\]
and satisfies $H^\dagger H = \sum_{I=5}^8 \phi_I^2$. From the generator definitions, we verify the trace identity:
\[
\Tr[(\Sigma_{h-}^I)^\dagger \Sigma_{h-}^J] = \Tr(\Gamma_- \Gamma^I \Gamma^J \Gamma_-) = 8\delta^{IJ}.
\]

Substituting the covariant derivative $D_\mu \Phi_- = \sum_I (D_\mu \phi_I) \Sigma_{h-}^I$ into the kinetic term yields:
\[
\Tr[(D_\mu \Phi_-)^\dagger D^\mu \Phi_-] = \sum_{I,J} (D_\mu \phi_I) (D^\mu \phi_J) \Tr[(\Sigma_{h-}^I)^\dagger \Sigma_{h-}^J] = 8 \sum_I (D_\mu \phi_I)^2.
\]
This establishes the equivalence with the canonical Higgs kinetic term:
\[
\frac{1}{8} \Tr[(D_\mu \Phi_-)^\dagger D^\mu \Phi_-] = (D_\mu H)^\dagger D^\mu H.
\]
Similarly, the Higgs potential reduces to the standard form:
\begin{equation}
    \mathcal{V}_h(\Phi_-, \Phi_+) \equiv \frac{1}{4} \lambda_h^2 \left( \frac{1}{16} \Tr(\Phi_-^\dagger \Phi_- + \Phi_+^\dagger \Phi_+) - v_h^2 \right)^2 = \frac{1}{4}\lambda_h^2(H^\dagger H - v_h^2)^2.
\end{equation}

%%%%%%%%%%%

\subsubsection{Yukawa Sector}

For the leptons ($\Psi=l$) in the chirality sector $s=-$, the Yukawa term is:
\be
\mathcal{L}_{\text{Yukawa}} \supset \bar{\Psi}_-^i \Phi_- (\tilde{\Gamma}_- \tilde{\lambda}_{ij}^l + \hat{\Gamma}_- \hat{\lambda}_{ij}^l) \Psi_-^j .
\ee

We now verify that this term correctly couples the left-handed doublet to the right-handed singlet and recovers the conventional Yukawa coupling matrices for leptons and quarks.

To demonstrate this, we consider the lepton case $\Psi = l$ and take the Higgs component $\Phi_- \rightarrow \phi_5 \Sigma^5_{h-}$. Using the projector $\Gamma_- = \text{diag}(P_L, P_R)$ and and the generator representation:
\be
 \Sigma_{h-}^5 = \begin{pmatrix} 0 & \gamma_5 P_R \\ -\gamma_5 P_L & 0 \end{pmatrix}, \quad P_{L,R} = \frac{1}{2}(1 \mp \gamma_5), \nn 
\ee 
we find that applying the projector $\tilde{\Gamma}_- = \text{diag}(0, I)$ gives
\be
\label{5th Higgs Field}
\Sigma_{h-}^5 \tilde{\Gamma}_- = \begin{pmatrix} 0 & \gamma_5 P_R \\ 0 & 0 \end{pmatrix} ,
\ee
This structure explicitly shows that the left-handed component of $\bar{\Psi}^i_-$ couples exclusively to the right-handed component of $\bar{\Psi}^j_-$.

In general, the Yukawa coupling matrix can be expressed in the weak isospin space as the follows:
\be
M_Y = \tilde{\lambda}^l I_2 + \hat{\lambda}^l \sigma_3.
\ee
For leptons, this matrix acts on the neutrino and electron isospin states via the projector $P_\nu = \frac{1}{2}(I + \sigma_3)$ and $P_e = \frac{1}{2}(I - \sigma_3)$, yielding the respective couplings: 
\be
 P_\nu M_Y P_\nu = (\tilde{\lambda}^l + \hat{\lambda}^l) P_\nu , \quad  P_e M_Y P_e = (\tilde{\lambda}^l - \hat{\lambda}^l) P_e , \nn
 \ee
 for the neutrino and electron components. The resulting coefficients correspond precisely to the mass terms defined in the paper:
\be
\lambda^\nu = \tilde{\lambda}^l + \hat{\lambda}^l; \quad \lambda^e = \tilde{\lambda}^l - \hat{\lambda}^l. \nn 
\ee
This analysis applies equally to the quark sector.

Consequently, with the full Higgs field, the formulation reproduces the standard Yukawa Lagrangian:
\be
\mathcal{L}_{\text{Yukawa}}=\bar{l}_L^i H \lambda_{ij}^e e_R^j
+ \bar{l}_L^i \tilde{H} \lambda_{ij}^\nu \nu_R^j + \bar{q}_L^i H \lambda_{ij}^d d_R^j
+ \bar{q}_L^i \tilde{H} \lambda_{ij}^u u_R^j+ \text{h.c.} .
\ee

\end{document}